\documentclass[12pt]{article}

\usepackage[utf8]{inputenc}
\usepackage{lmodern}
\usepackage[T1]{fontenc}

\usepackage[paper=a4paper, top=1in, bottom=1in, left=1in, right=1in]{geometry}
\usepackage{setspace}
\usepackage{listings}
\usepackage{xcolor}
\usepackage{bookmark}

\RequirePackage{latexrelease}
\RequirePackage{fixltx2e}

\usepackage{amsmath}
\DeclareMathOperator*{\argmin}{arg\,min} 
\usepackage{amssymb}
\usepackage{mathtools}
\usepackage{amsthm}
\usepackage[italicdiff]{physics}
\usepackage{mathtools}

\newtheorem{theorem}{Theorem}

\newtheorem{proposition}{Proposition}
\newtheorem{lemma}{Lemma}

\newtheorem{definition}{Definition}
\newtheorem{remark}{Remark}

\usepackage[toc,page,title]{appendix}

\usepackage{comment}

\usepackage[english]{babel}

\usepackage[sectionbib,round,longnamesfirst]{natbib}

\usepackage{hyperref}
\hypersetup{
    colorlinks=true,
    linkcolor=blue,
    filecolor=msendera,      
    urlcolor=cyan,
    citecolor=blue,
}

\usepackage[capitalise]{cleveref}
\crefname{figure}{Figure}{Figures}
\Crefname{figure}{Figure}{Figures}
\crefname{assumption}{Assumption}{Assumptions}
\crefname{footnote}{footnote}{footnotes}
\crefname{condition}{Condition}{Conditions}
\crefname{remark}{Remark}{Remarks}

\usepackage{tikz}
\usetikzlibrary{intersections}
\usepackage{pgfplots}
    \pgfplotsset{compat=1.15}
\usepackage{graphicx}
\usepackage{standalone}
\usepackage{subcaption}
\usepackage[italicdiff]{physics}
\usepackage{float}

\usepackage{caption}
\usepgfplotslibrary{fillbetween}

\newcommand{\HT}{\hat{T}}
\newcommand{\htt}{\hat{\theta}}
\newcommand{\bx}{\boldsymbol{x}}
\newcommand{\CS}{\mathcal{S}}
\newcommand{\BT}{\boldsymbol{T}}
\newcommand{\TT}{\tilde{T}}

\usepackage[toc,page,title]{appendix}

\newcommand{\nocontentsline}[3]{}
\newcommand{\tocless}[2]{\bgroup\let\addcontentsline=\nocontentsline#1{#2}\egroup}

\let\oldappendix\appendix
\makeatletter
\renewcommand{\appendix}{%
  \addtocontents{toc}{\let\protect\numberline\protect\appendixnumberline}%
  \renewcommand{\@seccntformat}[1]{Appendix~\csname the##1\endcsname\quad}%
  \oldappendix
}
\makeatother

\begin{document}

\begin{titlepage} 
		\title{Communication as Voting\\[0.5em]
{\normalfont\large
\href{https://arxiv.org/pdf/2505.14639}
{\textcolor{blue}{[Latest version]}}}}
		\author{Kailin Chen\thanks{Toulouse School of Economics, kailin.chen@tse-fr.eu. For invaluable feedback and support, the author is indebted to Mehmet Ekmekci, Daniel Hauser, Johannes Hörner, Stephan Lauermann, Pauli Murto, Sven Rady, and Juuso Välimäki. For helpful comments, the author thanks numerous seminar and conference audiences, as well as Nemanja Antic, James Best, Günnur Ege Bilgin, Jacopo Bizzotto, Nina Bobkova, Gregorio Curello, Francesc Dilm\'e, Georgy Egorov, Tangren Feng, Zhengqing Gui, Nathan Hancart, Bård Harstad, Carl Heese, Ryota Iijima, Shaofei Jiang, Doron Levit, Zizhen Ma, Nadya Malenko, Margaret Meyer, Benny Moldovanu, Inés Moreno de Barreda, Marco Ottaviani, Harry Pei, Fedor Sandomirskiy, Peter Norman Sørensen, Francesco Squintani, Egor Starkov, Wing Suen, and Wenhao Wu.}}
		\date{\today}
		\maketitle
		\begin{abstract} 
			\noindent \singlespacing This paper studies information aggregation in informal elections by examining a cheap-talk model with multiple senders and one receiver. Each sender observes a noisy private signal about the unknown state and sends a message; the receiver observes the message tally and chooses the final policy. We show that any positive probability of a disagreement state, in which the senders and the receiver have opposing policy preferences, precludes full information aggregation even as the number of senders grows without bound. The key observation is that, in equilibrium, communication can reveal only information that induces the same preferred policy for both the receiver and the senders. Furthermore, there is a discontinuity in the scope for information transmission: for an open set of parameter values, existing models permit full information aggregation, but perturbing them by introducing an arbitrarily small positive probability of a disagreement state leads to no information transmission once the sender population is sufficiently large. Finally, introducing a mediator can improve information transmission and restore efficiency.

			\vspace{0.5in}


		\end{abstract}
		\setcounter{page}{0}
		\thispagestyle{empty}

	\end{titlepage}

\pagenumbering{arabic}

\section{Introduction}
\label{sec:1}

Information relevant to a policy decision is often dispersed across a large population of individuals, and thus needs to be gathered, through a process such as a vote or a survey, before it can be incorporated into the decision. The modern Condorcet jury theorem \citep[e.g.,][]{feddersen1997voting,feddersen1998convicting} shows that formal elections---in which the vote tally directly determines the policy choice---are an effective means of aggregating such information: under any qualified-majority rule, the electoral outcome converges in probability to the full-information outcome as the number of voters grows large.

Unlike formal elections, many real-world processes that resemble voting are advisory and nonbinding: there is a decision-maker who observes the vote tally or message tally from the population but retains discretion over the final decision. Common examples of such processes include petitions, protests, polls, and surveys. Another example is a nonbinding vote taken by shareholders in a company on management decisions. These processes can be modeled as multi-sender cheap-talk games in which the receiver is not constrained by a prespecified decision rule; that is, the senders' message tally (corresponding to the vote tally in a formal election) does not mechanically determine the policy choice. Following \citet{ekmekci2024information}, we refer to these games as \textit{informal elections}.

This paper studies the efficiency of information transmission in informal elections. Specifically, we ask whether the logic of the modern Condorcet jury theorem extends to such settings: to what extent can informal elections aggregate information? Can the receiver fully learn the unknown state as the number of senders grows large?

Our main finding is that information aggregation fails if there is any conflict between the senders and the receiver that persists under full information. The intuition rests on the observation that equilibrium communication can reveal only information that induces the same preferred policy for the receiver and the senders. Otherwise, either a sender or the receiver would have an incentive to deviate. If information were fully aggregated, uncertainty would be eliminated. Any conflict that persists under full information would then lead to opposing policy preferences in some circumstances, contradicting this equilibrium requirement. Furthermore, we identify a discontinuity in the scope for information transmission relative to the predictions of existing models in which conflict arises only under uncertainty. Specifically, for an open set of parameter values, these models permit full information aggregation, but an arbitrarily small perturbation that introduces conflict persisting under full information leads to no information transmission once the sender population is sufficiently large.

We consider a model with $N$ senders and a single receiver who must decide whether to implement a proposal or maintain the status quo. All players' payoffs from the implementation of the proposal depend on an unknown state. The senders have common interests, but their interests conflict with those of the receiver. Each sender privately observes a noisy signal about the state, then either approves or rejects the proposal. The receiver observes the approval tally and makes the final decision.

Our model is related to the broader literature on communication from multiple senders with noisy private information, including \citet{wolinsky2002eliciting} and \citet{morgan2008information}. We take as our benchmark the setting studied by \citet{levit2011nonbinding} and \citet{battaglini2017public}, which we refer to collectively as LMB. The LMB model features two states, high and low, with the receiver and the senders agreeing on the preferred policy in each state. However, this alignment is only ordinal: the receiver's payoff differs in magnitude from that of the senders. Hence, when the state is uncertain, a conflict of interest emerges: the senders and the receiver have different thresholds of doubt for favoring the proposal over the status quo. \Cref{tab:1} illustrates this payoff structure. LMB obtain an all-or-nothing result: with conflict below a certain threshold, asymptotically full information aggregation is always possible; with conflict above the threshold, communication unravels---i.e., no information is transmitted.

\begin{table}
\centering
\begin{subtable}[b]{0.5\textwidth}
  \centering
  \begin{tabular}{| c| c| c|}
  \hline
   & high & low \\
  \hline
  Senders & $1$ & $-1$ \\
  \hline
  Receiver & $u\geq 1$ & $-1$ \\
  \hline
  \end{tabular}
  \caption{Proposal}
\end{subtable}
\hfill
\begin{subtable}[b]{0.45\textwidth}
  \centering
  \begin{tabular}{| c| c| c|}
  \hline
   & high & low \\
  \hline
  Senders & $0$ & $0$ \\
  \hline
  Receiver & $0$ & $0$ \\
  \hline
  \end{tabular}
  \caption{Status quo}
\end{subtable}
\caption{Example payoff structure (binary-state benchmark). The left table presents the senders' and receiver's payoffs from the implementation of the proposal in each state, while the right table presents their payoffs from the status quo. The value of $u$ determines the level of conflict between the senders' interests and the receiver's: the senders favor the proposal if they believe the probability of the high state exceeds $1/2$, whereas the receiver favors it if her belief exceeds $1/(1+u)$.}
\label{tab:1}
\end{table}

In this paper, we enlarge the LMB model by introducing a medium state in which the receiver prefers the proposal while the senders prefer the status quo. Thus, whereas the low and high states are agreement states, in which all players have the same policy preference, and conflict occurs only under uncertainty, the medium state is a disagreement state, in which conflict stems from opposing policy preferences and thus persists even under full information. This second form of conflict arises naturally in many real-world applications. For example, in a shareholder vote concerning the acquisition of an asset, the asset may yield moderate returns, in which case shareholders would oppose the acquisition but management would favor it because they will derive additional private benefits from it. Likewise, a protest or petition may concern a project with unknown future environmental costs; if these costs are moderate, the population may oppose the project while political leaders favor it, because the prospect of electoral turnover leads them to discount the costs more heavily. \Cref{tab:2} illustrates the payoff structure in this extended model.

\begin{table}
\centering
\begin{subtable}[b]{0.5\textwidth}
  \centering
  \begin{tabular}{| c| c| c| c| }
  \hline
   & high & medium & low \\
  \hline
  Senders & $1$ & $<0$ & $-1$ \\
  \hline
  Receiver & $u>1$ & $>0$ & $-1$ \\
  \hline
  \end{tabular}
  \caption{Proposal}
\end{subtable}
\hfill
\begin{subtable}[b]{0.45\textwidth}
  \centering
  \begin{tabular}{| c| c| c| c| }
  \hline
   & high & medium & low \\
  \hline
  Senders & $0$ & $0$ & $0$\\
  \hline
  Receiver & $0$ & $0$ & $0$\\
  \hline
  \end{tabular}
  \caption{Status quo}
\end{subtable}
\caption{Example payoff structure (with high, medium, and low states). The medium state is a disagreement state, whereas the low and high states are agreement states. }
\label{tab:2}
\end{table}

The addition of the disagreement state has stark consequences for information transmission. Our first main result, \cref{thm:3}, shows that information aggregation always fails in the presence of the disagreement state. Specifically, the probability that the receiver chooses a suboptimal policy even in the agreement states remains bounded away from zero as the number of senders grows large. Hence, even when the information dispersed across the senders would suffice to identify the state (i.e., even when the number of signals available is arbitrarily high), only a limited amount of this information can ever be transmitted to the receiver.

The failure of information aggregation follows from an \textit{equilibrium no-conflict} property, which can be understood as follows. When there is only one sender, the sender's incentive compatibility constraint and the receiver's obedience constraint ensure that, in equilibrium, the receiver's decision will be mutually
acceptable. We show that the same conclusion holds when there are multiple senders: in any informative equilibrium (i.e., one in which the senders' information affects the outcome), given the senders' equilibrium strategies, the senders and the receiver agree on the policy choice at every realized approval tally. Equivalently, the receiver's equilibrium strategy (i.e., her best response to the senders' equilibrium strategies) maximizes the expected payoffs of both the senders and the receiver across all receiver strategies. The equilibrium no-conflict property imposes an endogenous use-compatibility constraint: the senders' messages can reveal only information that induces the same preferred policy for both the receiver and the senders.

Information aggregation fails when there is a disagreement state because fully revealing communication would violate the equilibrium no-conflict property: some values of the approval tally would reveal that the realized state is the disagreement state, and in these cases the receiver and the senders would prefer opposing policies. Intuitively, when conflict persists under full information, highly informative communication is no longer harmless to the senders, because it may enable the receiver to choose a policy unfavorable to them. Such conflict thus creates an incentive for the senders to misrepresent their information, which gives rise to an endogenous bouncreatingeness of equilibrium communication.

The disagreement state can do more than prevent full information aggregation: under certain conditions, it causes communication to unravel completely. Our next main result, \cref{thm:4}, identifies a discontinuity in the scope for information transmission relative to the predictions of the LMB benchmark. Specifically, for an open set of parameter values under which the benchmark permits full information aggregation, perturbing it by assigning an arbitrarily small positive probability to the disagreement state leads to no information transmission once the sender population is sufficiently large. Hence, LMB's predictions regarding informative communication are not robust to the introduction of conflict that persists under full information.


The discontinuity arises because the two forms of conflict that we have described impose opposing requirements on the informativeness of communication in our game. As the conflict that exists only under uncertainty intensifies, communication must become more informative to reduce the uncertainty and mitigate the effects of the conflict. By contrast, the conflict that would persist under full information renders highly informative communication unsustainable, as already discussed, because of the equilibrium no-conflict property. When the first form of conflict reaches an intermediate level, these opposing requirements become incompatible, causing communication to unravel if the disagreement state occurs with any positive probability, however small.

Although the disagreement state always precludes full information aggregation and may even cause communication to unravel completely, a level of informative communication is nevertheless possible under certain conditions. We characterize these conditions and provide comparative statics on the amount of information transmitted. In particular, the informativeness of each sender’s message weakly decreases as either form of conflict intensifies.


After establishing our main results, we extend them to a more general model, with signal and message spaces of arbitrary finite size (rather than binary signals and messages as in the main model). Here, the equilibria of interest are those in which the senders' strategy is \emph{monotone}. We show that, when the number of senders is sufficiently large, the senders endogenously pool all but one signal realization on a common message, so that the equilibria in this model effectively reduce to equilibria of the main model (with binary signals and messages). The proof of this result develops a novel large-deviations approach that may be of independent interest to theorists studying communication and voting.

Finally, as our main results show that the presence of a disagreement state drastically limits cheap-talk communication, we ask whether introducing a mediator with commitment power can improve information transmission and restore efficiency. We assume that the mediator does not observe the senders' signals but can commit to a communication mechanism that collects their messages and provides a recommendation to the receiver. (The role of such a mediator may be played, for example, by a shareholder representative in the context of a nonbinding shareholder vote, or a polling firm or survey platform in the context of a poll or survey.) We show that, as the number of senders grows large, the mediator can asymptotically implement any outcome on the Pareto frontier between the receiver's first-best outcome and the senders' optimal Bayesian-persuasion outcome, using a sequence of mechanisms having a simple and intuitive step form.

This paper contributes to the cheap-talk literature and, more broadly, to the literature on information transmission with multiple senders. One branch of that literature \citep[e.g.,][]{milgrom1986relying, krishna2001model, battaglini2004policy, gentzkow2016competition, gentzkow2017bayesian, meyer2019robustness} examines situations in which senders possess identical information but have conflicting interests; these papers emphasize how competition affects information revelation. By contrast, we study a setting in which senders have aligned preferences but possess heterogeneous information; they attempt to coordinate their messages by drawing on their noisy private signals. This setting shares qualitative features with models in voting theory \citep[e.g.,][]{feddersen1997voting,feddersen1998convicting,myerson1998extended,dekel2000sequential, duggan2001bayesian} which examine information aggregation among voters under prespecified voting rules. The voting literature shows that dispersed information is effectively aggregated under a qualified-majority rule, but that aggregation generally fails under the unanimity rule.\footnote{Notably, \citet{razin2003signaling} studies majority voting
between two candidates, with the winner choosing policy in
response to information conveyed by the election results.
When candidates are sufficiently polarized, information fails to aggregate fully in certain symmetric equilibria, even as the electorate grows arbitrarily large. In our setting with a fixed receiver,
no sequence of symmetric equilibria aggregates information
whenever the disagreement state has positive probability,
however small.} We establish a connection to this result by showing that in our model, the receiver's equilibrium strategy resembles a near-unanimity rule, which undermines coordination among the senders and thereby prevents information aggregation. Unlike in formal voting, however, this near-unanimity rule is not imposed exogenously; it arises endogenously from the receiver’s best response and the senders’ equilibrium incentives.

This paper is also related to the literature on learning from rich data \citep[e.g.,][]{moscarini2002law,mu2021blackwell,frick2023belief,frick2024multidimensional,frick2024welfare}, which applies similar information-theoretic tools to evaluate learning efficiency with a large number of signals across a range of environments. We examine situations in which information sources (senders) are biased and strategically misrepresent their information, which undermines the power of large numbers and prevents complete learning.

Finally, because we study communication with a coarse message space, this paper is related to the literature on coarse information design \citep[e.g.,][]{lyu2023coarse}. It also contributes to the literature on mediated communication \citep[e.g.,][]{goltsman2009mediation,salamanca2021value,corrao2023bounds} and to the literature on mechanisms for eliciting information from multiple agents \citep[e.g.,][]{wolinsky2002eliciting,gerardi2009aggregation,best2025divide}. The latter two connections are discussed in detail in \cref{sec:8}.

\section{Model}
\label{sec:2}

\subsection{Model Setting}
\label{sec:2.1}

A single \textit{receiver} (she) must decide whether to implement a proposal or maintain the status quo. Before making the decision, she observes messages from a population of $N>1$ privately informed \textit{senders} (each referred to as he). Each sender simultaneously either approves or rejects the proposal. The receiver observes the approval tally $T\in\{0,\ldots,N\}$ and decides whether to implement the proposal. 

If the receiver maintains the status quo, all players’ payoffs are zero. If she implements the proposal, payoffs depend on an unknown state $\theta\in\{L,M,H\}$ (low, medium, or high). The receiver's payoff is $U_R(\theta)$, while every sender has the same payoff $U_S(\theta)$. The states $L$ and $H$ are \textit{agreement states}: both the receiver and the senders prefer the status quo in state $L$ and the proposal in state $H$, i.e.,
\begin{equation*}
    U_R(H)>0\text{ and } U_S(H)>0,
\end{equation*}
\begin{equation*}
    U_R(L)<0\text{ and } U_S(L)<0.
\end{equation*}
The state $M$ is a \textit{disagreement state}: the receiver prefers the proposal while the senders prefer the status quo, i.e.,
\begin{equation}
\label{disagreement}
    U_{R}(M)>0\text{ and } U_{S}(M)<0.
\end{equation}

The receiver and the senders share a common prior belief $q^0=(q^0_{L},q^0_M,q^0_{H})\in\Delta^3$ about the unknown state. We assume
$q^0_L>0$ and $q^0_H>0$ throughout. In \cref{sec:5} we study the benchmark case
$q^0_M=0$, while in \cref{sec:6} we study the case $q^0_M>0$.

Before choosing whether to accept or reject the proposal, each sender $i\in\{1,\ldots,N\}$ privately observes a noisy signal $s^i\in\{g,b\}$ (good or bad). Conditional on the state, the senders' signals are independent and identically distributed (i.i.d.), with
\begin{equation*}
    \rho_{\theta}=\mathbb{P}[s^i=g\mid \theta]\quad\;\forall \theta\in\{L,M,H\}.
\end{equation*}
The probability of receiving signal $g$ is strictly increasing in the state:
\begin{equation}
\label{eqn:1}
    0<\rho_{L}<\rho_M<\rho_H<1.
\end{equation}

\begin{remark}
    For simplicity, our main model has exactly three states, a binary message space (approve or reject), and a binary signal space. However, our analysis also applies when the state space is a finite set $\{\theta_1,\ldots,\theta_n\}\subset \mathbb{R}$, and the players' preferences are defined by cutoffs $\htt_R \neq \htt_S$, with the receiver preferring the proposal if $\theta>\htt_R$ and the senders preferring the proposal if $\theta>\htt_S$. Furthermore, in \cref{sec:7} we relax the assumptions that the message space and signal space are binary.
\end{remark}

\subsection{Sources of Conflict}
\label{sec:2.2}

Our model features two distinct forms of conflict between the receiver's preferences and the senders'. The first arises from the disagreement state $M$, in which the receiver prefers the proposal while the senders prefer the status quo. The magnitude of this conflict is captured by $q^0_M$, the prior probability of $M$.

The second form of conflict arises in the agreement states, $L$ and $H$, as follows. If the disagreement state $M$ is excluded, so that $\theta\in\{L,H\}$, then the receiver and the senders would always agree if they knew the state. However, when the state is unknown, differences between the players' payoff magnitudes can generate conflicting interests, even if the receiver and the senders hold a common belief. 
Specifically, given a belief $(q_L,q_{H})\in\Delta^2$ regarding $L$ and $H$, the receiver prefers the proposal if 
\begin{equation*}
    \frac{q_H}{q_{L}}>-\frac{U_R(L)}{U_R(H)},
\end{equation*}
while the senders prefer the proposal if 
\begin{equation*}
    \frac{q_H}{q_{L}}>-\frac{U_S(L)}{U_S(H)}.
\end{equation*}
We refer to the quantities $-U_R(L)/U_R(H)$ and $-U_S(L)/U_S(H)$ as the players' \emph{thresholds of doubt}. In the remainder of the paper, we assume that the receiver's threshold of doubt is less than or equal to that of the senders: 
\begin{equation}
\label{eqn:2}
    -\frac{U_R(L)}{U_R(H)}\leq-\frac{U_S(L)}{U_S(H)}.
\end{equation}
This condition is illustrated in \cref{fig:1}. It implies that, under any common belief about the agreement states $L$ and $H$, the receiver is (weakly) more willing to implement the proposal than the senders are.\footnote{We impose \eqref{eqn:2} only for expositional clarity. Our main results—\cref{thm:1}, \cref{thm:5}, and the failure-of-information-aggregation conclusion of \cref{thm:3}—do not depend on it.}
The magnitude of the players' conflict of interest is captured by the ratio of their thresholds of doubt,
\begin{equation*}
    \frac{-U_S(L)/U_S(H)}{-U_R(L)/U_R(H)}=\frac{U_S(L)}{U_S(H)}\cdot \frac{U_R(H)}{U_R(L)}\geq 1.
\end{equation*}

\begin{figure}
    \centering
    \begin{tikzpicture}
     \draw[black,-latex] (0,0)--(13,0);
     \draw[]  (13,0) -- (13,0) node[right]{$\frac{q_{H}}{q_{L}}$};
     \draw[]  (0,0.1) -- (0,-0.1) node[below]{$0$};
     \draw[]  (4,0.1) -- (4,-0.1) node[below]{$-\frac{U_R(L)}{U_R(H)}$};
     \draw[]  (9,0.1) -- (9,-0.1) node[below]{$-\frac{U_S(L)}{U_S(H)}$};
          \draw[] (2,0.5) node[above]{\footnotesize Receiver: status quo };
     \draw[] (2,0.1) node[above]{\footnotesize Senders: status quo };
          \draw[] (11,0.5) node[above]{\footnotesize Receiver: proposal };
     \draw[] (11,0.1) node[above]{\footnotesize Senders: proposal };
     \draw[] (6.5,0.5) node[above]{\footnotesize Receiver: proposal };
     \draw[] (6.5,0.1) node[above]{\footnotesize Senders: status quo };
\end{tikzpicture}
    \caption{Thresholds of doubt. The receiver prefers the proposal if $q_H/q_{L}>-U_R(L)/U_R(H)$, while the senders prefer the proposal if $q_H/q_{L}>-U_S(L)/U_S(H)$. When $q_H/q_{L}\in(-U_R(L)/U_R(H),-U_S(L)/U_S(H))$, the receiver and the senders disagree. }
    \label{fig:1}
\end{figure}

Note that the second form of conflict arises only under uncertainty, whereas the first persists even under full information. The assumptions \eqref{disagreement} and \eqref{eqn:2} jointly imply that, under any common belief about the state $\theta\in\{L,M,H\}$, the receiver is more willing to implement the proposal than the senders are.

\subsection{Strategy and Equilibrium}
\label{sec:2.3}

We focus on symmetric equilibria in which all the senders adopt a common strategy $\bx=(x_b,x_g)\in[0,1]^2$, meaning that each sender approves the proposal with probability $x_b$ upon receiving signal $b$ and $x_g$ upon receiving signal $g$. Without loss of generality, we assume that each sender is more likely to approve the proposal after receiving signal $g$ than after receiving signal $b$, so that $x_g\geq x_b$.

Babbling equilibria, in which $x_g=x_b$ (i.e.~the senders ignore their signals), always exist. In these equilibria, the approval tally conveys no information, and the receiver makes her decision based only on her prior belief. 

Now suppose the senders choose a strategy $\bx$ with $x_g>x_b$. In state $\theta$, the probability that the approval tally $T$ equals a given value $t$ is
\begin{equation*}
    \mathbb{P}[T=t\mid \theta;\bx]=\binom{N}{t} [\underbrace{\rho_{\theta}x_g+(1-\rho_{\theta})x_b}_{\text{prob.~of approving}}]^{t}[\underbrace{1-\rho_{\theta}x_g-(1-\rho_{\theta})x_b}_{\text{prob.~of rejecting}}]^{N-t}.
\end{equation*}
The likelihood ratios
\begin{equation*}
    \frac{\mathbb{P}[T=t\mid M;\bx]}{\mathbb{P}[T=t\mid L;\bx]},\quad\frac{\mathbb{P}[T=t\mid H;\bx]}{\mathbb{P}[T=t\mid L;\bx]}, \quad \text{ and } \quad \frac{\mathbb{P}[T=t\mid H;\bx]}{\mathbb{P}[T=t\mid M;\bx]}
\end{equation*}
are strictly increasing in $t$, because when the realized state is higher, the senders receive signal $g$ more frequently and are therefore more likely to approve the proposal:
\begin{equation}
\label{eqn:3}
    \rho_{L}x_g+(1-\rho_{L})x_b<\rho_{M}x_g+(1-\rho_{M})x_b<\rho_{H}x_g+(1-\rho_{H})x_b.
\end{equation}
Hence, for each $t'<t$, the receiver's posterior belief conditional on the approval tally $T$ being equal to $t$ first-order stochastically dominates her belief conditional on $T=t'$. Since the receiver's payoff satisfies the single-crossing property, it follows from \citet{milgrom1994monotone} that the receiver's pure-strategy best response must be a cutoff strategy $\HT\in\{0,\ldots,N+1\}$: she chooses the proposal if and only if $T\geq\HT$.\footnote{The cutoff $\HT=0$ means that the receiver always implements the proposal (regardless of the approval tally), while $\HT=N+1$ means that she always maintains the status quo.} Throughout the paper, we assume that the receiver breaks ties in favor of the proposal.\footnote{The existence of informative equilibria does not depend on the receiver randomizing at $\hat{T}$. It can be shown that, whenever an informative equilibrium exists in which the receiver randomizes, there also exists an informative equilibrium without randomization.}

In the sections below, we examine perfect Bayesian equilibria characterized by pairs $(\bx,\HT)$. We say a senders' strategy $\bx$ is \textit{informative} if $x_g>x_b$, i.e., the senders' messages depend on their private information. We say a receiver's strategy $\HT$ is \emph{responsive} if $\HT\in\{1,\ldots,N\}$, i.e., the receiver's decision depends on the approval tally. Finally, we say an equilibrium $(\bx,\HT)$ is \textit{informative} if the senders' strategy $\bx$ is informative and the receiver's strategy $\HT$ is responsive.\footnote{There exist equilibria that are neither babbling nor informative, in which the senders' strategy is informative but the receiver's cutoff is nonresponsive. These equilibria are payoff-equivalent to babbling equilibria. For simplicity, we omit them from consideration.}

\section{Characterization of Equilibria}
\label{sec:3}


\subsection{Senders' and Receiver's Best Responses}

We begin by characterizing each sender's inference when making his decision. Consider a strategy profile $(\bx,\HT)$ with $\HT\in\{1,\ldots,N\}$. Under this strategy profile, we say a sender is \textit{pivotal} if exactly $\HT-1$ of the other $N-1$ senders approve the proposal. A sender's decision determines whether $T\ge\HT$ (and thus affects the outcome) if and only if he is pivotal. Hence it is optimal for each sender to condition on being pivotal when making his decision. The probability that a given sender is pivotal in state $\theta$ is
\begin{equation*}
    \mathbb{P}[\mathrm{piv}\mid \theta;\bx,\HT]=\binom{N-1}{\HT-1} [\underbrace{\rho_{\theta}x_g+(1-\rho_{\theta})x_{b}}_{\text{prob.\ of approving}}]^{\HT-1}[\underbrace{1-\rho_{\theta}x_g-(1-\rho_{\theta})x_{b}}_{\text{prob.\ of rejecting}}]^{N-\HT}.
\end{equation*}
When a sender receives signal $s\in\{g,b\}$, he strictly prefers approving the proposal to rejecting it if his expected payoff from the proposal, conditional on his signal and on being pivotal, is positive, i.e.,
\begin{equation*}
    \mathbb{E}\left[U_S(\theta)\mid s,\mathrm{piv};\bx,\HT\right]> 0.
\end{equation*}
Equivalently, this condition can be written as
\begin{equation*}
    \sum_{\theta\in\{L,M,H\}}\underbrace{q^{0}_{\theta}}_{\text{prior}} \cdot \underbrace{\mathbb{P}[s\mid \theta]}_{\text{signal}} \cdot \underbrace{\mathbb{P}[\mathrm{piv}\mid \theta;\bx,\HT]}_{\text{being pivotal}} \cdot U_S(\theta)> 0,
\end{equation*}
which yields the following condition on the payoff-weighted likelihood ratio:
\begin{equation*}
    L_S(s,\mathrm{piv};\bx,\HT):=\frac{q^0_{H}\cdot \mathbb{P}[s\mid H]\cdot \mathbb{P}[\mathrm{piv}\mid H;\bx,\HT]\cdot U_S(H)}{-\sum_{\theta\in\{L,M\}} q^0_{\theta}\cdot \mathbb{P}[s\mid\theta]\cdot \mathbb{P}[\mathrm{piv}\mid\theta;\bx,\HT] \cdot U_S(\theta)}>1.
\end{equation*}
The numerator of $L_S(s,\mathrm{piv};\bx,\HT)$ captures the sender's expected gain from approving the proposal in state \(H\), and the denominator captures his expected loss from approving it in states \(L\) and \(M\). If $L_S(s,\mathrm{piv};\bx,\HT)<1$, a sender receiving signal $s$ strictly prefers to reject the proposal. If $L_S(s,\mathrm{piv};\bx,\HT)=1$, a sender receiving signal $s$ is indifferent between approving the proposal and rejecting it.

An informative strategy $\bx$ is a \textit{senders' (symmetric) best response} to a cutoff $\HT$ chosen by the receiver if, for each sender, choosing $\bx$ is optimal given that the receiver chooses the cutoff strategy $\HT$ and all other senders choose $\bx$.
By the preceding argument, $\bx$ is a senders' best response to $\HT$ if and only if, for each $s\in\{g,b\}$,
\begin{equation}
\label{eqn:4}
    \begin{cases}
    x_s=1\quad\quad\quad\text{when}\; L_S(s,\mathrm{piv};\bx,\HT)>1,\\
    x_s\in[0,1] \,\,\quad\;\text{when}\; L_S(s,\mathrm{piv};\bx,\HT)=1,\\
    x_s=0\quad\quad\quad\text{when}\; L_S(s,\mathrm{piv};\bx,\HT)<1.
    \end{cases}
\end{equation} 
By \eqref{eqn:1}, each sender is more likely to receive signal $g$ in higher states than in lower states. Hence,
\begin{equation*}
    L_S(g,\mathrm{piv};\bx,\HT)>L_S(b,\mathrm{piv};\bx,\HT).
\end{equation*}
That is, conditional on being pivotal, each sender is strictly more willing to approve the proposal after receiving \(g\) than after receiving \(b\).
Therefore, if $\bx$ is a senders' best response to $\HT$, it must satisfy
\begin{equation}
\label{eqn:5}
    \begin{cases}
    x_g=1\;\;\text{if}\;\;x_{b}>0,\\
    x_{b}=0\;\;\text{if}\;\;x_{g}<1.
    \end{cases}
\end{equation}

Next we characterize the receiver's best response, i.e., her optimal cutoff $\HT$, given an informative strategy $\bx$. When the approval tally $T$ equals $t$, the receiver chooses the proposal if and only if, conditional on this event, her expected payoff from the proposal is nonnegative, i.e.,
\begin{equation*}
    \mathbb{E}[U_R(\theta)\mid T=t;\bx]\geq 0.
\end{equation*}
Equivalently, this condition can be written as
\begin{equation*}
    \sum_{\theta\in\{L,M,H\}}\underbrace{q^{0}_{\theta}}_{\text{prior}} \cdot \underbrace{\mathbb{P}[T=t\mid \theta;\bx]}_{\text{$t$ approvals}} \cdot U_{R}(\theta)\geq 0,
\end{equation*}
which yields the following condition on the payoff-weighted likelihood ratio:
\begin{equation*}
    L_R(t;\bx):=\frac{\sum_{\theta\in\{M,H\}} q^0_{\theta}\cdot \mathbb{P}[T=t\mid\theta;\bx]\cdot U_R(\theta)}{-q^0_{L}\cdot \mathbb{P}[T=t\mid L;\bx] \cdot U_R(L)}\geq 1.
\end{equation*}
The numerator of $L_R(t;\bx)$ captures the receiver's expected gain from choosing the proposal in states \(M\) and \(H\), and the denominator captures her expected loss from choosing it in state \(L\).

By \eqref{eqn:3}, each sender is more likely to approve the proposal in higher states than in lower ones. Hence, the likelihood ratios
\begin{equation*}
    \frac{\mathbb{P}[T=t\mid H;\bx]}{\mathbb{P}[T=t\mid L;\bx]}\text{ and }\frac{\mathbb{P}[T=t\mid M;\bx]}{\mathbb{P}[T=t\mid L;\bx]}
\end{equation*}
are strictly increasing in $t$. This implies that $L_R(t;\bx)$ is strictly increasing in $t$. Therefore, the receiver's best response is given by
\begin{equation}
\label{eqn:6}
    \HT=\min\{t\in\{0,\ldots,N\} \mid L_R(t;\bx)\geq 1\},
\end{equation}
with the convention that the minimum of the empty set is $N+1$.

An informative equilibrium is thus a pair $(\bx,\HT)$ satisfying \eqref{eqn:4} and \eqref{eqn:6}, with $x_{g}>x_{b}$ and $\HT\in\{1,\ldots,N\}$. The next lemma shows that, in every informative equilibrium, senders always reject the proposal upon receiving signal $b$.

\begin{lemma}
\label{lem:1}
In every informative equilibrium $(\bx,\HT)$, $x_{b}=0$.
\end{lemma}
Let us sketch the proof. Suppose for the sake of contradiction that $x_{b}>0$. By \eqref{eqn:5}, this means $x_g=1$ and $x_{b}\in(0,1)$. Hence any sender who rejects the proposal must have received signal $b$, and so, by \eqref{eqn:4}, conditional on being pivotal, he is indifferent between accepting and rejecting. Thus, when there are $\HT-1$ approvals from $N-1$ senders (the pivotal event), the receiver weakly prefers the proposal if the remaining pivotal sender rejects it, because that sender must have received signal $b$ and be indifferent, and the receiver is at least as willing to implement the proposal as the senders. Hence, the receiver weakly prefers the proposal to the status quo when $T=\HT-1$. But this contradicts the minimality of $\HT$ in \eqref{eqn:6}.

By \cref{lem:1}, in every informative equilibrium, the senders' equilibrium strategy takes the form \(\bx=(0,x_g)\). Henceforth, we identify this strategy with the scalar \(x:=x_g\in(0,1]\).
We let $x=0$ represent all the babbling equilibria, while every informative equilibrium is associated with some $x\in(0,1]$. Roughly speaking, $x$ measures the informativeness of a sender's message. It follows from \eqref{eqn:6} that, for two informative equilibria $(x,\HT)$ and $(x',\HT')$, if $x<x'$ then $\HT\leq\HT'$. That is, if the senders increase their willingness to approve the proposal, the receiver responds by raising her cutoff.

\subsection{Information Aggregation and Transmission}
This paper studies the efficiency of information transmission. Specifically, we ask whether the information dispersed across the senders suffices to identify the state as the number of senders grows large. In addition we ask whether, when information aggregation fails, some information can still be transmitted.
\begin{definition}
For each $N>1$, let $\Gamma_N$ denote an equilibrium of the game with $N$ senders. The sequence of equilibria $\{\Gamma_N\}_{N>1}$ \textbf{aggregates information} if
\begin{equation*}
    \lim_{N\to\infty} \left[\mathbb{P}(\mathrm{proposal}\mid L;\Gamma_N)+\mathbb{P}(\mathrm{status\ quo}\mid H;\Gamma_N)\right]=0.
\end{equation*}
\end{definition}
Our definition of information aggregation requires only that the receiver choose the correct policy in the agreement states $L$ and $H$. Note, however, that in our game, information aggregation is equivalent to complete learning by the receiver: any sequence of equilibria that separates $L$ from $H$ through the approval tally also separates the intermediate state $M$ from both $L$ and $H$ as $N\to\infty$.\footnote{In any informative equilibrium $(x,\HT)$, the probability that a sender will approve the proposal in state
\(\theta\) is \(p_\theta:=\rho_\theta x\). Hence, the gaps in the expected number of approvals between $M$
and each of $L$ and $H$ are fixed positive fractions of the gap
between $L$ and $H$. Equivalently, $p_M-p_L$ and $p_H-p_M$ are
fixed positive fractions of $p_H-p_L$. Therefore, if the approval tally asymptotically separates
\(L\) from \(H\), it also separates \(M\) from both \(L\) and \(H\).}

As a benchmark, consider a voting game in which the policy choice is not left to the receiver, but rather is determined by a prespecified qualified-majority rule; that is, the proposal is adopted if and only if the proportion of senders approving it exceeds a fixed interior threshold. As shown by \citet{feddersen1997voting} and \citet{duggan2001bayesian}, in such a game there always exists a sequence of equilibria that aggregates information. Our question is therefore whether information aggregation is possible when the receiver is \emph{not} constrained by a prespecified rule.

If no sequence of equilibria aggregates information, we seek to identify conditions under which partial information transmission is possible, i.e., conditions under which an informative equilibrium exists.
\begin{definition}
\textbf{Information transmission persists} if there exists $N_1$ such that, for every $N>N_1$, an informative equilibrium exists. \textbf{Information transmission fails} if there exists $N_2$ such that, for every $N>N_2$, no informative equilibrium exists.
\end{definition}

\section{Non-Conflict and Pareto Dominance}\label{sec:4}

\subsection{An Equilibrium No-Conflict Property}

In this section we show that in every informative equilibrium, once the senders' equilibrium strategy is fixed, the senders and the receiver agree on the policy choice at every realized approval tally. Consequently, if the senders follow their equilibrium strategy, then the receiver's equilibrium strategy maximizes the expected payoffs of both the senders and the receiver over all receiver strategies.

\begin{theorem}[equilibrium no-conflict property]
    \label{thm:1}
    Fix an informative equilibrium $(x,\HT)$. If the senders follow their equilibrium strategy, namely, $x_g=x$ and $x_b=0$, then the receiver's equilibrium strategy $\HT$ maximizes every player’s expected payoff over all receiver strategies.
\end{theorem}

\begin{proof}
Fix an informative equilibrium $(x,\HT)$. Since the cutoff strategy $\HT$ is the receiver's best response to $x$, it maximizes her expected payoff. Because the senders all have identical preferences, it suffices to show that, given $x$, $\HT$ also maximizes the expected payoff of an arbitrary sender.

We first show that, conditional on the approval tally being $T=\HT$, each sender weakly prefers the proposal over the status quo:
\begin{equation}
\label{eqn:7}
\mathbb{E}\!\left[U_S(\theta)\mid T=\HT\right]\ge 0.
\end{equation}
Let
\[
\CS:=\{S\subseteq \{1,\ldots,N\}\mid |S|=\HT\}, \qquad
\CS_i:=\{S\in\CS\mid i\in S\}.
\]
For each subset $S\subseteq \{1,\ldots,N\}$, let $A(S)$ be the event that the set of senders approving the proposal is exactly $S$. The events $\{A(S)\}_{S\in\CS}$ are disjoint, and their union is the event $\{T=\HT\}$.

For a fixed sender $i$, the event $\bigcup_{S\in\CS_i}A(S)$ is the event that sender $i$ approves and exactly $\HT-1$ of the other $N-1$ senders approve (i.e., sender $i$ approves and is pivotal). Since sender $i$ plays a best response in equilibrium, his expected payoff from approving when pivotal must be nonnegative:
\begin{equation}
\label{eqn:8}
\mathbb{E}\!\left[U_S(\theta)\,\middle|\, \bigcup_{S\in\CS_i}A(S)\right]\ge 0.
\end{equation}

By the law of total expectation,
\[
\mathbb{P}(T=\HT)\,\mathbb{E}\!\left[U_S(\theta)\mid T=\HT\right]
=
\sum_{S\in\CS}\mathbb{P}(A(S))
\mathbb{E}\!\left[U_S(\theta)\mid A(S)\right],
\]
\[
\mathbb{P}\!\left(\bigcup_{S\in\CS_i}A(S)\right)
\mathbb{E}\!\left[U_S(\theta)\,\middle|\, \bigcup_{S\in\CS_i}A(S)\right]
=
\sum_{S\in\CS_i}\mathbb{P}(A(S))
\mathbb{E}\!\left[U_S(\theta)\mid A(S)\right]\quad \forall i.
\]
Summing the second identity over all $i \in \{1,\ldots,N\}$ and using the fact that each set $S\in\CS$ has size $\HT$, we obtain
\[
\mathbb{P}(T=\HT)\,\mathbb{E}\!\left[U_S(\theta)\mid T=\HT\right]
=
\frac{1}{\HT}\sum_{i=1}^N
\mathbb{P}\!\left(\bigcup_{S\in\CS_i}A(S)\right)
\mathbb{E}\!\left[U_S(\theta)\,\middle|\, \bigcup_{S\in\CS_i}A(S)\right].
\]
Hence \eqref{eqn:7} follows from \eqref{eqn:8}. Applying the same argument to the event $\{T=\HT-1\}$ and to the event that sender $i$ is pivotal and rejects yields
\begin{equation}
\label{eqn:9}
\mathbb{E}\!\left[U_S(\theta)\mid T=\HT-1\right]\le 0.
\end{equation}

Finally, let
\[
\Lambda_S(t):=
\frac{q^0_H\,\mathbb{P}(T=t\mid H)\,U_S(H)}
{-\sum_{\theta\in\{L,M\}}q_{\theta}^0\,\mathbb{P}(T=t\mid \theta)\,U_S(\theta)}.
\]
Then
\[
\mathbb{E}\!\left[U_S(\theta)\mid T=t\right]\ge(\le) \,0
\quad\Longleftrightarrow\quad
\Lambda_S(t)\ge (\le) \,1.
\]
By \eqref{eqn:3}, the likelihood ratios
\[
\frac{\mathbb{P}(T=t\mid H)}{\mathbb{P}(T=t\mid L)}
\qquad\text{and}\qquad
\frac{\mathbb{P}(T=t\mid H)}{\mathbb{P}(T=t\mid M)}
\]
are strictly increasing in $t$. Hence $\Lambda_S(t)$ is strictly increasing in $t$. Combining this with \eqref{eqn:7} and \eqref{eqn:9}, we obtain
\[
\mathbb{E}\!\left[U_S(\theta)\mid T=t\right]\le 0 \quad \forall\, t\le \HT-1,
\qquad
\mathbb{E}\!\left[U_S(\theta)\mid T=t\right]\ge 0 \quad \forall\, t\ge \HT.
\]
Therefore, the cutoff strategy $\HT$ is sender-optimal over all receiver strategies.
\end{proof}

\begin{remark}
    \label{rem:2}
    The proof of \cref{thm:1} does not require the senders to have identical information structures or to use a common strategy. Hence the same conclusion extends to asymmetric environments in which the senders may have heterogeneous information structures and use different strategies, provided that (\romannumeral 1) each sender's information structure satisfies the monotone likelihood ratio property (MLRP), and (\romannumeral 2) the equilibrium satisfies a monotonicity requirement, with each sender being weakly more likely to approve the proposal after receiving a higher signal than after receiving a lower signal.
\end{remark}

As we noted in the introduction, the equilibrium no-conflict property can be understood as follows. If there were only one sender, then sender incentive compatibility and receiver obedience would imply that, in every equilibrium sustaining informative communication, both the sender and the receiver weakly prefer the policy implemented following each on-path recommendation. \Cref{thm:1} extends this familiar logic to decentralized communication featuring multiple senders. 

The crucial consequence of \cref{thm:1} is that the scope for communication in our model is determined by the conflicts of interest between the receiver and the senders. In \cref{sec:5,sec:6} we analyze in detail how the two forms of conflict identified in \cref{sec:2.2} shape the possibility of information aggregation or transmission.

\subsection{Pareto Dominance}

\begin{proposition}
    \label{prop:1}
   Each player's expected payoff in any informative equilibrium is weakly higher than his or her expected payoff in any babbling equilibrium.
\end{proposition}

\begin{proof}
    Let $\HT'\in\{0,N+1\}$ denote the receiver's cutoff strategy in a babbling equilibrium (i.e., she either always chooses the proposal or always chooses the status quo, regardless of the observed approval tally). Fix an informative equilibrium $(x,\HT)$. Now, given that the senders are following the strategy $x$, suppose that the receiver instead uses $\HT'$. Then, because $\HT'$ implements a constant policy, every player's expected payoff is the same as in the babbling equilibrium. By \cref{thm:1}, given the sender strategy $x$, the cutoff $\HT$ maximizes every player's expected payoff over all receiver strategies, including $\HT'$. Therefore, under the informative equilibrium $(x,\HT)$, every player's expected payoff is weakly higher than it is in the babbling equilibrium.
\end{proof}

Moreover, informative equilibria are weakly Pareto-ranked by the value of $x$.
\begin{proposition}
    \label{prop:2}
    For any two informative equilibria $\Gamma_1=(x_1,\HT_1)$ and $\Gamma_2=(x_2,\HT_2)$, if $x_2>x_1$, then each player's expected payoff is weakly higher under $\Gamma_2$ than under $\Gamma_1$.
\end{proposition}

\begin{proof}
For each informative equilibrium, given the senders' equilibrium strategy,
\cref{thm:1} implies that the receiver's equilibrium strategy maximizes every
player's expected payoff over all receiver strategies, i.e., over all decision
rules based on the approval tally. Let $T_i$ denote the approval tally generated
by the senders' equilibrium strategy in $\Gamma_i$, for $i=1,2$. Since $x_2>x_1$, each sender's message under $\Gamma_1$ can be obtained by
garbling his message under $\Gamma_2$. Hence each sender's message is Blackwell
more informative under $\Gamma_2$ than under $\Gamma_1$, and so $T_2$ is
Blackwell more informative than $T_1$. By Blackwell's theorem, it follows that for every player,
the maximal expected payoff over all decision rules based on $T_2$ is weakly
higher than that based on $T_1$, which completes the proof.

\end{proof}

Let $\Gamma_{\max}=(x_{\max},\HT_{\max})$ denote an equilibrium with the
largest value of $x$.
We refer to $\Gamma_{\max}$ as the \textit{most informative equilibrium}.\footnote{Note that an informative equilibrium may not exist, and in this case the most informative equilibrium is a babbling equilibrium.}
By \cref{prop:1,prop:2}, $\Gamma_{\max}$ weakly Pareto-dominates all other
equilibria.

\section{Without the Disagreement State}
\label{sec:5}

This section reviews the benchmark case studied by \citet{levit2011nonbinding}, \citet{battaglini2017public}, and \citet{ekmekci2024information}, in which the disagreement state $M$ is excluded (i.e., $q^0_M=0$). In this case, as discussed in \cref{sec:2.2}, the senders and the receiver would always prefer the same policy if they knew the state; their interests conflict only under uncertainty. The conflict stems from the differences between their payoffs in the agreement states $L$ and $H$ and is summarized by $\frac{U_S(L)}{U_S(H)}\cdot \frac{U_R(H)}{U_R(L)}$. The papers cited above show that in this setting, information transmission is all-or-nothing: it occurs if and only if the conflict is below a critical threshold.

\begin{theorem}[{\citealt{levit2011nonbinding}; \citealt{battaglini2017public}}]
    \label{thm:2}
    Assume that $q^{0}_M=0$.
    \mbox{}
    \begin{enumerate}
        \item If 
            \begin{equation}
                \label{eqn:10}
                \frac{U_S(L)}{U_S(H)}\cdot \frac{U_R(H)}{U_R(L)}>\frac{\rho_H}{\rho_L}\cdot\frac{1-\rho_L}{1-\rho_H},
            \end{equation}
           then no informative equilibrium exists for any $N$.
            
        \item If
            \begin{equation}
                \label{eqn:11}
                \frac{U_S(L)}{U_S(H)}\cdot \frac{U_R(H)}{U_R(L)}<\frac{\rho_H}{\rho_L}\cdot\frac{1-\rho_L}{1-\rho_H},
            \end{equation}
            then there exists a sequence of equilibria that aggregates information.
    \end{enumerate}
\end{theorem}

To see the first part of the theorem, fix an informative equilibrium $(x,\HT)$. By \cref{thm:1}, if the senders follow their equilibrium strategy, then they prefer the proposal when $T=\HT$, and the receiver prefers the status quo when $T=\HT-1$. Since $q^0_M=0$, these conditions imply that
\[
\frac{q^0_H}{q^0_L}\cdot\frac{\mathbb{P}(T=\HT\mid H)}{\mathbb{P}(T=\HT\mid L)}
\ge -\frac{U_S(L)}{U_S(H)},
\qquad
\frac{q^0_H}{q^0_L}\cdot\frac{\mathbb{P}(T=\HT-1\mid H)}{\mathbb{P}(T=\HT-1\mid L)}
\le -\frac{U_R(L)}{U_R(H)}.
\]
Combining them yields
\begin{equation}
    \label{eqn:12}
    \frac{\mathbb{P}(T=\HT\mid H)}{\mathbb{P}(T=\HT\mid L)}
\cdot
\frac{\mathbb{P}(T=\HT-1\mid L)}{\mathbb{P}(T=\HT-1\mid H)}
\ge
\frac{U_S(L)}{U_S(H)}\cdot \frac{U_R(H)}{U_R(L)}.
\end{equation}
Since moving from $T=\HT-1$ to $T=\HT$ amounts to replacing one rejection
with one approval, canceling the common factors implies that the left-hand
side of \eqref{eqn:12} equals
\[
\frac{\mathbb{P}(\text{approval}\mid H)}{\mathbb{P}(\text{approval}\mid L)}
\cdot
\frac{\mathbb{P}(\text{rejection}\mid L)}{\mathbb{P}(\text{rejection}\mid H)}=\frac{\rho_H x}{\rho_L x}\cdot \frac{1-\rho_L x}{1-\rho_H x}.
\]
This quantity is increasing in $x$ and is therefore maximized at $x=1$, i.e., when the senders approve upon receiving signal $g$ and reject upon receiving signal $b$ (in other words, they report truthfully). Its maximum value is
\[
\frac{\rho_H}{\rho_L}\cdot\frac{1-\rho_L}{1-\rho_H}.
\]
Hence, if \eqref{eqn:10} holds, the right-hand side of \eqref{eqn:12}
exceeds the maximum possible value of its left-hand side. Therefore, \eqref{eqn:12} cannot be satisfied, and no informative
equilibrium exists.

The argument above also implies that, as $\frac{U_S(L)}{U_S(H)}\cdot\frac{U_R(H)}{U_R(L)}$ approaches the critical threshold $\frac{\rho_H}{\rho_L}  \cdot  \frac{1-\rho_L}{1-\rho_H}$ from
below, 
any informative equilibrium $(x,\HT)$ that exists must be almost truth-telling, with \(x\approx 1\). In such equilibria, communication is highly informative and so largely resolves uncertainty, which mitigates the effects of the players' conflict of interest. The following result from \citet{ekmekci2024information} establishes the existence of such equilibria whenever the population of senders is sufficiently large.

\begin{lemma}[almost truth-telling]
\label{lem:2}
Assume that $q^{0}_M=0$ and \eqref{eqn:11} holds. Then for each $\epsilon>0$ there exists $N_{\epsilon}$ such that, for every $N>N_{\epsilon}$, the game with $N$ senders has an informative equilibrium $(x_N,\HT_N)$ with $x_N\ge 1-\epsilon$.
\end{lemma}

\Cref{lem:2} implies that if \eqref{eqn:11} holds, then one can select
informative equilibria $(x_N,\HT_N)$, for all sufficiently large $N$,
in such a way that $x_N\to 1$. Consequently, by the law of large numbers, as $N\to\infty$, the approval tally
identifies the state, so the receiver fully learns
the state, i.e., information is aggregated.

\section{With the Disagreement State}
\label{sec:6}

Throughout this section, we assume $q^{0}_M>0$, i.e.~the disagreement state $M$ occurs with positive probability, so that conflict persists even under full information.

\subsection{Failure of Information Aggregation}

\begin{theorem}
    \label{thm:3}
    Assume $q^{0}_M>0$. Then there exists a constant $c>0$ such that, for every $N$ and every symmetric equilibrium $\Gamma_N$ with $N$ senders,
    \begin{equation}\label{agreement-error}
    \mathbb{P}(\mathrm{proposal }\mid L;\Gamma_N)+\mathbb{P}(\mathrm{status\ quo}\mid H;\Gamma_N)>c.
    \end{equation}   
    In particular, no sequence of symmetric equilibria aggregates information.
    Furthermore, there exists a constant $T_0>0$ such that, for every $N$ and every informative equilibrium $\Gamma_N=(x_N,\HT_N)$ with $N$ senders,
    \begin{equation}
    \label{eqn:13}
    Nx_N<T_0 \text{ and } \HT_N<T_0.
    \end{equation}
   \end{theorem}

\cref{thm:3} shows that if the disagreement state occurs with any positive probability, then the receiver cannot fully learn the state even from an arbitrarily large number of senders. In fact, even in the agreement states $L$ and $H$, she chooses a suboptimal policy with positive probability. Hence, in the informal elections examined in this paper, the presence of the disagreement state precludes information aggregation.

The intuition for the failure of information aggregation when
$q^{0}_M>0$ comes from \cref{thm:1}, which says that equilibrium
communication can reveal only information that induces the same preferred
policy for the receiver and the senders. Suppose for the sake of contradiction that the receiver fully learns the state as $N\to\infty$. Then, since $q^{0}_M>0$, there must exist a range of realized values of $T$ for which the receiver infers that the state is $M$ and therefore prefers the proposal, whereas the senders prefer the status quo. This contradicts \cref{thm:1}.\footnote{Under the conditions of \cref{rem:2}, the
same argument applies when senders have heterogeneous information
structures and use different strategies. In addition, we can show that complete learning remains impossible
in even more general environments, as long as the senders agree ordinally with one another in each state---that is, when all the senders prefer the proposal in state $H$ and the status quo in states $L$ and $M$, although
their payoff intensities may differ.}

The first inequality in \eqref{eqn:13} implies that for every sequence of symmetric equilibria $\{(x_N,\HT_N)\}_N$,
\begin{equation}
\label{eqn:14}
    \lim_{N\to\infty}x_N=0.
\end{equation}
That is, as $N\to\infty$, each sender's message conveys vanishingly little information to the receiver in every informative equilibrium. (If this were not the case, then by the law of large numbers, the receiver could identify the state.) Furthermore, the first inequality in \eqref{eqn:13} specifies the rate at which $x_N\to 0$: it implies that $x_N=O(1/N)$.

The second inequality in \eqref{eqn:13} captures an important connection between our result and the voting literature: it says that the receiver maintains the status quo only if all but a bounded number of senders (independent of the total population $N$) reject the proposal. Hence the receiver's equilibrium strategy resembles a unanimity or near-unanimity rule for choosing the status quo.\footnote{When $\HT=1$, the status quo is maintained if and only if all the senders reject the proposal.} Our conclusion that information aggregation always fails is thus reminiscent of the failure of information aggregation under the unanimity rule in \citet{feddersen1998convicting} and \citet{duggan2001bayesian}. The distinction is that, in those papers, the voting rule is exogenously fixed, whereas here the near-unanimity rule arises endogenously from the receiver's best response and the senders' equilibrium incentives.

Finally, we can show that if
\(
\frac{U_S(L)}{U_S(H)}\cdot
\frac{U_R(H)}{U_R(L)}
<
\frac{\rho_H}{\rho_L}
\), then in the most informative equilibrium, the receiver's error
probability in states \(L\) and \(H\) (i.e., the left-hand side of \eqref{agreement-error}) vanishes as \(q_M^0 \to 0\).
We discuss the threshold \(\rho_H/\rho_L\) below.

\subsection{Conditions for Information Transmission}
\label{sec:6.2}

Having shown that the presence of the disagreement state $M$ rules out complete learning, we now identify the conditions under which partial learning, i.e.~information transmission, remains possible.

\begin{theorem}
    \label{thm:4}
    \mbox{}
    \begin{enumerate}
        \item If 
            \begin{equation*}
                \frac{U_S(L)}{U_S(H)} \cdot \frac{U_R(H)}{U_R(L)} > \frac{\rho_H}{\rho_L},
            \end{equation*}
            then information transmission fails whenever $q^0_M > 0$.
            
        \item If
            \begin{equation*}
                \frac{U_S(L)}{U_S(H)} \cdot \frac{U_R(H)}{U_R(L)} < \frac{\rho_H}{\rho_L},
            \end{equation*}
            then, for each fixed value of \(q^0_H/q^0_L\), there exists a corresponding threshold \(\hat q>0\) such that information transmission persists if \(q^0_M<\hat q\) and fails if \(q^0_M>\hat q\). Furthermore, $\hat{q}$ is weakly decreasing as a function of $\frac{U_S(L)}{U_S(H)}\cdot\frac{U_R(H)}{U_R(L)}$ when this ratio is varied by changing one of $U_S(L)$, $U_S(H)$, $U_R(L)$, and $U_R(H)$ while holding the other three fixed.
    \end{enumerate}
\end{theorem}

\cref{fig:2} illustrates \cref{thm:4}. Broadly, the theorem says that information transmission persists if both forms of conflict—captured respectively by $\frac{U_S(L)}{U_S(H)}\cdot\frac{U_R(H)}{U_R(L)}$ and $q^0_M$—are small. Comparing \cref{thm:2,thm:4}, we see that introducing the disagreement state $M$ drastically alters the conditions under which information transmission persists. In particular, the presence of $M$ generates a discontinuity relative to the $q^0_M=0$ benchmark: suppose that
\begin{equation}
\label{eqn:15}
    \frac{U_S(L)}{U_S(H)}\cdot\frac{U_R(H)}{U_R(L)}\in\left(\frac{\rho_H}{\rho_L}, \frac{\rho_H}{\rho_L}\cdot\frac{1-\rho_L}{1-\rho_H}\right).
\end{equation}
Then, if $q^0_M=0$, the second part of \cref{thm:2} implies that information transmission persists; in fact, the receiver fully learns the state as $N\to\infty$. However, if $q^0_M>0$, the first part of \cref{thm:4} says that information transmission fails: babbling is the unique equilibrium outcome for all sufficiently large $N$. Thus, perturbing the $q_M^0=0$ benchmark by assigning any positive probability to $M$ changes the large-population prediction from complete learning to complete unraveling.\footnote{Keeping $q_H^0/q_L^0$ fixed while varying $q_M^0$ provides one convenient perturbation path, but is not required.}

\begin{figure}
    \centering
    \begin{tikzpicture}
\begin{axis}[
        xmin=-0.1,
        xmax=1.3,
        ymin=-0.25,
        ymax=1.5,
        axis lines=center,
        xtick={},
        ytick={},
        xticklabel=\empty,
        yticklabel=\empty,
        xtick style={draw=none},
        ytick style={draw=none},
        xlabel={$\frac{U_S(L)}{U_S(H)}\cdot\frac{U_R(H)}{U_R(L)}$},
        ylabel={$q^0_{M}$},
        xlabel style={below right},
        ylabel style={above left},
        smooth,
        domain=0:1.2,
        samples=100,
        no markers,
]
\addplot [name path=B, domain=0:0.6, color=red, line width=0.4mm, dashed]{3*(x-0.6)^2};

\addplot [name path=A, domain=0:0.8, color=blue, line width=0.5mm]{0};

\path[fill=lightgray] (axis cs:0,0) -- 
    plot[domain=0:0.6, samples=20] (axis cs:\x,{3*(\x-0.6)^2}) -- 
    (axis cs:0.6,0) -- cycle;

\node[draw=black, fill=blue!90, circle, minimum size=3pt, inner sep=0pt] at (axis cs:0,0) {};

\node[draw=black, fill=blue!90, circle, minimum size=3pt, inner sep=0pt] at (axis cs:0.6,0) {};
\node[above] at (axis cs:0.6,0) {$\frac{\rho_H}{\rho_L}$};
\node[above, red] at (axis cs:0.28,0.34) {\large $\hat{q}$};

\node[draw=black, fill=blue!90, circle, minimum size=3pt, inner sep=0pt] at (axis cs:0.8,0) {};
\node[below] at (axis cs:0.8,0) {$\frac{\rho_H}{\rho_L}\cdot\frac{1-\rho_L}{1-\rho_H}$};
\node[below] at (axis cs:0.05,0) {1};
\node[above] at (axis cs:-0.05,0) {0};
\end{axis}
\end{tikzpicture}
    \caption{Conditions for information transmission. When $q^0_M=0$, information transmission persists if $\frac{U_S(L)}{U_S(H)}\cdot\frac{U_R(H)}{U_R(L)}<\frac{\rho_H}{\rho_L}\cdot\frac{1-\rho_L}{1-\rho_H}$ (i.e., in the region indicated in blue). When $q^0_M>0$, information transmission persists if $\frac{U_S(L)}{U_S(H)}\cdot\frac{U_R(H)}{U_R(L)}<\frac{\rho_H}{\rho_L}$ and $q^0_M<\hat{q}$ (i.e., in the shaded region).}
    \label{fig:2}
\end{figure}

This discontinuity arises because the two forms of conflict impose
opposing requirements on the informativeness of equilibrium communication.
As we saw in \cref{sec:5}, when $q^0_M=0$, any informative equilibrium must satisfy
\begin{equation}
\label{eqn:c1}
    \frac{\mathbb{P}(\text{approval}\mid H)}
    {\mathbb{P}(\text{approval}\mid L)}
    \cdot
    \frac{\mathbb{P}(\text{rejection}\mid L)}
    {\mathbb{P}(\text{rejection}\mid H)}
    =
    \frac{\rho_H}{\rho_L}
    \cdot
    \frac{1-\rho_L x}{1-\rho_H x}
    \ge
    \frac{U_S(L)}{U_S(H)}
    \cdot
    \frac{U_R(H)}{U_R(L)}.
\end{equation}
We derived the critical threshold
$\frac{\rho_H}{\rho_L}\cdot\frac{1-\rho_L}{1-\rho_H}$ in
\cref{thm:2} by considering the senders' strategy $x=1$, under which the senders report truthfully and so each message is maximally
informative. Intuitively, as the conflict that arises from uncertainty
intensifies toward the critical threshold, communication must
approach maximal informativeness to reduce uncertainty and thereby
mitigate the effects of the conflict. Furthermore, when this conflict is at any fixed level satisfying \eqref{eqn:15}, $x$ is bounded away from zero (i.e.~communication must be somewhat informative), because
$\frac{\rho_H}{\rho_L}\cdot\frac{1-\rho_L x}{1-\rho_H x}$ is minimized at
$x=0$, where it equals $\frac{\rho_H}{\rho_L}$. In other words, this form of conflict places a lower bound on informativeness.

By contrast, conflict arising from the disagreement state places an upper bound on informativeness. In particular, if state $M$ occurs with any positive probability, \eqref{eqn:14} says that along every sequence of informative equilibria, the senders' individual messages become nearly uninformative as the number of senders grows large. When \eqref{eqn:15} holds, the bounds imposed by the two forms of conflict become incompatible for all sufficiently large $N$, which generates the discontinuity.

Letting $x\to0$ in \eqref{eqn:c1}, we find that information transmission can persist only if
\begin{equation*}
    \frac{\rho_H}{\rho_L}
    \ge
    \frac{U_S(L)}{U_S(H)}
    \cdot
    \frac{U_R(H)}{U_R(L)}.
\end{equation*}
This establishes the first part of \cref{thm:4}. For the second part, recall that \cref{thm:3} provides a bound $T_0$ on the receiver's cutoff. It therefore suffices to examine the limiting equilibrium conditions for each fixed cutoff $\HT<T_0$. These conditions yield the threshold $\hat q$.

\subsection{Most Informative Equilibrium}
\label{sec:6.3}

In this section we show that in the most informative equilibrium $\Gamma_{\max}=(x_{\max},\HT_{\max})$, the informativeness of each sender's message weakly declines as the conflict between the senders and the receiver increases. 

\begin{proposition}
\label{prop:3}
Fix $N$. With $q^0_H/q^0_L$ held fixed, $x_{\max}$ is weakly decreasing
as a function of $q^0_M$. Moreover, $x_{\max}$ is weakly decreasing as a function of
\(
\frac{U_S(L)}{U_S(H)}\cdot\frac{U_R(H)}{U_R(L)}
\)
when this ratio is varied by changing one of $U_S(L)$, $U_S(H)$, $U_R(L)$, and $U_R(H)$ while holding the other three fixed.
\end{proposition}

The proof of \cref{prop:3} is relegated to the appendix.

Next we quantify the total information transmitted to the receiver under $\Gamma_{\max}$. This is captured by the quantity
\begin{equation*}
    I:=\limsup_{N\to\infty}\frac{V_{R}^{\max}(N)-V_{R}^0(N)}{V_{R}^{\mathrm{full}}(N)-V_{R}^0(N)},
\end{equation*}
where, for each $N$, (\romannumeral 1)~$V_{R}^{\max}(N)$ denotes the receiver's expected payoff under $\Gamma_{\max}$, (\romannumeral 2)~$V_{R}^{0}(N)$ denotes her expected payoff in the babbling equilibrium, and (\romannumeral 3)~$V_{R}^{\mathrm{full}}(N)$ denotes her expected payoff under full information.

\begin{figure}
    \centering
    \resizebox{6in}{2.0in}{
    \begin{tikzpicture}
\begin{axis}[
        xmin=-35,
        xmax=365,
        ymin=-35,
        ymax=200,
        axis lines=center,
        xtick={\empty},
        ytick={\empty},
        xlabel={$\frac{U_S(L)}{U_S(H)}\cdot\frac{U_R(H)}{U_R(L)}$},
        ylabel={$I$},
        xlabel style={below right},
        ylabel style={above left},
        smooth,
        domain=0:350,
        samples=100,
        no markers,
]
\addplot [domain=0:240, color=blue, line width=0.3mm]{150};
\addlegendentry{$q^0_M=0$}
\addplot [domain=0:100, color=red, line width=0.3mm]{200/(0.01*x+1)-100};
\addlegendentry{$q^0_M>0$}
\addplot [domain=100:350, color=red, line width=0.3mm]{0};
\addplot [domain=240:365, color=blue, line width=0.3mm]{0};
\node[below] at (axis cs:150,0)  {$\frac{\rho_H}{\rho_L}$};
\node[below] at (axis cs:240,0)  {$\frac{\rho_H}{\rho_{L}}\cdot\frac{1-\rho_{L}}{1-\rho_{H}}$};
\draw[dashed] (axis cs:240,0) -- (axis cs:240,150);
\node[font=\scriptsize,above] at (axis cs:100,155)  {\textbf{full information}};
\node[font=\scriptsize,above] at (axis cs:280,5)  {\textbf{no information}};
\node[font=\scriptsize,above] at (axis cs:125,55)  {\textbf{partial information}};

\node[label={[label distance=0.4mm]-60:$1$}] at (axis cs:0,0) {};
\node[label={[label distance=0.4mm]120:$0$}] at (axis cs:0,0) {};
\node[label={[label distance=0.4mm]180:$1$}] at (axis cs:0,150) {};

\draw[thick, black] (axis cs:100,-1.5) -- (axis cs:100,3.5);
\draw[thick, black] (axis cs:150,-1.5) -- (axis cs:150,3.5);
\draw[thick, black] (axis cs:240,-1.5) -- (axis cs:240,3.5);
\end{axis}
\begin{scope}[xshift=10cm]
\begin{axis}[
        xmin=-35,
        xmax=365,
        ymin=-35,
        ymax=200,
        axis lines=center,
        xtick={\empty},
        ytick={\empty},
        xlabel={$\frac{U_{S}(L)}{U_{S}(H)}\cdot\frac{U_{R}(H)}{U_{R}(L)}$},
        ylabel={$I$},
        xlabel style={below right},
        ylabel style={above left},
        smooth,
        domain=0:350,
        samples=100,
        no markers,
]
\addplot [domain=0:240, color=blue, line width=0.3mm]{150};
\addlegendentry{$q^0_M=0$}
\addplot [domain=0:150, color=red, line width=0.3mm]{148*(1 - (x/150)^30)};
\addlegendentry{$q^0_M\approx0$}
\addplot [domain=150:350, color=red, line width=0.3mm]{0};
\addplot [domain=240:365, color=blue, line width=0.3mm]{0};
\node[below] at (axis cs:150,0)  {$\frac{\rho_H}{\rho_L}$};
\node[below] at (axis cs:240,0)  {$\frac{\rho_H}{\rho_{L}}\cdot\frac{1-\rho_{L}}{1-\rho_{H}}$};
\draw[dashed] (axis cs:240,0) -- (axis cs:240,150);
\draw[dashed] (axis cs:150,0) -- (axis cs:150,150);
\node[font=\scriptsize,above] at (axis cs:100,155)  {\textbf{full information}};
\node[font=\scriptsize,above] at (axis cs:280,5)  {\textbf{no information}};

\node[label={[label distance=0.4mm]-60:$1$}] at (axis cs:0,0) {};
\node[label={[label distance=0.4mm]120:$0$}] at (axis cs:0,0) {};
\node[label={[label distance=0.4mm]180:$1$}] at (axis cs:0,150) {};

\draw[thick, black] (axis cs:150,-1.5) -- (axis cs:150,3.5);
\draw[thick, black] (axis cs:240,-1.5) -- (axis cs:240,3.5);
\end{axis}
\end{scope}
\end{tikzpicture}
    }
    \caption{Total information transmitted to the receiver. In both panels, the blue curve shows $I$ plotted against $\frac{U_S(L)}{U_S(H)}\cdot\frac{U_R(H)}{U_R(L)}$ for $q^0_M=0$. In the left panel, the red curve gives $I$ for $q^0_M>0$. In the right panel, the red curve gives $I$ for $q^0_M\approx 0$.}
    \label{fig:3}
\end{figure}

The left panel of \cref{fig:3} shows the total information transmitted to the receiver when $q^0_M=0$ and when $q^0_M>0$.  When $q^0_M=0$, as previously established, information transmission is all-or-nothing and depends on whether
\begin{equation*}
    \frac{U_S(L)}{U_S(H)}\cdot\frac{U_R(H)}{U_R(L)}<\frac{\rho_H}{\rho_L}\cdot\frac{1-\rho_L}{1-\rho_H}.
\end{equation*}
When $q^0_M>0$, complete learning is not possible even when
there is no conflict arising from the agreement
states, i.e., when
\begin{equation*}
    \frac{U_S(L)}{U_S(H)}\cdot\frac{U_R(H)}{U_R(L)}=1.
\end{equation*}
As such conflict grows, i.e.~as $\frac{U_S(L)}{U_S(H)}\cdot\frac{U_R(H)}{U_R(L)}$ increases, $x_{\max}$ weakly decreases and less information is transmitted. For values of $\frac{U_S(L)}{U_S(H)}\cdot\frac{U_R(H)}{U_R(L)}$ approaching $\frac{\rho_H}{\rho_L}$, no information is transmitted, and babbling is the unique equilibrium outcome for all sufficiently large \(N\).

The right panel of \cref{fig:3} illustrates the limiting case as $q^0_M$ approaches zero. Here, when
\begin{equation*}
    \frac{U_S(L)}{U_S(H)}\cdot\frac{U_R(H)}{U_R(L)}<\frac{\rho_H}{\rho_L},
\end{equation*}
almost full information is transmitted. However, when 
\begin{equation*}
    \frac{U_S(L)}{U_S(H)}\cdot\frac{U_R(H)}{U_R(L)}\in \left(\frac{\rho_H}{\rho_L}, \frac{\rho_H}{\rho_L}\cdot\frac{1-\rho_L}{1-\rho_H}\right),
\end{equation*}
babbling remains the unique equilibrium outcome for all sufficiently large \(N\). This highlights the discontinuity in information transmission described in \cref{sec:6.2}.

\subsection{Bounds on the Benefits of Population Growth}

The preceding subsections show that, in the presence of the disagreement state \(M\), the informativeness of the senders' messages remains limited as \(N\to\infty\); in fact, under some conditions, informativeness unravels completely for all sufficiently large \(N\). We now
establish a sharper limitation on sender welfare: we show that the welfare gains from
expanding the sender population are exhausted at a finite population size.

For each \(N\), let \(V_S^{\max}(N)\) denote the common expected payoff of
the senders in the most informative equilibrium with $N$ senders.

\begin{proposition}
\label{prop:finite_sender_optimum}
There exists a finite population size \(N^*\geq2\) such that
\[
    V_S^{\max}(N^*)
    =
    \sup_{N\geq2}V_S^{\max}(N).
\]
\end{proposition}

Thus, once the sender population reaches \(N^*\), increasing it further cannot raise sender welfare. This implies, for example, that in applications such as protests and polls, enlarging the pool of participants indefinitely may not benefit citizens or respondents.

Note that \cref{prop:finite_sender_optimum} need not hold in the absence of
the disagreement state \(M\), i.e., when \(q_M^0=0\). In particular, if
\eqref{eqn:11} holds, then by \cref{thm:2} there exists a sequence of
equilibria that aggregates information. Consequently, as \(N\to\infty\),
\(V_S^{\max}(N)\) converges
to the senders' first-best payoff, and hence to the supremum of their equilibrium
payoffs. This supremum need not be attained at any finite $N$.

\Cref{prop:finite_sender_optimum} builds on \cref{thm:3}, which implies that increasing $N$ cannot make the message tally $T$ arbitrarily more informative. Indeed, $Nx$, which determines the scale of the mean approval tally in every state, is bounded by \eqref{eqn:13}. Under symmetry, therefore, increasing $N$ dilutes the information conveyed by each sender's message, as
reflected in \(x=O(1/N)\). The proposition shows that for sufficiently large $N$, this dilution does not benefit the senders.

We can extend \cref{prop:finite_sender_optimum} to monotone asymmetric
equilibria, in which each sender \(i\) chooses strategy
\((x_{b,i},x_{g,i})\), with
\[
    0\leq x_{b,i}\leq x_{g,i}\leq1.
\]
(Here \(x_{s,i}\) is the probability that sender \(i\) approves the proposal
after receiving signal \(s\in\{b,g\}\).)
Let \(W_S(N)\) denote the senders' highest expected payoff over all monotone asymmetric equilibria with \(N\) senders. Then \(W_S(N)\) is
weakly increasing in \(N\). Indeed, any equilibrium with \(N\) senders can be embedded in an equilibrium with \(N+1\) senders without changing any player's payoff. In the latter equilibrium, the additional sender uses a signal-independent strategy: either he always rejects and the receiver's cutoff is unchanged, or he always approves and the receiver's cutoff is raised by one.

It can be shown that \(W_S(N)\) becomes constant when $N$ is sufficiently large. Moreover, there exists a finite \(\widehat N\) such that, for every \(N>\widehat N\), a sender-optimal outcome can be supported in an equilibrium in which \(\widehat N\) senders use informative strategies (i.e., strategies satisfying \(x_{g,i}>x_{b,i}\)), while the remaining \(N-\widehat N\) senders use signal-independent strategies. Thus, once the population is sufficiently large, additional senders do not contribute private information to the message tally.

The argument for the asymmetric case builds on an extension of
\cref{thm:3}, which says that for every \(N\) and every monotone asymmetric equilibrium,
the aggregate informativeness
\[
    \sum_{i=1}^{N}(x_{g,i}-x_{b,i})
\]
is bounded above by a constant independent of \(N\). This is analogous to the
bound on \(Nx\) in the symmetric case. However, asymmetry allows for a
sender-optimal outcome in which this bounded amount of informativeness
is concentrated among finitely many senders, rather than spread across the entire
population. 

\citet{dekel2000sequential} and
\citet{chakraborty2003efficient} obtain a similar structure in
common-value voting games: their
welfare-optimal equilibria assign informative strategies to a subset
of voters, with the remaining voters voting independently of their signals.

Note that the conclusion of \cref{prop:finite_sender_optimum} is specific to sender welfare.
Whether we restrict attention to symmetric equilibria or allow monotone
asymmetric equilibria, the supremum of the receiver's equilibrium payoffs need not be
attained at any finite population; it may be approached only as
\(N\to\infty\). This is because, although the senders and the receiver rank all symmetric
equilibria in the same way for any fixed \(N\), their rankings need not agree
across equilibria with different population sizes.

\section{Beyond the Binary Setting}
\label{sec:7}

In this section we extend the basic model to allow for nonbinary signal and message spaces. Each sender $i\in\{1,\ldots,N\}$ now receives a private signal
\[
s^i\in\{s_1,\ldots,s_J\}\subset \mathbb{R},
\qquad
s_1<\cdots<s_J,
\]
with $J\geq2$. Signals are conditionally i.i.d.\ across senders, given the state. There exists $\alpha>0$ such that
\begin{equation*}
   \mathbb{P}[s_j\mid\theta]>\alpha \quad \forall j\in\{1,\ldots,J\}\;\text{and}\;\theta\in\{L,M,H\},
\end{equation*}
so no signal rules out any state. We generalize \eqref{eqn:1} by imposing the strict monotone likelihood ratio property (MLRP):
\begin{equation*}
    \frac{\mathbb{P}[s_j\mid H]}{\mathbb{P}[s_j\mid M]},\;\frac{\mathbb{P}[s_j\mid H]}{\mathbb{P}[s_j\mid L]},\;\text{ and }\;\frac{\mathbb{P}[s_j\mid M]}{\mathbb{P}[s_j\mid L]}\;\text{are strictly increasing in $j$}.
\end{equation*}
Each sender $i$ sends a message
\[
z^i\in\{z_1,\ldots,z_K\}\subset\mathbb{R},
\qquad
z_1<\cdots<z_K,
\]
with $K\geq2$. We let $T_k$ denote the number of senders who send $z_k$,
and define
\[
\Delta^K(N)
:=
\left\{
(t_1,\ldots,t_K)\in\{0,\ldots,N\}^K
\;\middle|\;
\sum_{k=1}^K t_k=N
\right\}.
\]
The receiver observes the message-count vector
$\BT=(T_1,\ldots,T_K)\in\Delta^K(N)$ and chooses between the proposal
and the status quo.

This nonbinary framework applies to many real-world settings. For example, in a shareholder vote, in addition to approving or rejecting a proposal, shareholders can abstain. In public petitions, citizens can join pro-reform or pro-status quo rallies, or remain silent. In a survey, respondents may rate a policy on a numerical scale rather than simply approving or rejecting it.

In studying this extended model, we again examine symmetric equilibria: we assume all the senders adopt the same strategy $\boldsymbol{P}=\{p_{j,k}\}_{J\times K}$, where $p_{j,k}$ denotes the probability of sending message $z_k$ upon observing signal $s_j$.
The receiver's strategy is characterized by a function
\begin{equation*}
    \psi:\Delta^K(N) \to [0,1].
\end{equation*}
That is, under the strategy $\psi$, after observing $\BT$, she chooses the proposal with probability $\psi(\BT)$.

We say the senders' strategy is \textit{monotone} if higher signals render higher messages
more likely, i.e., if higher signals shift the distribution over messages upward in the MLRP order:\footnote{This is equivalent to
$\frac{p_{j',k}}{p_{j,k}}\leq\frac{p_{j',k'}}{p_{j,k'}}$ whenever both
$p_{j,k}$ and $p_{j,k'}$ are positive.}
\begin{equation*}
    p_{j',k}\cdot p_{j,k'}\leq p_{j,k} \cdot p_{j',k'}
    \;\;\text{for each $j<j'$ and $k<k'$}.
\end{equation*}
Consequently, over messages sent with positive probability, the induced message
distributions satisfy weak MLRP:
\begin{equation*}
    \frac{\mathbb{P}[z_k\mid H]}{\mathbb{P}[z_k\mid M]},\;
    \frac{\mathbb{P}[z_k\mid H]}{\mathbb{P}[z_k\mid L]},\;
    \text{and }\;
    \frac{\mathbb{P}[z_k\mid M]}{\mathbb{P}[z_k\mid L]}
    \;\text{are weakly increasing in $k$}.
\end{equation*}

The receiver's strategy is \textit{monotone} if replacing a lower message with a
higher one weakly increases the probability that she chooses the proposal---i.e., if, for
each $\tilde{\boldsymbol{t}}=(t_1,\ldots,t_K)\in\Delta^K(N-1)$ and $m<m'$,
\begin{equation*}
    \psi(t_1,\ldots,t_m+1,\ldots,t_{m'},\ldots,t_K)
    \leq
    \psi(t_1,\ldots,t_m,\ldots,t_{m'}+1,\ldots,t_K).
\end{equation*}

Given a monotone sender strategy, the monotone comparative statics result of \citet{milgrom1994monotone}, together with the rule that ties are broken in favor of the proposal, implies that the receiver’s best response is monotone at all positive-probability message counts; at zero-probability message counts, we can always select a monotone completion. Conversely, if the
receiver uses a monotone strategy, any best response by the senders can be replaced,
without changing the senders' payoffs, by a monotone best response. Thus, whenever
one side uses a monotone strategy, it is without loss to take the other side's best response
to be monotone.

A \emph{monotone equilibrium} is one in which the senders' strategy is monotone. In applications, it is natural to restrict attention to monotone equilibria: intuitively, more favorable signals should increase sender support for the proposal, and an increase in sender support should raise the probability that the receiver will choose the proposal.\footnote{There is a growing literature on monotone equilibria in communication games, including \citet{cho1990strategic}, \citet{krishna2001model}, \citet{chen2008selecting}, \citet{ivanov2010communication}, and \citet{kolotilin2021relational}, among others. Most work on communication games in which each sender receives a noisy signal \citep[e.g.,][]{austen1990information,austen1993interested,morgan2008information,hagenbach2010strategic,galeotti2013strategic,currarini2020strategic} focuses on binary signals and messages, as in our basic model, which ensures monotonicity. Monotone equilibria are also studied in sequential voting games \citep{dekel2000sequential}.} By contrast, non-monotone equilibria rely on counterintuitive coordination patterns and may be unstable in the sense of \citet{fey1997stability}, and their predictions are not robust to small perturbations in the model parameters.

Fix a monotone equilibrium \((\boldsymbol P,\psi)\). We say a message \(z_k\) is \emph{active} if it is sent with positive probability, i.e.,
\[
    \sum_{j=1}^J p_{j,k}>0.
\]
Two active messages \(z_k\) and \(z_\ell\) are \emph{equivalent}, and we write
\(z_k\sim_\psi z_\ell\), if for every
\(\tilde{\boldsymbol t}=(t_1,\ldots,t_K)\in\Delta^K(N-1)\),
\[
    \psi(t_1,\ldots,t_k+1,\ldots,t_\ell,\ldots,t_K)
    =
    \psi(t_1,\ldots,t_k,\ldots,t_\ell+1,\ldots,t_K).
\]
That is, conditional on any message count from the other \(N-1\) senders, replacing
\(z_k\) by \(z_\ell\) does not change the probability that the receiver will choose the proposal. 

For our analysis, we simplify the message space for each monotone equilibrium as follows. First we discard the inactive messages. We then merge each equivalence class of active messages, assign its total probability conditional on each signal to the lowest-indexed message in the class, and relabel the resulting message space in increasing order as \(z_1,z_2,\ldots\). This reduction preserves payoffs and outcomes. Since \(\psi\) is monotone, each equivalence class is an interval in the message order, so the reduction also preserves monotonicity.

Throughout the remainder of this section, all monotone equilibria are represented in the reduced message space. Thus we can restrict our attention to equilibria in which no two active messages are equivalent and the lowest message \(z_1\) is active.

\begin{proposition}
\label{prop:4}
Suppose $q^0_M>0$. Then, for each $\epsilon>0$, there exists $N_{\epsilon}$ such that for all $N>N_{\epsilon}$, in every monotone equilibrium with $N$ senders, each sender chooses $z_1$ with probability one upon observing any signal $s\in\{s_1,\ldots,s_{J-1}\}$, and chooses $z_1$ with probability greater than $1-\epsilon$ upon observing signal $s_J$:
\begin{eqnarray*}
    p_{j,1}&=&1\,\;\forall\;j\in\{1,\ldots,J-1\},\\
    p_{J,1}&>&1-\epsilon.
\end{eqnarray*}
\end{proposition}

\cref{prop:4} implies that, for all sufficiently large $N$, in every non-babbling
monotone equilibrium, the senders map every signal other than $s_J$ to the lowest
message, $z_1$. Hence active messages other than $z_1$ are sent only after
signal $s_J$. It follows that all active messages other than $z_1$ are equivalent.
In accordance with our reduction of the message space, this means at most two messages can be active:
$z_1$ and, if it exists, a single non-lowest message, which we label $z_2$. Furthermore, the signal space is effectively binary: all that matters is whether $s_J$ is observed. Thus all non-babbling monotone
equilibria are equivalent to equilibria in our main model, and so our results for that model (from Sections~\ref{sec:4} and \ref{sec:6}) extend to the present setting.

The proof of \cref{prop:4} proceeds in three steps. First, we extend \cref{lem:1} by showing that the senders who observe signal $s_1$ send message $z_1$ with probability one. Second, we use MLRP and the monotonicity of the receiver's equilibrium strategy to establish that the senders' equilibrium strategy has the following non-crossing property: if $z_k$ is sent with positive probability after signal $s_j$ (i.e., $p_{j,k}>0$), then no higher message is sent with positive probability after any lower signal (i.e., if $j'<j$ and $k'>k$, then $p_{j',k'}=0$). 
Equivalently, the support of the message distribution following a lower signal lies weakly below that following a higher signal.
Third, we establish an analog of \eqref{eqn:14}: we show that the informativeness of each sender's message vanishes as $N\to\infty$. Together, these three steps show that the senders' equilibrium strategy must take the form described in \cref{prop:4} as $N$ grows large.

The main technical difficulty in the final step is that the receiver's best response is no longer characterized by a single cutoff as it was in our original model. Instead, it is characterized by a partition of the multidimensional space of message-count vectors into regions in which the receiver chooses the proposal and regions in which she chooses the status quo. Hence there are multiple pivotal events, i.e., multiple message-count vectors at which a sender may change the receiver's decision by replacing one message with another. This multiplicity of pivotal events makes it difficult to characterize the senders' beliefs conditional on being pivotal, and hence to analyze their equilibrium strategy. 

We overcome this difficulty using large deviations theory. The key observation is that, by MLRP, the message distributions under \(L\)
and \(M\) are harder to distinguish than those under \(L\) and \(H\). At a
pivotal event, the receiver is close to indifference while weighing \(L\)
against the alternative \(\{M,H\}\). This near-indifference is more
likely to reflect the comparison between \(L\) and \(M\) than that between
\(L\) and \(H\). We then show that, if messages remain informative as $N\to\infty$, being pivotal is exponentially less likely in state $H$ than in state $M$. Consequently, conditional on being pivotal, each sender assigns vanishingly small probability to $H$. Since the senders prefer the status quo in both $L$ and $M$, for all sufficiently large $N$, the information conveyed by being pivotal dominates any private signal. Each sender therefore strictly prefers to induce the status quo and sends $z_1$ regardless of his signal. But this renders his message uninformative---a contradiction. 

A more detailed sketch of the large-deviation argument above appears in Appendix~\ref{ap:a}. This approach may be of independent interest to theorists studying communication and voting.

\section{Mediated Communication}
\label{sec:8}

\subsection{Mediation Mechanisms}

The preceding sections show that the presence of a disagreement state strongly limits direct communication between the senders and the receiver: information aggregation always fails, and communication may unravel completely. We now ask whether information transmission can be improved by the use of a mediator. The mediator does not observe the senders' signals, but can commit ex ante to a communication mechanism that collects messages from the senders and delivers a policy recommendation to the receiver. 

Mediated communication arises naturally in many applications. In a petition or a protest, the mediator may be a petition platform or advocacy organization that aggregates citizens' views and conveys a recommendation to the policymaker. In a nonbinding shareholder vote, the mediator may be a shareholder representative; in a survey or poll, it may be a survey platform or a polling firm. The mediator's ability to commit is often sustained by reputational concerns, because deviating from the announced procedure would damage its credibility in future interactions.

In the analysis below, we restrict our attention to direct and anonymous mechanisms. That is, each sender reports a signal in $\{g,b\}$ to the mediator, and the mechanism depends only on the number of senders who report $g$. Formally, we define a mechanism with $N$ senders as a function
\[
    \phi_N:\{0,\ldots,N\}\to[0,1]
\]
mapping $t \in \{0,\ldots,N\}$ to the probability that the mediator recommends the proposal after receiving $t$ reports of the signal $g$.

A mechanism is \textit{feasible} if truthful reporting is incentive compatible for the senders and the receiver is willing to follow the mediator's recommendation. The revelation principle implies that we can restrict our attention to feasible direct mechanisms without loss of generality \citep{myerson1982optimal,forges1986approach}. Note that, because we assume a symmetric environment, our restriction to anonymous mechanisms is also without loss of generality: averaging any feasible direct mechanism over all permutations
of sender identities preserves feasibility and the state-contingent recommendation
probabilities, while making the mechanism depend only on the number of senders who
report $g$.

\subsection{Asymptotic Outcomes of Mediation}

We now examine the outcome of mediated communication as $N\to\infty$. As a benchmark, consider a mediator who observes all the senders' signals. This mediator asymptotically learns the state as $N \to \infty$. Therefore, whenever she is aligned with either the senders or the receiver, she recommends the proposal in state $H$ and the status quo in state $L$. The only remaining question is what she recommends in the disagreement state $M$.

If the mediator is aligned with the receiver, she recommends the proposal in state $M$. This implements the receiver's first-best outcome. If the mediator is aligned with the senders, her preferred recommendation in state $M$ is the status quo. However, the receiver must be willing to follow the mediator's recommendation. If the mediator recommends the status quo with certainty in state $L$ and
with probability $p$ in state $M$, the receiver is willing to follow a status-quo
recommendation only if her expected payoff from the proposal falls below $0$,\footnote{Here, following the standard convention in the Bayesian-persuasion
literature, we assume that the receiver follows the mediator's recommendation
when indifferent.} i.e.,
\[
    q_L^0 U_R(L)+p q_M^0 U_R(M)\leq 0.
\]
Hence the mediator can recommend the status quo in state $M$ with probability at most
\[
    p^*
    =
    \max\left\{
        p\in[0,1]
        \,\middle|\,
        q_L^0 U_R(L)+p q_M^0 U_R(M)\leq 0
    \right\}.
\]
If she recommends the status quo in state $M$ with probability $p^*$, the outcome is the senders' optimal Bayesian-persuasion outcome: the best outcome the senders could obtain if they pooled their dispersed information, learned the state, and committed to a state-contingent recommendation rule for the receiver.

More generally, for each $p\in[0,p^*]$, suppose the mediator recommends the status quo in state $L$, the proposal in state $H$, and, in state $M$, the status quo with probability $p$. As $p$ varies from $0$ to $p^*$, the induced allocations trace the Pareto frontier from the receiver-optimal outcome to the sender-optimal Bayesian-persuasion outcome. The parameter $p$ governs the division of the payoff trade-off in state $M$ between the senders and the receiver.

The theorem below shows that, even without directly observing the senders' signals, the mediator can asymptotically implement any allocation on the Pareto frontier of the full-information benchmark.

\begin{theorem}
    \label{thm:5}
    For every $p\in[0,p^*]$, there exists a sequence of feasible mechanisms $\{\phi_N\}_{N>1}$ satisfying\footnote{When \(q_M^0=0\), we omit the implementation requirement for state \(M\).}
    \begin{align*}
        &\lim_{N\to\infty}\mathbb{P}\!\left[\mathrm{proposal}\mid H;\phi_N\right] = 1,\\
        &\lim_{N\to\infty}\mathbb{P}\!\left[\mathrm{proposal}\mid M;\phi_N\right] = 1-p,\\
        &\lim_{N\to\infty}\mathbb{P}\!\left[\mathrm{proposal}\mid L;\phi_N\right] = 0.
    \end{align*}
\end{theorem}

\noindent \textbf{Without the disagreement state.} We begin with the case $q^0_M=0$, in which the disagreement state $M$ is absent. In this case, the Pareto frontier collapses to a single allocation: the status quo is chosen in state $L$ and the proposal is chosen in state $H$.

Let $T'$ denote the number of senders who receive signal $g$. As in the proof of \cref{thm:1}, there exists a cutoff $\TT_N$ such that
\[
    \mathbb{E}\left[U_S(\theta)\mid T'=t;N\right]\ge 0
    \iff t\ge \TT_N.
\]
Consider the following \textit{cutoff mechanism}:
\[
    \phi_N(t)=
    \begin{cases}
        1 & \text{if } t\ge \TT_N,\\
        0 & \text{if } t<\TT_N.
    \end{cases}
\]
That is, the mediator recommends the proposal if and only if at least $\TT_N$ senders report $g$. Now, suppose the receiver follows the mediator's recommendation. Under the cutoff $\TT_N$, if all other senders report truthfully, then truthful reporting by sender $i$ leads the mechanism
to select the policy that is optimal for the senders conditional on the realized signal
tally $T'$. Hence, truthful reporting is incentive compatible. As $N \to \infty$, the
law of large numbers implies that $T'$ identifies the state with probability approaching
one. The cutoff mechanism therefore asymptotically implements the senders' first-best outcome,
which coincides with the receiver's first-best outcome. Consequently, for all sufficiently large
$N$, the receiver is willing to follow the mediator's recommendation, so the mechanism is feasible.

More generally, the mediator can approach the common first-best outcome by using a simple binary voting mechanism with an arbitrary qualified-majority rule: each sender approves or rejects the proposal, and the mediator recommends the proposal whenever the approval share exceeds a fixed interior cutoff. This follows directly from the modern Condorcet jury theorem \citep[e.g.,][]{feddersen1997voting,feddersen1998convicting}.

\medskip
\noindent \textbf{With the disagreement state.} We now turn to the case $q^0_M>0$ and consider a mediator who aims to implement the receiver's first-best outcome, which corresponds to $p=0$ in \cref{thm:5}.

\begin{figure}
    \centering
    \pgfmathdeclarefunction{gauss}{3}{%
\pgfmathparse{1/(#3*sqrt(1))*exp(-((#1-#2)^2)/(2*#3^2))}%
}

\begin{tikzpicture}
\begin{axis}[
  no markers, 
  domain=0:17, 
  samples=100,
  ymin=0,
  axis lines=center, 
  every axis y label/.style={at=(current axis.above origin),anchor=south},
  every axis x label/.style={at=(current axis.right of origin),anchor=west},
  height=4cm, 
  width=15cm,
  xtick=\empty, 
  ytick=\empty,
   xlabel={$T$},
  enlargelimits=false, 
  clip=false, 
  axis on top,
  grid = major,
  hide y axis
  ]
\addplot [very thick,blue] {gauss(x, 4, 1};
\addlegendentry{$L$}
\addplot [very thick,brown] {gauss(x, 8, 1};
\addlegendentry{$M$}
\addplot [very thick,red] {gauss(x, 12, 1)};
\addlegendentry{$H$}
\node[below,blue] at (axis cs:4,0)  {$N\rho_L$}; 
\node[below,brown] at (axis cs:8,0)  {$N\rho_M$}; 
\node[below,red] at (axis cs:12,0)  {$N\rho_H$}; 
\node[below,green!50!black] at (axis cs:6,0)  {$\tilde{T}_N$};
\draw[blue, dashed] (axis cs:4,0) -- (axis cs:4,1);
\draw[brown, dashed] (axis cs:8,0) -- (axis cs:8,1);
\draw[red, dashed] (axis cs:12,0) -- (axis cs:12,1);
\draw[very thick, green!50!black](axis cs:6,-0.04) -- (axis cs:6,0.06);
\end{axis}
\end{tikzpicture}
    \caption{The distribution of the number of senders receiving $g$ in each state when $N$ is large. A cutoff mechanism $\TT'_N$ with $\TT'_N\in[N\rho_L,N\rho_M]$ approaches the receiver's first-best but is not incentive compatible.}
    \label{fig:4}
\end{figure}

We show first that for large $N$, any cutoff mechanism with cutoff $\TT'_N$ that approaches the receiver's first-best is not
incentive compatible. Conditional on state $\theta$, the number of senders receiving
signal $g$ follows a binomial distribution, $\operatorname{Bin}(N,\rho_\theta)$, and is
approximately normal with mean $N\rho_\theta$ and variance
$N\rho_\theta(1-\rho_\theta)$, as illustrated in \cref{fig:4}.  Because the receiver
prefers the status quo in state $L$ and the proposal in states $M$ and $H$, any cutoff $\TT'_N$ that approaches the receiver's first-best must satisfy $\TT'_N\in[N\rho_L,N\rho_M]$. For any such cutoff,
$\TT'_N/N\leq\rho_M<\rho_H$, so the binomial likelihood ratio between states $H$ and
$M$ is close to zero:
\[
    \frac{\mathbb{P}[\TT'_N\mid H;N]}
         {\mathbb{P}[\TT'_N\mid M;N]}
    \approx 0,
\]
as illustrated in \cref{fig:4}. The corresponding likelihood ratio for the pivotal event is therefore also close to
zero:
\[
    \frac{\mathbb{P}[\TT'_N-1\mid H;N-1]}
         {\mathbb{P}[\TT'_N-1\mid M;N-1]}
    \approx 0.
\]
Because the prior odds $q^0_H/q^0_M$ and the signal likelihood ratio $\rho_H/\rho_M$ are fixed, even a sender who receives signal $g$ assigns negligible probability to state $H$
conditional on being pivotal. Hence this sender prefers to report $b$, since he prefers the status quo in both states
$L$ and $M$, which violates incentive compatibility.

Furthermore, no qualified-majority rule implements the receiver's first-best outcome, because every such rule yields the senders' first-best outcome by the Condorcet jury theorem.

Now, consider the following \textit{step mechanism}:
\[
    \phi_N(t)=
    \begin{cases}
        1 & \text{if } t\ge \TT^{\beta}_N,\\
        \mu_N & \text{if } t\in[\TT^{\alpha}_N,\TT^{\beta}_N),\\
        0 & \text{if } t<\TT^{\alpha}_N.
    \end{cases}
\]
Under this mechanism, the probability of recommending the
proposal is weakly increasing in the number of senders reporting $g$, with upward
jumps at the two thresholds $\TT^{\alpha}_N$ and $\TT^{\beta}_N$.
Equivalently, after receiving the reports, the mediator randomizes between two cutoff
mechanisms, selecting $\TT^{\alpha}_N$ with probability $\mu_N$ and
$\TT^{\beta}_N$ with probability $1-\mu_N$. A sender can
be pivotal in either of two events: (\romannumeral 1) the realized cutoff is
$\TT^{\alpha}_N$ and exactly $\TT^{\alpha}_N-1$ of the other $N-1$ senders report
$g$; or (\romannumeral 2) the realized cutoff is $\TT^{\beta}_N$ and exactly
$\TT^{\beta}_N-1$ of the other $N-1$ senders report $g$.

We show that there exists a sequence of step mechanisms $\{\phi_N\}_{N>1}$ that asymptotically implements the receiver's first-best outcome, with
\[
    \frac{\TT_N^{\alpha}}{N}\in(\rho_L,\rho_M), \qquad \frac{\TT_N^{\beta}}{N}\to \rho_H, \qquad \mu_N\to 1,
\]
as illustrated in \cref{fig:5}. The reasoning is as follows. As shown earlier, under truthful reporting, the cutoff mechanism given by $\TT_N^{\alpha}$ would implement the receiver's first-best outcome with probability approaching one as $N\to\infty$, but it is not incentive compatible: conditional on being pivotal under $\TT_N^{\alpha}$, each sender assigns asymptotically negligible probability to state $H$ and therefore prefers the status quo regardless of his signal. To restore truthful reporting, the mediator introduces a second cutoff $\TT_N^{\beta}$, which generates a countervailing pivotal event: conditional on being pivotal under $\TT_N^{\beta}$, each sender infers that the state is $H$ with probability approaching one and therefore prefers the proposal regardless of his signal. By randomizing between the two cutoffs, the mediator can balance these opposing incentives and thereby induce truthful reporting. Finally, by choosing $\TT_N^{\beta}$ so that $\TT_N^{\beta}/N$ approaches $\rho_H$, the mediator can make the likelihood of the pivotal
event under $\TT_N^{\beta}$ arbitrarily large relative to that of the pivotal event under
$\TT_N^{\alpha}$. Consequently, only a vanishing weight on $\TT_N^{\beta}$ is
needed to offset the incentive to misreport at $\TT_N^{\alpha}$, and so to sustain truthful
reporting. This allows the mediator to choose $\mu_N\to 1$, so the step mechanism
asymptotically places all weight on $\TT_N^{\alpha}$ and therefore approaches the receiver's first-best outcome.

\begin{figure}
    \centering
    \pgfmathdeclarefunction{gauss}{3}{%
\pgfmathparse{1/(#3*sqrt(1))*exp(-((#1-#2)^2)/(2*#3^2))}%
}

\begin{tikzpicture}
\begin{axis}[
  no markers, 
  domain=0:17, 
  samples=100,
  ymin=0,
  axis lines=center, 
  every axis y label/.style={at=(current axis.above origin),anchor=south},
  every axis x label/.style={at=(current axis.right of origin),anchor=west},
  height=4cm, 
  width=15cm,
  xtick=\empty, 
  ytick=\empty,
   xlabel={$T$},
  enlargelimits=false, 
  clip=false, 
  axis on top,
  grid = major,
  hide y axis
  ]
\addplot [very thick,blue] {gauss(x, 4, 0.8)};
\addlegendentry{$L$}
\addplot [very thick,brown] {gauss(x, 8, 0.8)};
\addlegendentry{$M$}
\addplot [very thick,red] {gauss(x, 12, 0.8)};
\addlegendentry{$H$}
\node[below,blue] at (axis cs:4,0)  {$N\rho_L$}; 
\node[below,brown] at (axis cs:8,0)  {$N\rho_M$}; 
\node[below,red] at (axis cs:12,0)  {$N\rho_H$}; 
\node[above,green!50!black] at (axis cs:6,1.21)  {$\tilde{T}^{\alpha}_N$};
\node[above,green!50!black] at (axis cs:11.5,1.21)  {$\tilde{T}^{\beta}_N$};
\draw[blue, dashed] (axis cs:4,0) -- (axis cs:4,1.21);
\draw[brown, dashed] (axis cs:8,0) -- (axis cs:8,1.21);
\draw[red, dashed] (axis cs:12,0) -- (axis cs:12,1.21);
\draw[green!50!black, dashed] (axis cs:6,0) -- (axis cs:6,1.21);
\draw[green!50!black, dashed] (axis cs:11.5,0) -- (axis cs:11.5,1.21);
\end{axis}
\end{tikzpicture}
    \caption{A step mechanism approaching the receiver's first-best. The mediator chooses $\TT_N^{\alpha}\in(N\rho_L,N\rho_M)$, $\TT_N^{\beta}\approx N\rho_H$, and $\mu_N\approx 1$.}
    \label{fig:5}
\end{figure}

\medskip
The case in which the mediator is aligned with the receiver ($p=0$) resembles the case in which the receiver can commit to a mechanism based on the senders' reports, as examined by \citet{wolinsky2002eliciting}, \citet{gerardi2009aggregation}, and \citet{feng2019getting}. In the LMB setting (where there is no disagreement state), \citet{kattwinkel2024optimal} show that the receiver's optimal mechanism is an \textit{interval mechanism}: the receiver chooses the status quo if and only if the number of senders reporting $b$ is in an intermediate range, neither too high nor too low. Moreover, \citet{best2025divide} show that the interval mechanism is asymptotically optimal in a related setting with unconditionally independent signals. By contrast, in our setting (where there is a disagreement state), any interval mechanism that approaches the receiver's first-best fails to be incentive compatible for large $N$ and so cannot be optimal.\footnote{We can show that, for large $N$, if an interval mechanism approaches the receiver's first-best and incentivizes a sender to report signal $b$ truthfully, then it cannot also incentivize him to report signal $g$ truthfully.}

The argument above provides helpful perspective on our main model (with cheap-talk communication). There, the receiver cannot commit to the step mechanism described above, which induces truthful reporting. Instead, her best response in equilibrium is always a cutoff rule, which limits the amount of information that can be transmitted, as shown by \cref{thm:1,thm:3}.

By arguments analogous to those for $p=0$, any allocation on the Pareto frontier satisfying $p\in(0,p^*]$ can be asymptotically implemented by a sequence of step mechanisms $\{\phi_N\}_{N>1}$ satisfying
\[
    \frac{\TT_N^{\alpha}}{N}\in(\rho_L,\rho_M), \qquad \frac{\TT_N^{\beta}}{N}\in(\rho_M,\rho_H), \qquad \mu_N\to 1-p.
\]
The thresholds and mixing probability can be chosen jointly so that the two pivotal incentives exactly offset each other and truthful reporting is sustained. We note that the mediator can also asymptotically implement any allocation on the Pareto frontier by adopting a binary voting mechanism and randomizing over two qualified-majority rules.

Our results can also be connected to the growing literature studying mediated communication between one sender and one receiver \citep[e.g.,][]{goltsman2009mediation,salamanca2021value,corrao2023mediation}. In particular, \citet{corrao2023bounds} examine a setting in which the sender's payoff is state-independent and the mediator is aligned with the sender. Unlike Bayesian persuasion, in which a sender with commitment power can communicate directly with the receiver, mediation requires the mediator to elicit the sender's information. Thus, the gap between the sender's Bayesian-persuasion payoff and his payoff under optimal mediation measures the cost of elicitation, whereas the gap between his Bayesian-persuasion payoff and his maximal cheap-talk payoff measures the value of commitment. \citet{corrao2023bounds} show that whenever the sender attains his Bayesian-persuasion payoff under optimal mediation, he must also attain that payoff under cheap-talk communication. That is, no cost of elicitation implies no value of commitment.

Although our setting also features state-dependent sender payoffs, it differs crucially from that of \citet{corrao2023bounds} in that there are multiple senders. This means each sender's incentive depends not only on his own information but also on his assessment of the other senders' information and reports. This strategic interdependence gives the mediator additional instruments for eliciting information through the mediation mechanism. As $N\to\infty$, these instruments become powerful enough for the cost of elicitation to vanish; that is, they enable the mediator to asymptotically attain the senders'
Bayesian-persuasion payoff, corresponding to $p=p^*$ in
\cref{thm:5}. By contrast, with
direct cheap-talk communication (our main model), when $q_M^0>0$, the senders' maximal payoff remains strictly below their
Bayesian-persuasion payoff. Thus, in our multi-sender environment,
a vanishing cost of elicitation is compatible with a strictly positive
value of commitment.

\section{Concluding Remarks}

This paper studies cheap-talk communication between a single receiver and multiple senders, primarily in large populations. In our model, the senders' message spaces are coarse, and the receiver observes the aggregate message tally. The senders' and receiver's preferences are aligned in the low and high states, but opposed in the medium state. We show that cheap-talk communication in this setting is subject to sharp limitations: information aggregation fails, and communication may unravel completely. By contrast, mediated communication can be used to asymptotically implement any Pareto-efficient outcome on the frontier between the receiver's first-best and the senders' optimal Bayesian-persuasion outcome.

A final remark is that, under certain circumstances, cheap talk with decentralized information (i.e., with information dispersed across a crowd) may outperform cheap talk with information centralized in one individual (i.e., with a single sender observing multiple signals). In the latter case, the sender has a stronger informational advantage, which amplifies adverse selection and reduces the receiver's willingness to trust his recommendation when the conflict of interest is large. By contrast, when there are multiple senders, dispersed information fails to aggregate, creating noise in the senders' messages. As shown by \citet{blume2007noisy}, such noise can serve as a commitment device that mitigates the conflict and improves communication. A detailed discussion can be found in Appendix~\ref{ap:B}.

{
\normalsize{}
\nocite{*}
\bibliographystyle{abbrvnat}
\bibliography{reference}

\begin{thebibliography}{46}
\providecommand{\natexlab}[1]{#1}
\providecommand{\url}[1]{\texttt{#1}}
\expandafter\ifx\csname urlstyle\endcsname\relax
  \providecommand{\doi}[1]{doi: #1}\else
  \providecommand{\doi}{doi: \begingroup \urlstyle{rm}\Url}\fi

\bibitem[Austen-Smith(1990)]{austen1990information}
D.~Austen-Smith.
\newblock Information transmission in debate.
\newblock \emph{American Journal of Political Science}, 34:\penalty0 124--152, 1990.

\bibitem[Austen-Smith(1993)]{austen1993interested}
D.~Austen-Smith.
\newblock Interested experts and policy advice: Multiple referrals under open rule.
\newblock \emph{Games and Economic Behavior}, 5:\penalty0 3--43, 1993.

\bibitem[Battaglini(2004)]{battaglini2004policy}
M.~Battaglini.
\newblock Policy advice with imperfectly informed experts.
\newblock \emph{B.E. Journal of Theoretical Economics}, 4\penalty0 (1), 2004.

\bibitem[Battaglini(2017)]{battaglini2017public}
M.~Battaglini.
\newblock Public protests and policy making.
\newblock \emph{Quarterly Journal of Economics}, 132:\penalty0 485--549, 2017.

\bibitem[Best et~al.(2025)Best, Quigley, Saeedi, and Shourideh]{best2025divide}
J.~Best, D.~Quigley, M.~Saeedi, and A.~Shourideh.
\newblock Divide or confer: Aggregating information without verification.
\newblock Working paper, 2025.

\bibitem[Blume et~al.(2007)Blume, Board, and Kawamura]{blume2007noisy}
A.~Blume, O.~J. Board, and K.~Kawamura.
\newblock Noisy talk.
\newblock \emph{Theoretical Economics}, 2:\penalty0 395--440, 2007.

\bibitem[Chakraborty and Ghosh(2003)]{chakraborty2003efficient}
A.~Chakraborty and P.~Ghosh.
\newblock Efficient equilibria and information aggregation in common interest voting games.
\newblock Working paper, 2003.

\bibitem[Chen et~al.(2008)Chen, Kartik, and Sobel]{chen2008selecting}
Y.~Chen, N.~Kartik, and J.~Sobel.
\newblock Selecting cheap-talk equilibria.
\newblock \emph{Econometrica}, 76:\penalty0 117--136, 2008.

\bibitem[Cho and Sobel(1990)]{cho1990strategic}
I.-K. Cho and J.~Sobel.
\newblock Strategic stability and uniqueness in signaling games.
\newblock \emph{Journal of Economic Theory}, 50:\penalty0 381--413, 1990.

\bibitem[Corrao(2023)]{corrao2023mediation}
R.~Corrao.
\newblock Mediation markets: The case of soft information.
\newblock Working paper, 2023.

\bibitem[Corrao and Dai(2023)]{corrao2023bounds}
R.~Corrao and Y.~Dai.
\newblock The bounds of mediated communication.
\newblock Working paper, 2023.

\bibitem[Currarini et~al.(2020)Currarini, Ursino, and Chand]{currarini2020strategic}
S.~Currarini, G.~Ursino, and A.~K.~S. Chand.
\newblock Strategic transmission of correlated information.
\newblock \emph{Economic Journal}, 130:\penalty0 2175--2206, 2020.

\bibitem[Dekel and Piccione(2000)]{dekel2000sequential}
E.~Dekel and M.~Piccione.
\newblock Sequential voting procedures in symmetric binary elections.
\newblock \emph{Journal of Political Economy}, 108:\penalty0 34--55, 2000.

\bibitem[Duggan and Martinelli(2001)]{duggan2001bayesian}
J.~Duggan and C.~Martinelli.
\newblock A {Bayesian} model of voting in juries.
\newblock \emph{Games and Economic Behavior}, 37:\penalty0 259--294, 2001.

\bibitem[Ekmekci and Lauermann(2024)]{ekmekci2024information}
M.~Ekmekci and S.~Lauermann.
\newblock Information aggregation in large protests: A continuum model.
\newblock Working paper, 2024.

\bibitem[Feddersen and Pesendorfer(1997)]{feddersen1997voting}
T.~Feddersen and W.~Pesendorfer.
\newblock Voting behavior and information aggregation in elections with private information.
\newblock \emph{Econometrica}, 65:\penalty0 1029--1058, 1997.

\bibitem[Feddersen and Pesendorfer(1998)]{feddersen1998convicting}
T.~Feddersen and W.~Pesendorfer.
\newblock Convicting the innocent: The inferiority of unanimous jury verdicts under strategic voting.
\newblock \emph{American Political Science Review}, 92:\penalty0 23--35, 1998.

\bibitem[Feng and Wu(2019)]{feng2019getting}
T.~Feng and Q.~Wu.
\newblock Getting information from your enemies.
\newblock Working paper, 2019.

\bibitem[Fey(1997)]{fey1997stability}
M.~Fey.
\newblock Stability and coordination in {Duverger}'s law: A formal model of preelection polls and strategic voting.
\newblock \emph{American Political Science Review}, 91\penalty0 (1):\penalty0 135--147, 1997.

\bibitem[Forges(1986)]{forges1986approach}
F.~Forges.
\newblock An approach to communication equilibria.
\newblock \emph{Econometrica}, 54\penalty0 (6):\penalty0 1375--1385, 1986.

\bibitem[Frick et~al.(2023)Frick, Iijima, and Ishii]{frick2023belief}
M.~Frick, R.~Iijima, and Y.~Ishii.
\newblock Belief convergence under misspecified learning: A martingale approach.
\newblock \emph{Review of Economic Studies}, 90:\penalty0 781--814, 2023.

\bibitem[Frick et~al.(2024{\natexlab{a}})Frick, Iijima, and Ishii]{frick2024multidimensional}
M.~Frick, R.~Iijima, and Y.~Ishii.
\newblock Multidimensional screening with rich consumer data.
\newblock Working paper, 2024{\natexlab{a}}.

\bibitem[Frick et~al.(2024{\natexlab{b}})Frick, Iijima, and Ishii]{frick2024welfare}
M.~Frick, R.~Iijima, and Y.~Ishii.
\newblock Welfare comparisons for biased learning.
\newblock \emph{American Economic Review}, 114:\penalty0 1612--1649, 2024{\natexlab{b}}.

\bibitem[Galeotti et~al.(2013)Galeotti, Ghiglino, and Squintani]{galeotti2013strategic}
A.~Galeotti, C.~Ghiglino, and F.~Squintani.
\newblock Strategic information transmission networks.
\newblock \emph{Journal of Economic Theory}, 148:\penalty0 1751--1769, 2013.

\bibitem[Gentzkow and Kamenica(2017{\natexlab{a}})]{gentzkow2016competition}
M.~Gentzkow and E.~Kamenica.
\newblock Competition in persuasion.
\newblock \emph{Review of Economic Studies}, 84:\penalty0 300--322, 2017{\natexlab{a}}.

\bibitem[Gentzkow and Kamenica(2017{\natexlab{b}})]{gentzkow2017bayesian}
M.~Gentzkow and E.~Kamenica.
\newblock Bayesian persuasion with multiple senders and rich signal spaces.
\newblock \emph{Games and Economic Behavior}, 104:\penalty0 411--429, 2017{\natexlab{b}}.

\bibitem[Gerardi et~al.(2009)Gerardi, McLean, and Postlewaite]{gerardi2009aggregation}
D.~Gerardi, R.~McLean, and A.~Postlewaite.
\newblock Aggregation of expert opinions.
\newblock \emph{Games and Economic Behavior}, 65\penalty0 (2):\penalty0 339--371, 2009.

\bibitem[Goltsman et~al.(2009)Goltsman, H{\"o}rner, Pavlov, and Squintani]{goltsman2009mediation}
M.~Goltsman, J.~H{\"o}rner, G.~Pavlov, and F.~Squintani.
\newblock Mediation, arbitration and negotiation.
\newblock \emph{Journal of Economic Theory}, 144:\penalty0 1397--1420, 2009.

\bibitem[Hagenbach and Koessler(2010)]{hagenbach2010strategic}
J.~Hagenbach and F.~Koessler.
\newblock Strategic communication networks.
\newblock \emph{Review of Economic Studies}, 77:\penalty0 1072--1099, 2010.

\bibitem[Ivanov(2010)]{ivanov2010communication}
M.~Ivanov.
\newblock Communication via a strategic mediator.
\newblock \emph{Journal of Economic Theory}, 145:\penalty0 869--884, 2010.

\bibitem[Kattwinkel and Winter(2024)]{kattwinkel2024optimal}
D.~Kattwinkel and A.~Winter.
\newblock Optimal decision mechanisms for committees: Acquitting the guilty.
\newblock Working paper, 2024.

\bibitem[Kolotilin and Li(2021)]{kolotilin2021relational}
A.~Kolotilin and H.~Li.
\newblock Relational communication.
\newblock \emph{Theoretical Economics}, 16:\penalty0 1391--1430, 2021.

\bibitem[Krishna and Morgan(2001)]{krishna2001model}
V.~Krishna and J.~Morgan.
\newblock A model of expertise.
\newblock \emph{Quarterly Journal of Economics}, 116:\penalty0 747--775, 2001.

\bibitem[Levit and Malenko(2011)]{levit2011nonbinding}
D.~Levit and N.~Malenko.
\newblock Nonbinding voting for shareholder proposals.
\newblock \emph{Journal of Finance}, 66:\penalty0 1579--1614, 2011.

\bibitem[Lyu et~al.(2023)Lyu, Suen, and Zhang]{lyu2023coarse}
Q.~Lyu, W.~Suen, and Y.~Zhang.
\newblock Coarse information design.
\newblock Working paper, 2023.

\bibitem[Meyer et~al.(2019)Meyer, Moreno~de Barreda, and Nafziger]{meyer2019robustness}
M.~Meyer, I.~Moreno~de Barreda, and J.~Nafziger.
\newblock Robustness of full revelation in multisender cheap talk.
\newblock \emph{Theoretical Economics}, 14\penalty0 (4):\penalty0 1203--1235, 2019.

\bibitem[Milgrom and Roberts(1986)]{milgrom1986relying}
P.~Milgrom and J.~Roberts.
\newblock Relying on the information of interested parties.
\newblock \emph{RAND Journal of Economics}, 17:\penalty0 18--32, 1986.

\bibitem[Milgrom and Shannon(1994)]{milgrom1994monotone}
P.~Milgrom and C.~Shannon.
\newblock Monotone comparative statics.
\newblock \emph{Econometrica}, 62:\penalty0 157--180, 1994.

\bibitem[Morgan and Stocken(2008)]{morgan2008information}
J.~Morgan and P.~C. Stocken.
\newblock Information aggregation in polls.
\newblock \emph{American Economic Review}, 98:\penalty0 864--896, 2008.

\bibitem[Moscarini and Smith(2002)]{moscarini2002law}
G.~Moscarini and L.~Smith.
\newblock The law of large demand for information.
\newblock \emph{Econometrica}, 70:\penalty0 2351--2366, 2002.

\bibitem[Mu et~al.(2021)Mu, Pomatto, Strack, and Tamuz]{mu2021blackwell}
X.~Mu, L.~Pomatto, P.~Strack, and O.~Tamuz.
\newblock From {Blackwell} dominance in large samples to {R{\'e}nyi} divergences and back again.
\newblock \emph{Econometrica}, 89:\penalty0 475--506, 2021.

\bibitem[Myerson(1982)]{myerson1982optimal}
R.~B. Myerson.
\newblock Optimal coordination mechanisms in generalized principal--agent problems.
\newblock \emph{Journal of Mathematical Economics}, 10\penalty0 (1):\penalty0 67--81, 1982.

\bibitem[Myerson(1998)]{myerson1998extended}
R.~B. Myerson.
\newblock Extended {Poisson} games and the {Condorcet} jury theorem.
\newblock \emph{Games and Economic Behavior}, 25:\penalty0 111--131, 1998.

\bibitem[Razin(2003)]{razin2003signaling}
R.~Razin.
\newblock Signaling and election motivations in a voting model with common values and responsive candidates.
\newblock \emph{Econometrica}, 71:\penalty0 1083--1119, 2003.

\bibitem[Salamanca(2021)]{salamanca2021value}
A.~Salamanca.
\newblock The value of mediated communication.
\newblock \emph{Journal of Economic Theory}, 192:\penalty0 105191, 2021.

\bibitem[Wolinsky(2002)]{wolinsky2002eliciting}
A.~Wolinsky.
\newblock Eliciting information from multiple experts.
\newblock \emph{Games and Economic Behavior}, 41:\penalty0 141--160, 2002.

\end{thebibliography}
}

\begin{appendix}

\newcommand{\tdt}{\tilde{t}}
\newcommand{\btt}{\boldsymbol{\tdt}}

\section{A Large Deviations Approach}
\label{ap:a}

We sketch the final step of the proof of \cref{prop:4} to illustrate the use of large deviations theory to study communication and voting games. We aim to show that, along every sequence of monotone equilibria, the senders' equilibrium strategies must become vanishingly uninformative as $N\to\infty$.

For simplicity, suppose that each sender can send a message $z\in\{z_1,z_2,z_3\}\subset\mathbb R$, where $z_1<z_2<z_3$. Fix a monotone sender strategy $\boldsymbol P$ and suppose that it is \emph{informative}, in the sense that at least two rows of $\boldsymbol P$ are linearly independent. We also assume that all three messages are active
under \(\boldsymbol P\). We start with the case in which all senders use $\boldsymbol P$ for each $N$. Let $\psi_N:\Delta^3(N)\to\{0,1\}$ denote the receiver's best response for each $N$, which is monotone.

For each state $\theta\in\{L,M,H\}$, define
\[
    g_\theta(z_k)
    :=
    \mathbb P[z=z_k\mid\theta;\boldsymbol P],
    \qquad k=1,2,3,
\]
and let
\[
    G_\theta
    =
    \bigl(g_\theta(z_1),g_\theta(z_2),g_\theta(z_3)\bigr)
\]
denote the distribution of a sender's message in state $\theta$. 
Since (i) the primitive signal distributions satisfy the strict MLRP, and (ii)
$\boldsymbol P$ is monotone and informative, the induced message distributions
$\{G_L,G_M,G_H\}$ satisfy the weak MLRP and are pairwise distinct.

\paragraph{Pivotal events.}

Let
\[
    E_N
    =
    \left\{
    \tilde{\boldsymbol t}
    =
    (\tilde t_1,\tilde t_2,\tilde t_3)
    \in\Delta^3(N-1)
    \;\middle|\;
    \psi_N(\tilde t_1+1,\tilde t_2,\tilde t_3)
    \neq
    \psi_N(\tilde t_1,\tilde t_2,\tilde t_3+1)
    \right\}.
\]
Thus, $E_N$ is the set of message counts of the other $N-1$ senders for which the remaining sender can change the receiver's decision by sending $z_3$ rather than $z_1$. Since $\psi_N$ is monotone,
\[
\psi_N(\tilde t_1+1,\tilde t_2,\tilde t_3)
\leq
\psi_N(\tilde t_1,\tilde t_2+1,\tilde t_3)
\leq
\psi_N(\tilde t_1,\tilde t_2,\tilde t_3+1).
\]
Hence, each \textit{pivotal event}---i.e., each message count of the other $N-1$ senders for which the remaining sender can change the receiver's decision by sending one message rather than another---must belong to $E_N$. We therefore call $E_N$ \emph{the set of pivotal events}. 

Because $\boldsymbol P$ is informative, for all sufficiently large $N$,
the receiver chooses the status quo at message counts near $G_L$ and the
proposal at message counts near $G_M$ and $G_H$. Hence, $E_N$ is
nonempty and has positive probability in every state.

The key claim is
\begin{equation}
\label{eqn:16}
    \lim_{N\to\infty}
    \frac{\mathbb P[E_N\mid H;\boldsymbol P]}
         {\mathbb P[E_N\mid L;\boldsymbol P]}
    =0
    \quad\text{and}\quad
    \lim_{N\to\infty}
    \frac{\mathbb P[E_N\mid H;\boldsymbol P]}
         {\mathbb P[E_N\mid M;\boldsymbol P]}
    =0,
\end{equation}
i.e., the probability of being pivotal is vanishingly smaller in state $H$ than in either state $L$ or $M$. We next explain why this claim holds.

\paragraph{From message counts to message frequencies.}

For each
$\tilde{\boldsymbol t}\in\Delta^3(N-1)$, define its empirical frequency by
\[
    \bar\gamma(\tilde{\boldsymbol t})
    =
    \left(
    \frac{\tilde t_1}{N-1},
    \frac{\tilde t_2}{N-1},
    \frac{\tilde t_3}{N-1}
    \right)
    \in\Delta^3.
\]
Define the Kullback--Leibler divergence from $\gamma$ to $G_\theta$ by
\[
KL(\gamma,G_\theta)
=
\sum_{k=1}^3
\gamma_k\log\frac{\gamma_k}{g_\theta(z_k)},
\]
with the convention $0\log 0=0$.

For any two states $\theta,\theta'\in\{L,M,H\}$, 
\begin{equation}
\label{eqn:17}
\begin{split}
    \frac{\mathbb P[\tilde{\boldsymbol t}\mid\theta;\boldsymbol P]}
         {\mathbb P[\tilde{\boldsymbol t}\mid\theta';\boldsymbol P]}
    &=
    \prod_{k=1}^{3}
    \left[
    \frac{g_\theta(z_k)}{g_{\theta'}(z_k)}
    \right]^{\tilde t_k}                                                     
    =
    \exp\left\{
    (N-1)
    \left[
    KL\bigl(\bar\gamma(\tilde{\boldsymbol t}),G_{\theta'}\bigr)
    -
    KL\bigl(\bar\gamma(\tilde{\boldsymbol t}),G_\theta\bigr)
    \right]
    \right\}.
\end{split}
\end{equation}
More generally, uniformly over
$\tilde{\boldsymbol t}\in\Delta^3(N-1)$,
\begin{equation}
\label{eqn:18}
\mathbb P[\tilde{\boldsymbol t}\mid\theta;\boldsymbol P]
    =
    \exp\left\{
    -(N-1)KL\bigl(\bar\gamma(\tilde{\boldsymbol t}),G_\theta\bigr)
    +O(\log N)
    \right\}. 
\end{equation}
Note that $O(\log N)$ comes from the multinomial coefficient. Thus, after normalizing message counts by $N-1$, the probability of a message-count vector $\tilde{\boldsymbol t}$ in state $\theta$ is governed, at the exponential scale, by the large-deviation cost $KL(\bar\gamma(\tilde{\boldsymbol t}),G_\theta)$ of its empirical frequency
$\bar\gamma(\tilde{\boldsymbol t})$: a lower cost corresponds to an exponentially
higher probability.

\paragraph{From pivotal events to pivotal frequencies.}

Let
\[
    F
    =
    \left\{
    \gamma\in\Delta^3
    \;\middle|\;
    KL(\gamma,G_L)
    =
    \min\{KL(\gamma,G_M),KL(\gamma,G_H)\}
    \right\}
\]
be the \emph{set of pivotal frequencies}. The set $F$ is the large-deviation
boundary between frequencies that favor state $L$ and those that favor either
$M$ or $H$. By \eqref{eqn:17}, for any message-count vector $\tilde{\boldsymbol t}$ with $\bar\gamma(\tilde{\boldsymbol t})\in F$, its likelihood in state $L$ has the same exponential rate as its combined likelihood in states $M$ and $H$.

We use $F$ to characterize $E_N$ asymptotically. For
each $\epsilon>0$, let
\[
    F(\epsilon)
    =
    \left\{
    \gamma\in\Delta^3
    \;\middle|\;
    \left|
    KL(\gamma,G_L)
    -
    \min\{KL(\gamma,G_M),KL(\gamma,G_H)\}
    \right|
    <\epsilon
    \right\}.
\]
We claim that, for every $\epsilon>0$, there exists
$\widetilde N(\epsilon)$ such that, if $N>\widetilde N(\epsilon)$,
\begin{equation}
\label{eqn:19}
    E_N
    \subset
    \left\{
    \tilde{\boldsymbol t}\in\Delta^3(N-1)
    \;\middle|\;
    \bar\gamma(\tilde{\boldsymbol t})\in F(\epsilon)
    \right\}.
\end{equation}
To see this, take $N$ sufficiently large and first suppose that
\begin{equation}
\label{eqn:20}
    KL(\bar\gamma(\tilde{\boldsymbol t}),G_L)
    \leq
    \min\{
    KL(\bar\gamma(\tilde{\boldsymbol t}),G_M),
    KL(\bar\gamma(\tilde{\boldsymbol t}),G_H)
    \}
    -\epsilon. 
\end{equation}
By \eqref{eqn:17},
\begin{equation*}
    \frac{\mathbb P[\tilde{\boldsymbol t}\mid L;\boldsymbol P]}
         {\mathbb P[\tilde{\boldsymbol t}\mid M;\boldsymbol P]}
    \geq \exp\{(N-1)\epsilon\}
    \quad\text{and}\quad
    \frac{\mathbb P[\tilde{\boldsymbol t}\mid L;\boldsymbol P]}
         {\mathbb P[\tilde{\boldsymbol t}\mid H;\boldsymbol P]}
    \geq \exp\{(N-1)\epsilon\}.
\end{equation*}
Therefore, conditional on the other $N-1$ senders generating the message-count
vector $\tilde{\boldsymbol t}$, the posterior odds of $M$ and $H$ relative to
$L$ are exponentially small. Since the remaining sender's message multiplies these posterior odds only by factors bounded independently of $N$, after observing either $(\tilde t_1+1,\tilde t_2,\tilde t_3)$ or $(\tilde t_1,\tilde t_2,\tilde t_3+1)$, the receiver's posterior probability of $L$ approaches $1$, and she always chooses the status quo.

Similarly, if
\begin{equation}
\label{eqn:21}
    KL(\bar\gamma(\tilde{\boldsymbol t}),G_L)
    \geq
    \min\{
    KL(\bar\gamma(\tilde{\boldsymbol t}),G_M),
    KL(\bar\gamma(\tilde{\boldsymbol t}),G_H)
    \}
    +\epsilon,
\end{equation}
then, conditional on the other $N-1$ senders generating
$\tilde{\boldsymbol t}$, either $M$ or $H$ is exponentially more likely than
$L$. After observing either $(\tilde t_1+1,\tilde t_2,\tilde t_3)$ or $(\tilde t_1,\tilde t_2,\tilde t_3+1)$, the receiver's posterior probability of $L$ approaches $0$, and she always chooses the proposal. Thus,  for sufficiently large $N$, if either \eqref{eqn:20} or \eqref{eqn:21} holds, we have $\tilde{\boldsymbol t}\notin E_N$, proving \eqref{eqn:19}.

Thus, although $E_N$ is a subset of the growing lattice $\Delta^3(N-1)$, the inclusion in \eqref{eqn:19} characterizes it asymptotically by the fixed set $F$: as $N$ grows large, every pivotal message-count vector has an empirical frequency arbitrarily close to $F$.
Moreover, \eqref{eqn:17} and \eqref{eqn:18} express likelihood ratios and message-count probabilities, respectively, at the exponential scale in terms of the large-deviation costs associated with empirical frequencies. Together, these results reduce the original problem on the growing lattice $\Delta^3(N-1)$ to a large-deviation problem on the fixed simplex $\Delta^3$. We can therefore reduce the analysis to the large-deviation costs $KL(\gamma,G_\theta)$ for frequencies $\gamma\in F$, rather than tracking the exact probabilities $\mathbb P[\tilde{\boldsymbol t}\mid\theta;\boldsymbol P]$ of individual pivotal message-count vectors $\tilde{\boldsymbol t}\in E_N$.

\paragraph{The most likely pivotal frequency.}

We next identify the frequencies in $F$ that are most likely to occur.  For each $\theta\in\{M,H\}$, define\footnote{Note that $KL(\gamma,G_\theta)$ is continuous and strictly convex in $\gamma\in\Delta^3$. Moreover, the equality constraint in \eqref{eqn:22} is affine, and its intersection with $\Delta^3$ is closed and nonempty. Hence, the minimum is attained, and the minimizer is unique.}
\begin{equation}
\label{eqn:22}
    \gamma^*_{L,\theta}
    =
    \argmin_{\gamma\in\Delta^3}
    \left\{
    KL(\gamma,G_L)
    \;\middle|\;
    KL(\gamma,G_L)=KL(\gamma,G_\theta)
    \right\}.
\end{equation}
The equality constraint identifies frequencies with the same large-deviation costs $KL(\gamma,G_L)$ and $KL(\gamma,G_\theta)$ in states $L$ and $\theta$. By \eqref{eqn:17}, for such frequencies, the likelihoods in states $L$ and $\theta$ have the same exponential rate. Among these frequencies, $\gamma^*_{L,\theta}$ minimizes the common large-deviation cost and is therefore the most likely at the exponential scale in both states. Any other frequency satisfying the equality constraint has a strictly higher common cost and is exponentially less likely in both states.

The common value
\begin{equation}
\label{eqn:23}
    C_{L,\theta}
    :=
    KL(\gamma^*_{L,\theta},G_L)
    =
    KL(\gamma^*_{L,\theta},G_\theta)
\end{equation}
is the Chernoff information between $G_L$ and
$G_\theta$, which measures the statistical distinguishability.\footnote{Equivalently,
$
    C_{L,\theta}
    =
    -\min_{\lambda\in[0,1]}
    \log\left[
    \sum_{k=1}^{3}
    g_L(z_k)^\lambda g_\theta(z_k)^{1-\lambda}
    \right],
$} Intuitively, $C_{L,\theta}$ captures one half of the statistical distance between $G_L$ and $G_\theta$: it is the common large-deviation cost of reaching their closest ``midpoint.''

The weak MLRP and the nontriviality of the $L$--$M$ and $M$--$H$
comparisons imply
\begin{equation}
\label{eqn:24}
    C_{L,M}<C_{L,H},
\end{equation}
which says that, in terms of Chernoff information, $L$ is statistically closer
to its neighboring state $M$ than to $H$.\footnote{This comparison was first established by
\citet{frick2023belief}.} By \eqref{eqn:23}, $\gamma^*_{L,M}$ therefore has a lower common large-deviation cost in states $L$ and $M$ than $\gamma^*_{L,H}$ has in states $L$ and $H$. Hence, $\gamma^*_{L,M}$ is exponentially more likely than $\gamma^*_{L,H}$ in state $L$, and it is also exponentially more likely in state $M$ than $\gamma^*_{L,H}$ is in state $H$.

We can also show that
\begin{equation}
\label{eqn:25}
    KL(\gamma^*_{L,M},G_L)
    =
    KL(\gamma^*_{L,M},G_M)
    <
    KL(\gamma^*_{L,M},G_H).
\end{equation}
Thus, $\gamma^*_{L,M}$ has the same large-deviation cost in states $L$ and $M$, and this cost is strictly lower than its cost in state $H$. Hence, $\gamma^*_{L,M}$ is exponentially more likely in states $L$ and $M$ than in state $H$. Intuitively, the MLRP implies the message distributions in the MLR order, with $G_M$ lying between $G_L$ and $G_H$. The frequency $\gamma^*_{L,M}$ is the closest ``midpoint'' between $G_L$ and $G_M$. Because $G_H$ is shifted further toward higher messages, reaching $\gamma^*_{L,M}$ requires a larger deviation in state $H$.

Equation \eqref{eqn:25} implies that $\gamma^*_{L,M}\in F$. For each
$\theta\in\{M,H\}$, let
\[
    F_\theta
    =
    \left\{
    \gamma\in\Delta^3
    \;\middle|\;
    KL(\gamma,G_L)=KL(\gamma,G_\theta)
    \right\}.
\]
By the definition of $F$, it follows that $F\subseteq F_M\cup F_H$. Moreover, \eqref{eqn:22} and \eqref{eqn:23} imply that
\[
    KL(\gamma^*_{L,M},G_L)
    =
    \min_{\gamma\in F_M}KL(\gamma,G_L)
    \quad
    \text{and}
    \quad
    KL(\gamma^*_{L,H},G_L)
    =
    \min_{\gamma\in F_H}KL(\gamma,G_L).
\]
By \eqref{eqn:24}, we have
\[
    KL(\gamma^*_{L,M},G_L)
    <
    KL(\gamma^*_{L,H},G_L).
\]
Since $\gamma^*_{L,M}\in F$ and every $\gamma\in F$ belongs to at least one of $F_M$ and $F_H$, it follows that
\begin{equation}
\label{eqn:26}
    \gamma^*_{L,M}
    =
    \argmin_{\gamma\in F}KL(\gamma,G_L)
    =
    \argmin_{\gamma\in F}
    \min\{KL(\gamma,G_M),KL(\gamma,G_H)\}.
\end{equation}
The equality between the two minimization problems in \eqref{eqn:26} follows from the definition of $F$. Thus, among all pivotal frequencies, $\gamma^*_{L,M}$ has the lowest large-deviation cost in state $L$, equivalently, the lowest cost against the alternative states. By \eqref{eqn:18}, it therefore maximizes, at the exponential scale, both the likelihood in state $L$ and the combined likelihood in states $M$ and $H$. We refer to $\gamma^*_{L,M}$ as the \emph{most likely
pivotal frequency}. \cref{fig:6} illustrates the set of pivotal frequencies and the most likely one.

\begin{figure}
    \centering
    \begin{tikzpicture}[x=1.30cm,y=1.30cm,font=\small,line cap=round,line join=round]
  \coordinate (Z1) at (0,0);
  \coordinate (Z2) at (3,5.1962);
  \coordinate (Z3) at (6,0);

  \coordinate (LMtop)   at (1.4500,2.5115);
  \coordinate (LMright) at (5.4011,1.0372);
  \coordinate (LHtop)   at (2.0906,3.6211);
  \coordinate (LHbase)  at (3.7200,0);
  \coordinate (X)       at (2.8200,2.0000);

  \coordinate (GL)  at (1.5200,1.8800);
  \coordinate (GM)  at (2.3400,2.6600);
  \coordinate (GH)  at (4.3200,1.9500);
  \coordinate (rLM) at (1.9432,2.3274);
  \coordinate (rLH) at (2.9640,1.6800);

  \draw[line width=0.8pt] (Z1)--(Z2)--(Z3)--cycle;
  \node[below left=1pt,font=\large]  at (Z1) {$z_1$};
  \node[above=2pt,font=\large]       at (Z2) {$z_2$};
  \node[below right=1pt,font=\large] at (Z3) {$z_3$};

  \draw[densely dashed,line width=0.65pt] (LMtop)--(LMright);
  \draw[densely dashed,line width=0.65pt] (LHtop)--(LHbase);

  \draw[red,line width=1.55pt] (LMtop)--(X)--(LHbase);
  \node[anchor=east,text=red,font=\Large] at (3.13,0.78) {$F$};

  \node[circle,fill,inner sep=1.75pt] at (GL) {};
  \node[anchor=north,xshift=-4pt,yshift=-3pt] at (GL) {$G_L$};

  \node[circle,fill,inner sep=1.75pt] at (GM) {};
  \node[anchor=south east,xshift=1pt,yshift=1pt] at (GM) {$G_M$};

  \node[circle,fill,inner sep=1.75pt] at (GH) {};
  \node[anchor=south east,xshift=0pt,yshift=1pt] at (GH) {$G_H$};

  \node[draw,fill=white,line width=0.7pt,inner sep=1.55pt] at (rLM) {};
  \node[anchor=north west,xshift=-5.4pt,yshift=-8pt,fill=white,inner sep=1pt]
    at (rLM) {$\gamma^*_{L,M}$};

  \node[draw,fill=white,line width=0.7pt,inner sep=1.55pt] at (rLH) {};
  \node[anchor=north west,xshift=10pt,yshift=-4pt,fill=white,inner sep=1pt]
    at (rLH) {$\gamma^*_{L,H}$};
\end{tikzpicture}
    \caption{Pivotal frequencies in the probability simplex. The vertices
$z_1$, $z_2$, and $z_3$ represent degenerate distributions assigning
probability one to the corresponding messages. The points $G_L$, $G_M$,
and $G_H$ are the induced message distributions in states $L$, $M$, and
$H$. The dashed lines represent $F_M$ and $F_H$, while their red portions form the set of pivotal frequencies $F$. The points $\gamma^*_{L,M}$ and $\gamma^*_{L,H}$ are the Chernoff points for the pairs $(L,M)$ and $(L,H)$. Among all frequencies in $F$, $\gamma^*_{L,M}$ has the smallest large-deviation cost relative to $G_L$, and hence is the most likely pivotal frequency.}
    \label{fig:6}
\end{figure}

\paragraph{Uniform separation from state $H$.}

We claim that\footnote{Since $F$ is a closed subset of the compact simplex $\Delta^3$, it is compact. Moreover, $KL(\gamma,G_H)$ is well-defined and continuous on $\Delta^3$. Hence, the minimum in \eqref{eqn:27} is attained.}
\begin{equation}
\label{eqn:27}
    C_{L,M}
    =
    KL(\gamma^*_{L,M},G_L)
    =
    KL(\gamma^*_{L,M},G_M)
    <
    \min_{\gamma\in F}KL(\gamma,G_H).
\end{equation}
Equation \eqref{eqn:27} says that every pivotal frequency has a strictly higher
large-deviation cost in state $H$ than the most likely pivotal frequency
$\gamma^*_{L,M}$ has in states $L$ and $M$. Therefore, by \eqref{eqn:18}, every
pivotal frequency is exponentially less likely in state $H$ than
$\gamma^*_{L,M}$ is in states $L$ and $M$. This separation will imply that the
probability of being pivotal is exponentially smaller in state $H$ than in
states $L$ and $M$.

To prove \eqref{eqn:27}, first note that, for every $\gamma\in F$,
\[
    KL(\gamma,G_H)
    \geq
    \min\{KL(\gamma,G_M),KL(\gamma,G_H)\}
    =
    KL(\gamma,G_L)
    \geq
    C_{L,M},
\]
where the equality follows from the definition of $F$ and the final inequality follows from \eqref{eqn:26}. It remains to show that this lower bound is strict. Suppose, for contradiction, that there exists $\hat\gamma\in F$ such that
\[
    KL(\hat\gamma,G_H)
    =
    \min\{KL(\hat\gamma,G_M),KL(\hat\gamma,G_H)\}
    =
    KL(\hat\gamma,G_L)
    =
    C_{L,M}.
\]
Then
\[
    C_{L,H}
    =
    \min_{\substack{\gamma\in\Delta^3:\\
    KL(\gamma,G_L)=KL(\gamma,G_H)}}
    KL(\gamma,G_L)
    \leq
    KL(\hat\gamma,G_L)
    =
    C_{L,M},
\]
which contradicts \eqref{eqn:24}. This proves \eqref{eqn:27}.

\paragraph{From pivotal frequencies back to pivotal events.}

By \eqref{eqn:27} and the continuity of $KL(\cdot,G_H)$, we can choose $\epsilon>0$ and $\delta>0$ such that
\begin{equation}
\label{eqn:28}
    \inf_{\gamma\in F(\epsilon)}
    KL(\gamma,G_H)
    >
    C_{L,M}+\delta.
\end{equation}
For all sufficiently large $N$, \eqref{eqn:19} implies that every
$\tilde{\boldsymbol t}\in E_N$ satisfies
$\bar\gamma(\tilde{\boldsymbol t})\in F(\epsilon)$. Hence, by
\eqref{eqn:18}, \eqref{eqn:28}, and the fact that the number of message-count
vectors grows only polynomially in $N$,
\[
    \mathbb P[E_N\mid H;\boldsymbol P]
    \leq
    \exp\left\{
    -(N-1)(C_{L,M}+\delta)+O(\log N)
    \right\}.
\]

On the other hand, since the empirical-frequency grid 
\begin{equation*}
\{\bar\gamma(\boldsymbol t)\mid \boldsymbol t\in \Delta^3(N-1)\}
\end{equation*}
is asymptotically dense in $\Delta^3$, we can show that, for all sufficiently large $N$, there exists $\tilde{\boldsymbol t}_N^*\in E_N$ such that
\[
    \lim_{N\to\infty}
    \bar\gamma(\tilde{\boldsymbol t}_N^*)
    =
    \gamma^*_{L,M}.
\]
Therefore, by \eqref{eqn:18}, we have
\[
\begin{split}
    \mathbb P[E_N\mid L;\boldsymbol P]
    &\geq
    \mathbb P[\tilde{\boldsymbol t}_N^*\mid L;\boldsymbol P]
    =
    \exp\left\{-(N-1)(C_{L,M}+o(1))\right\},\\
    \mathbb P[E_N\mid M;\boldsymbol P]
    &\geq
    \mathbb P[\tilde{\boldsymbol t}_N^*\mid M;\boldsymbol P]
    =
    \exp\left\{-(N-1)(C_{L,M}+o(1))\right\}.
\end{split}
\]
Consequently, for each $\theta\in\{L,M\}$,
\[
    \frac{\mathbb P[E_N\mid H;\boldsymbol P]}
         {\mathbb P[E_N\mid\theta;\boldsymbol P]}
    \leq
    \exp\left\{
    -(N-1)(\delta-o(1))+O(\log N)
    \right\}
    \longrightarrow 0.
\]
Thus, being pivotal is exponentially less likely in state $H$ than in either state $L$ or $M$, which proves \eqref{eqn:16}.

\paragraph{Implication for sender incentives.}

For each private signal $s$,
\[
\frac{\mathbb P[H\mid E_N,s;\boldsymbol P]}
     {\mathbb P[L\mid E_N,s;\boldsymbol P]}
=
\frac{q^0_H\,\mathbb P[s\mid H]}
     {q^0_L\,\mathbb P[s\mid L]}
\cdot
\frac{\mathbb P[E_N\mid H;\boldsymbol P]}
     {\mathbb P[E_N\mid L;\boldsymbol P]},
\]
and an analogous expression holds after replacing $L$ by $M$. Since the signal
space is finite and no signal rules out any state, \eqref{eqn:16} therefore implies, for each $s$,
\[
    \frac{\mathbb P[H\mid E_N,s;\boldsymbol P]}
         {\mathbb P[L\mid E_N,s;\boldsymbol P]}
    \longrightarrow 0,
    \qquad
    \frac{\mathbb P[H\mid E_N,s;\boldsymbol P]}
         {\mathbb P[M\mid E_N,s;\boldsymbol P]}
    \longrightarrow 0.
\]
Thus, as $N\to\infty$, conditional on being pivotal, each sender assigns asymptotically negligible probability to $H$, regardless of his private signal. Hence, when $N$ is sufficiently large, each sender strictly prefers the status quo regardless of his private signal, since he prefers the status quo in both \(L\) and \(M\). Therefore, when the other \(N-1\) senders use \(\boldsymbol P\) and the receiver chooses \(\psi_N\), the remaining sender strictly benefits from deviating by sending \(z_1\) after every signal because this deviation strictly lowers the probability that the receiver chooses the proposal on the set of pivotal events $E_N$. Consequently, for all sufficiently large \(N\), \(\boldsymbol P\) cannot be a best response to itself given \(\psi_N\).

Finally, suppose that, along a sequence of informative monotone equilibria, the
senders' strategies do not become vanishingly uninformative. By compactness, a subsequence of these strategies converges to an informative strategy $\boldsymbol P$. Since the induced message distributions and KL divergences vary continuously with the sender strategy, the preceding argument applies uniformly in a neighborhood of $\boldsymbol P$. Thus, for all sufficiently large \(N\), there cannot exist an informative monotone equilibrium in which the senders' strategy is sufficiently close to \(\boldsymbol{P}\), because such a strategy cannot be a best response to itself given the receiver's equilibrium strategy. This contradiction completes the proof sketch.

\section{A Centralized-Information Benchmark}
\label{ap:B}

In this appendix, we compare the basic decentralized-information model, in which
$N$ conditionally i.i.d.\ signals are dispersed across $N$ senders, each observing one
signal, with a centralized-information benchmark in which a single sender observes all
$N$ signals. We focus on the large-population limit as $N\to\infty$, in which the single sender in the centralized-information benchmark fully learns the state.

In this centralized-information benchmark, the revelation principle allows us to restrict attention to direct recommendations. In state \(H\), the sender recommends the proposal, and the receiver follows the recommendation because their preferences are aligned. In states \(L\) and \(M\), by contrast, adverse selection arises: the sender pools the two states and recommends the status quo. The receiver is willing to follow this recommendation only if
\begin{equation*}
    q^0_L U_R(L) + q^0_M U_R(M) \leq 0.
\end{equation*}
Thus, if
\begin{equation}
\label{eqn:29}
    q^0_M > -\frac{U_R(L)}{U_R(M)} q^0_L,
\end{equation}
the receiver disregards the sender's message and always chooses the proposal, and communication therefore unravels. The resulting expected payoffs of the sender and the receiver coincide with their expected payoffs in the babbling equilibrium of the basic decentralized-information model.

Suppose that
\begin{equation}
\label{eqn:30}
    \frac{U_S(L)}{U_S(H)}
    \cdot
    \frac{U_R(H)}{U_R(L)}
    <
    \frac{\rho_H}{\rho_L}.
\end{equation}
Under \eqref{eqn:30}, for each fixed value of the ratio \(q^0_H/q^0_L\),
\cref{thm:4} establishes a threshold
\[\hat{q}\left(\frac{q^0_H}{q^0_L}\right)>0\]
such that information transmission persists in the
decentralized-information model with \(N\) senders whenever
\begin{equation}
\label{eqn:31}
    q^0_M < \hat{q}\left(\frac{q^0_H}{q^0_L}\right).
\end{equation}
Equivalently, an informative equilibrium exists for every sufficiently large \(N\).
Consequently, for every prior \(q^0\) satisfying \eqref{eqn:29} and
\eqref{eqn:31}, communication unravels in the centralized-information benchmark
with a single fully informed sender, whereas information transmission persists in the
decentralized-information model with dispersed information.
Moreover, the centralized-information benchmark yields the same expected payoffs as
the babbling equilibrium. Hence, by \cref{prop:1}, every player obtains a weakly higher expected payoff
in any informative equilibrium of the decentralized-information model than in
the centralized-information benchmark. Moreover, because the receiver's
equilibrium strategy is responsive to the approval tally rather than constant,
her expected payoff is strictly higher.

We indeed show that the set of priors satisfying \eqref{eqn:29} and
\eqref{eqn:31} is nonempty. Thus, there is a nonempty set of priors for which the receiver is strictly better off, and every sender is weakly better off in the decentralized-information model than in the centralized-information benchmark.

\begin{proposition}
    \label{prop:5}
    Suppose that \eqref{eqn:30} holds. Then the set
    \begin{equation*}
        \left\{
        q^0=(q^0_L,q^0_M,q^0_H)\in\Delta^3
        \,\middle|\,
        -\frac{U_R(L)}{U_R(M)}q^0_L
        <q^0_M<\hat{q}\left(\frac{q^0_H}{q^0_L}\right)
        \right\}
    \end{equation*}
    is nonempty.
\end{proposition}

Intuitively, centralization allows the sender to learn the state, but
this informational advantage sharpens adverse selection. A status-quo
recommendation reveals that the state is either $L$ or $M$; when the
disagreement state is sufficiently likely, the receiver therefore
refuses to follow it. Under decentralization, by contrast, the message
tally does not fully reveal the state. The resulting endogenous noise
softens the receiver's inference and can sustain informative
communication. As in \citet{blume2007noisy}, noise therefore plays a
commitment role: it mitigates the conflict and improves communication.

\section{Omitted Proofs}

\subsection{Proof of Lemma~\ref{lem:1}}

Suppose, toward a contradiction, that $(\bx,\HT)$ is an informative equilibrium with
$x_b>0$. By \eqref{eqn:5}, $x_g=1$. Since $x_g>x_b$, we have $x_b\in(0,1)$.
Thus \eqref{eqn:4} implies that a sender who receives signal $b$ is indifferent between
approving and rejecting conditional on being pivotal:
\begin{equation}
\label{eqn:32}
    \sum_{\theta\in\{L,M,H\}}
q^0_\theta
\mathbb{P}[b\mid\theta]
\mathbb{P}[\mathrm{piv}\mid\theta;\bx,\HT]
U_S(\theta)
=0.
\end{equation}

For each $\theta\in\{L,M,H\}$, when $x_g=1$, the probabilities of approval and
rejection are
\[
\rho_\theta+(1-\rho_\theta)x_b\in(0,1),
\qquad
(1-\rho_\theta)(1-x_b)\in(0,1).
\]
Since $\HT\in\{1,\ldots,N\}$, this implies
\[
\mathbb{P}[T=\HT-1\mid\theta;\bx]>0
\qquad\text{for each }\theta\in\{L,M,H\}.
\]
Moreover, since $\mathbb{P}[b\mid\theta]=1-\rho_\theta$,
\[
\begin{aligned}
(1-\rho_\theta)\mathbb{P}[\mathrm{piv}\mid\theta;\bx,\HT]
&=
(1-\rho_\theta)\binom{N-1}{\HT-1}
[\rho_\theta+(1-\rho_\theta)x_b]^{\HT-1}
[(1-\rho_\theta)(1-x_b)]^{N-\HT}  \\
&=
\frac{\binom{N-1}{\HT-1}}
{(1-x_b)\binom{N}{\HT-1}}
\mathbb{P}[T=\HT-1\mid\theta;\bx].
\end{aligned}
\]
The coefficient is positive and independent of $\theta$. Hence \eqref{eqn:32} implies
\[
\sum_{\theta\in\{L,M,H\}}
q^0_\theta
\mathbb{P}[T=\HT-1\mid\theta;\bx]
U_S(\theta)
=0.
\]
Therefore,
\[
\begin{aligned}
& q^0_H\mathbb{P}[T=\HT-1\mid H;\bx]U_S(H)
+
q^0_L\mathbb{P}[T=\HT-1\mid L;\bx]U_S(L)  \\
&\qquad =
-
q^0_M\mathbb{P}[T=\HT-1\mid M;\bx]U_S(M)
\geq 0,
\end{aligned}
\]
where the inequality follows from $U_S(M)<0$. Since $U_S(H)>0$, this implies
\[
q^0_H\mathbb{P}[T=\HT-1\mid H;\bx]
\geq
-\frac{U_S(L)}{U_S(H)}
q^0_L\mathbb{P}[T=\HT-1\mid L;\bx].
\]
By \eqref{eqn:2},
\[
-\frac{U_R(L)}{U_R(H)}
\leq
-\frac{U_S(L)}{U_S(H)}.
\]
Hence
\[
q^0_H\mathbb{P}[T=\HT-1\mid H;\bx]
\geq
-\frac{U_R(L)}{U_R(H)}
q^0_L\mathbb{P}[T=\HT-1\mid L;\bx],
\]
and so
\[
q^0_H\mathbb{P}[T=\HT-1\mid H;\bx]U_R(H)
+
q^0_L\mathbb{P}[T=\HT-1\mid L;\bx]U_R(L)
\geq 0.
\]
Using also $U_R(M)>0$, we obtain
\[
\sum_{\theta\in\{L,M,H\}}
q^0_\theta
\mathbb{P}[T=\HT-1\mid\theta;\bx]
U_R(\theta)
\geq 0.
\]
Thus the receiver weakly prefers the proposal when the approval tally is $T=\HT-1$, which contradicts the minimality of $\HT$ in \eqref{eqn:6}. Therefore $x_b=0$.

\subsection{Proof of Theorem~\ref{thm:3}}

By \cref{lem:1}, in every informative equilibrium we can write $x_b=0$ and
$x_g=x\in(0,1]$. Thus, in state $\theta$, the probability that a sender approves is
$\rho_\theta x$, and
\begin{equation}
\label{eqn:33}
    \mathbb{P}[T=t\mid \theta;x]
    =
    \binom{N}{t}
    (\rho_\theta x)^t
    (1-\rho_\theta x)^{N-t}.
\end{equation}
All sequences below are indexed by integers $N\to\infty$. Whenever an infinite
subsequence is selected, we keep the same notation and continue to index it by $N$.
Whenever a property holds eventually, we delete finitely many terms and keep the same
notation.

For $a\in[0,1]$ and $b\in(0,1)$, write
\[
KL(a,b):=a\log\frac{a}{b}+(1-a)\log\frac{1-a}{1-b},
\]
with the convention $0\log 0=0$. By \eqref{eqn:33}, for every $x>0$ and
$t\in\{0,\ldots,N\}$,
\begin{equation}
\label{eqn:34}
\frac{\mathbb{P}[T=t\mid \theta;x]}
{\mathbb{P}[T=t\mid \theta';x]}
=
\exp\left\{
N\left[
KL\left(\frac{t}{N},\rho_{\theta'}x\right)
-
KL\left(\frac{t}{N},\rho_\theta x\right)
\right]
\right\}.
\end{equation}
Moreover, for $p,q\in(0,1)$,
\[
\frac{\partial}{\partial a}\left[KL(a,p)-KL(a,q)\right]
=
\log\frac{q(1-p)}{p(1-q)}.
\]
Hence, for each fixed $x>0$,
\begin{equation}
\label{eqn:35}
a\mapsto KL(a,\rho_Lx)-KL(a,\rho_Mx)
\quad\text{is increasing,}
\end{equation}
while
\begin{equation}
\label{eqn:36}
a\mapsto KL(a,\rho_Hx)-KL(a,\rho_Mx)
\quad\text{is decreasing.}
\end{equation}

We first show that, along every sequence of informative equilibria
$\Gamma_N=(x_N,\HT_N)$ with $N\to\infty$, we have $x_N\to0$. Fix such a sequence.
Suppose, toward a contradiction, that $x_N\not\to0$. Then there exists
$\epsilon>0$ such that $x_N\geq\epsilon$ for infinitely many $N$. By the subsequence
convention and compactness, assume
\[
x_N\to \bar x\in[\epsilon,1].
\]

First suppose that $\HT_N/N\geq\rho_Mx_N$ for infinitely many $N$. By the
subsequence convention, assume
\[
\frac{\HT_N}{N}\geq \rho_Mx_N
\qquad\text{for every }N.
\]
Using \eqref{eqn:34} and \eqref{eqn:35},
\[
\log
\frac{\mathbb{P}[T=\HT_N\mid M;x_N]}
{\mathbb{P}[T=\HT_N\mid L;x_N]}
\geq
N\,KL(\rho_Mx_N,\rho_Lx_N).
\]
Since $x_N\to\bar x>0$ and $\rho_L<\rho_M$,
\[
KL(\rho_Mx_N,\rho_Lx_N)
\to
KL(\rho_M\bar x,\rho_L\bar x)>0.
\]
Therefore,
\[
\frac{\mathbb{P}[T=\HT_N\mid M;x_N]}
{\mathbb{P}[T=\HT_N\mid L;x_N]}
\to\infty.
\]
The one-step likelihood ratio satisfies
\begin{equation}
\label{eqn:37}
\frac{\mathbb{P}[T=\HT_N-1\mid M;x_N]}
{\mathbb{P}[T=\HT_N-1\mid L;x_N]}
=
\frac{\rho_L}{\rho_M}
\frac{1-\rho_Mx_N}{1-\rho_Lx_N}
\frac{\mathbb{P}[T=\HT_N\mid M;x_N]}
{\mathbb{P}[T=\HT_N\mid L;x_N]}.
\end{equation}
The prefactor in \eqref{eqn:37} is bounded away from zero along the sequence. Hence
\[
\frac{\mathbb{P}[T=\HT_N-1\mid M;x_N]}
{\mathbb{P}[T=\HT_N-1\mid L;x_N]}
\to\infty.
\]
Since $q^0_MU_R(M)>0$, while $q^0_L>0$ and $U_R(L)<0$, this implies
\[
L_R(\HT_N-1;x_N)\to\infty.
\]
This contradicts \eqref{eqn:6}, which implies
$L_R(\HT_N-1;x_N)<1$ in an informative equilibrium. Therefore
$\HT_N/N\geq\rho_Mx_N$ cannot occur infinitely often along the current sequence.

Deleting finitely many terms, assume instead that
\[
\frac{\HT_N}{N}<\rho_Mx_N
\qquad\text{for every }N.
\]
Using \eqref{eqn:34} and \eqref{eqn:36},
\[
\log
\frac{\mathbb{P}[T=\HT_N\mid H;x_N]}
{\mathbb{P}[T=\HT_N\mid M;x_N]}
\leq
-N\,KL(\rho_Mx_N,\rho_Hx_N).
\]
Since $x_N\to\bar x>0$ and $\rho_M<\rho_H$,
\[
KL(\rho_Mx_N,\rho_Hx_N)
\to
KL(\rho_M\bar x,\rho_H\bar x)>0.
\]
Therefore,
\[
\frac{\mathbb{P}[T=\HT_N\mid H;x_N]}
{\mathbb{P}[T=\HT_N\mid M;x_N]}
\to0.
\]
For every $\theta\in\{L,M,H\}$ and every informative equilibrium,
\begin{equation}
\label{eqn:38}
\rho_\theta\mathbb{P}[\mathrm{piv}\mid \theta;x,\HT]
=
\frac{\HT}{Nx}
\mathbb{P}[T=\HT\mid\theta;x],
\end{equation}
where the coefficient $\HT/(Nx)$ is positive and independent of $\theta$. Thus
\[
\frac{\rho_H\mathbb{P}[\mathrm{piv}\mid H;x_N,\HT_N]}
{\rho_M\mathbb{P}[\mathrm{piv}\mid M;x_N,\HT_N]}
=
\frac{\mathbb{P}[T=\HT_N\mid H;x_N]}
{\mathbb{P}[T=\HT_N\mid M;x_N]}
\to0.
\]
Since the denominator of $L_S(g,\mathrm{piv};x_N,\HT_N)$ includes the positive contribution
from state $M$,
\[
L_S(g,\mathrm{piv};x_N,\HT_N)
\leq
\frac{q^0_HU_S(H)}{-q^0_MU_S(M)}
\frac{\mathbb{P}[T=\HT_N\mid H;x_N]}
{\mathbb{P}[T=\HT_N\mid M;x_N]}
\to0.
\]
This contradicts \eqref{eqn:4}, because $x_N>0$ requires
$L_S(g,\mathrm{piv};x_N,\HT_N)\geq1$. Therefore $x_N\to0$ along every sequence of
informative equilibria.

We next show that $Nx$ is uniformly bounded over all informative equilibria. Suppose,
toward a contradiction, that it is not. Then there exists a sequence of informative
equilibria $\Gamma_N=(x_N,\HT_N)$ such that $N\to\infty$ and
\[
Nx_N\to\infty.
\]
The preceding paragraph implies $x_N\to0$. For each $i\in\{L,H\}$,
\begin{equation}
\label{eqn:39}
KL(\rho_Mx_N,\rho_ix_N)
=
\left(
\rho_M\log\frac{\rho_M}{\rho_i}+\rho_i-\rho_M
\right)x_N
+
o(x_N).
\end{equation}
The coefficient in parentheses is strictly positive for both $i=L$ and $i=H$. Hence
\[
N\,KL(\rho_Mx_N,\rho_ix_N)\to\infty
\qquad\text{for }i\in\{L,H\}.
\]

If $\HT_N/N\geq\rho_Mx_N$ infinitely often, then the same argument leading to
\eqref{eqn:37} gives
\[
L_R(\HT_N-1;x_N)\to\infty,
\]
contradicting \eqref{eqn:6}. If, after deleting finitely many terms,
$\HT_N/N<\rho_Mx_N$ for every $N$, then the same argument using \eqref{eqn:38}
gives
\[
L_S(g,\mathrm{piv};x_N,\HT_N)\to0,
\]
contradicting \eqref{eqn:4}. Therefore there exists $B>0$ such that
\begin{equation}
\label{eqn:40}
Nx<B
\end{equation}
for every informative equilibrium. The sequence argument gives a uniform bound because,
if no such $B$ existed, one could select informative equilibria with $Nx$ arbitrarily
large; since $x\leq1$, this selected sequence must have $N\to\infty$, and the preceding
contradiction would apply.

We now show that $\HT$ is uniformly bounded. Since
\[
z\mapsto \frac{1}{z}\log\frac{1-\rho_Lz}{1-\rho_Hz}
\]
has a finite continuous extension to $[0,1]$, there exists $C>0$ such that, for all
$z\in[0,1]$,
\begin{equation}
\label{eqn:41}
\log\frac{1-\rho_Lz}{1-\rho_Hz}\leq Cz.
\end{equation}
Thus, for every $x\in[0,1]$ and every $N$,
\[
\left(\frac{1-\rho_Hx}{1-\rho_Lx}\right)^N
=
\exp\left\{
-N\log\frac{1-\rho_Lx}{1-\rho_Hx}
\right\}
\geq e^{-CNx}.
\]
Hence, by \eqref{eqn:40}, in every informative equilibrium,
\[
\begin{aligned}
\frac{\mathbb{P}[T=\HT-1\mid H;x]}
{\mathbb{P}[T=\HT-1\mid L;x]}
&=
\left(\frac{\rho_H}{\rho_L}\right)^{\HT-1}
\left(\frac{1-\rho_Hx}{1-\rho_Lx}\right)^{N-\HT+1} \\
&\geq
\left(\frac{\rho_H}{\rho_L}\right)^{\HT-1}
\left(\frac{1-\rho_Hx}{1-\rho_Lx}\right)^N \\
&\geq
\left(\frac{\rho_H}{\rho_L}\right)^{\HT-1}e^{-CB}.
\end{aligned}
\]
On the other hand, \eqref{eqn:6} implies $L_R(\HT-1;x)<1$. Since the $M$-term in
the numerator of $L_R(\HT-1;x)$ is positive and $q^0_HU_R(H)>0$,
\[
\frac{\mathbb{P}[T=\HT-1\mid H;x]}
{\mathbb{P}[T=\HT-1\mid L;x]}
<
\frac{-q^0_LU_R(L)}{q^0_HU_R(H)}.
\]
Combining the last two displays gives
\[
\left(\frac{\rho_H}{\rho_L}\right)^{\HT-1}
<
\frac{-q^0_LU_R(L)}{q^0_HU_R(H)}e^{CB}.
\]
Because $\rho_H/\rho_L>1$, this yields a uniform upper bound on $\HT$. Choosing
$T_0$ larger than the uniform bounds on $Nx$ and $\HT$ gives
\[
Nx<T_0
\qquad\text{and}\qquad
\HT<T_0
\]
for every informative equilibrium, proving \eqref{eqn:13}.

It remains to prove the uniform lower bound on mistakes. Let
\[
\eta:=e^{-CB}>0.
\]
By \eqref{eqn:40} and \eqref{eqn:41}, for every informative equilibrium and every
$t<\HT$,
\begin{equation}
\label{eqn:42}
\begin{aligned}
\frac{\mathbb{P}[T=t\mid H;x]}
{\mathbb{P}[T=t\mid L;x]}
&=
\left(\frac{\rho_H}{\rho_L}\right)^t
\left(\frac{1-\rho_Hx}{1-\rho_Lx}\right)^{N-t} \\
&\geq
\left(\frac{1-\rho_Hx}{1-\rho_Lx}\right)^N
\geq e^{-CB}
=
\eta .
\end{aligned}
\end{equation}
Therefore,
\[
\mathbb{P}[\mathrm{status\ quo}\mid H]
=
\sum_{t<\HT}\mathbb{P}[T=t\mid H;x]
\geq
\eta\sum_{t<\HT}\mathbb{P}[T=t\mid L;x]
=
\eta\,\mathbb{P}[\mathrm{status\ quo}\mid L].
\]
It follows that
\[
\begin{aligned}
&\mathbb{P}[\mathrm{proposal}\mid L]
+
\mathbb{P}[\mathrm{status quo}\mid H] \\
&\qquad\geq
1-\mathbb{P}[\mathrm{status quo}\mid L]
+
\eta\,\mathbb{P}[\mathrm{status\ quo}\mid L] \\
&\qquad=
1-(1-\eta)\mathbb{P}[\mathrm{status\ quo}\mid L]
\geq \eta .
\end{aligned}
\]

Finally, consider a non-informative equilibrium. If the receiver's cutoff is
$\HT=0$ or $\HT=N+1$, then her action is constant in the approval tally, and
\[
\mathbb{P}[\mathrm{proposal}\mid L]
+
\mathbb{P}[\mathrm{status\ quo}\mid H]
=1.
\]
If instead $x_g=x_b$, then the approval tally has the same distribution in states $L$
and $H$. Hence
\[
\mathbb{P}[\mathrm{proposal}\mid L]
=
\mathbb{P}[\mathrm{proposal}\mid H],
\]
and again
\[
\mathbb{P}[\mathrm{proposal}\mid L]
+
\mathbb{P}[\mathrm{status\ quo}\mid H]
=1.
\]
Thus, with $c:=\eta/2>0$, every symmetric equilibrium satisfies
\[
\mathbb{P}[\mathrm{proposal}\mid L;\Gamma_N]
+
\mathbb{P}[\mathrm{status\ quo}\mid H;\Gamma_N]
>c.
\]
Hence no sequence of symmetric equilibria aggregates information.

\subsection{Proof of Theorem~\ref{thm:4}}

\subsubsection{Failure Above the Threshold \texorpdfstring{$\rho_H/\rho_L$}{rhoH/rhoL}}

Fix $q^0_M>0$ and suppose
\[
\frac{U_S(L)}{U_S(H)}
\frac{U_R(H)}{U_R(L)}
>
\frac{\rho_H}{\rho_L}.
\]
By \cref{thm:3}, there exists $T_0>0$ such that \eqref{eqn:13} holds in every
informative equilibrium. In particular,
\[
Nx<T_0.
\]
Since
\[
\frac{\rho_H}{\rho_L}
\frac{1-\rho_Lx}{1-\rho_Hx}
\to
\frac{\rho_H}{\rho_L}
\qquad\text{as }x\to0,
\]
there exists $\bar{x}\in(0,1)$ such that, for every $x\in(0,\bar{x})$,
\begin{equation}
\label{eqn:43}
\frac{U_S(L)}{U_S(H)}
\frac{U_R(H)}{U_R(L)}
>
\frac{\rho_H}{\rho_L}
\frac{1-\rho_Lx}{1-\rho_Hx}.
\end{equation}
Choose $N_2\in\mathbb{N}$ such that $N_2>T_0$ and $T_0/N_2<\bar{x}$.

Suppose, toward a contradiction, that there exists an informative equilibrium for some
$N>N_2$. By \cref{lem:1}, we can write this equilibrium as $(x,\HT)$, with
$x_b=0$ and $x_g=x$. Since $Nx<T_0$ and $N>N_2$,
\begin{equation}
\label{eqn:44}
0<x<\frac{T_0}{N}<\frac{T_0}{N_2}<\bar{x}<1.
\end{equation}
All binomial probabilities appearing below are therefore strictly positive. Since
$x\in(0,1)$, \eqref{eqn:4} implies
\[
L_S(g,\mathrm{piv};x,\HT)=1.
\]
For each $\theta\in\{L,M,H\}$,
\[
\rho_\theta\mathbb{P}[\mathrm{piv}\mid\theta;x,\HT]
=
\frac{\HT}{Nx}\mathbb{P}[T=\HT\mid\theta;x].
\]
The coefficient is positive and independent of $\theta$. Therefore
$L_S(g,\mathrm{piv};x,\HT)=1$ implies
\[
\frac{
q^0_H\mathbb{P}[T=\HT\mid H;x]U_S(H)}
{-\sum_{\theta\in\{L,M\}}
q^0_\theta\mathbb{P}[T=\HT\mid\theta;x]U_S(\theta)}
=1.
\]
Since $q^0_M>0$, $U_S(M)<0$, and
$\mathbb{P}[T=\HT\mid M;x]>0$, we obtain
\[
\frac{
q^0_H\mathbb{P}[T=\HT\mid H;x]U_S(H)}
{-q^0_L\mathbb{P}[T=\HT\mid L;x]U_S(L)}
=
1+
\frac{
-q^0_M\mathbb{P}[T=\HT\mid M;x]U_S(M)}
{-q^0_L\mathbb{P}[T=\HT\mid L;x]U_S(L)}
>1.
\]
Thus
\[
\frac{
q^0_H\mathbb{P}[T=\HT\mid H;x]}
{q^0_L\mathbb{P}[T=\HT\mid L;x]}
>
-\frac{U_S(L)}{U_S(H)}.
\]

By the minimality of $\HT$ in \eqref{eqn:6},
\[
L_R(\HT-1;x)<1.
\]
Since the $M$-term in the numerator of $L_R(\HT-1;x)$ is positive, this implies
\[
\frac{
q^0_H\mathbb{P}[T=\HT-1\mid H;x]U_R(H)}
{-q^0_L\mathbb{P}[T=\HT-1\mid L;x]U_R(L)}
<1,
\]
and hence
\[
\frac{
q^0_H\mathbb{P}[T=\HT-1\mid H;x]}
{q^0_L\mathbb{P}[T=\HT-1\mid L;x]}
<
-\frac{U_R(L)}{U_R(H)}.
\]
Combining the last two displays yields
\[
\frac{U_S(L)}{U_S(H)}
\frac{U_R(H)}{U_R(L)}
<
\frac{\mathbb{P}[T=\HT\mid H;x]}
{\mathbb{P}[T=\HT\mid L;x]}
\frac{\mathbb{P}[T=\HT-1\mid L;x]}
{\mathbb{P}[T=\HT-1\mid H;x]}.
\]
Because $x_b=0$ and $x_g=x$,
\[
\mathbb{P}[T=t\mid\theta;x]
=
\binom{N}{t}(\rho_\theta x)^t(1-\rho_\theta x)^{N-t}.
\]
Therefore,
\[
\frac{\mathbb{P}[T=\HT\mid H;x]}
{\mathbb{P}[T=\HT\mid L;x]}
\frac{\mathbb{P}[T=\HT-1\mid L;x]}
{\mathbb{P}[T=\HT-1\mid H;x]}
=
\frac{\rho_H}{\rho_L}
\frac{1-\rho_Lx}{1-\rho_Hx}.
\]
Hence every informative equilibrium with $N>N_2$ must satisfy
\begin{equation}
\label{eqn:45}
\frac{U_S(L)}{U_S(H)}
\frac{U_R(H)}{U_R(L)}
<
\frac{\rho_H}{\rho_L}
\frac{1-\rho_Lx}{1-\rho_Hx}.
\end{equation}
However, \eqref{eqn:44} implies $x<\bar{x}$, so \eqref{eqn:45} contradicts
\eqref{eqn:43}. Thus, for every $N>N_2$, no informative equilibrium exists. Therefore,
information transmission fails.

\subsubsection{The \texorpdfstring{$q^0_M$}{qM}-Cutoff Below the Threshold \texorpdfstring{$\rho_H/\rho_L$}{rhoH/rhoL}}
\label{ap:d2}

Suppose
\begin{equation}
\label{eqn:46}
\frac{U_S(L)}{U_S(H)}
\frac{U_R(H)}{U_R(L)}
<
\frac{\rho_H}{\rho_L}.
\end{equation}
Fix the ratio $q^0_L/q^0_H$ and write
\[
\lambda_L:=\frac{q^0_L}{q^0_H},
\qquad
\lambda_M:=\frac{q^0_M}{q^0_H}.
\]
Holding $q^0_L/q^0_H$ fixed is equivalent to holding $\lambda_L$ fixed, and
$q^0_M$ is strictly increasing in $\lambda_M$. Also write
\[
a:=-\frac{U_S(L)}{U_S(H)},
\qquad
b:=-\frac{U_S(M)}{U_S(H)},
\qquad
r:=-\frac{U_R(L)}{U_R(H)},
\qquad
m:=\frac{U_R(M)}{U_R(H)}.
\]
Then $a,b,r,m>0$, \eqref{eqn:2} gives $r\leq a$, and \eqref{eqn:46} is equivalent to
\[
\frac{a}{r}<\frac{\rho_H}{\rho_L}.
\]

Fix an integer $k\geq1$. We first characterize the limiting conditions for informative
equilibria with $\HT=k$. Consider $x=\mu/N$, where $\mu\geq0$ is fixed. The sender's
indifference condition $L_S(g,\mathrm{piv};x,k)=1$, after dividing by the $H$-term and taking
$N\to\infty$, becomes
\begin{equation}
\label{eqn:47}
\lambda_M b
\left(\frac{\rho_M}{\rho_H}\right)^k
e^{(\rho_H-\rho_M)\mu}
+
\lambda_L a
\left(\frac{\rho_L}{\rho_H}\right)^k
e^{(\rho_H-\rho_L)\mu}
=1.
\end{equation}
The limiting receiver conditions at $k$ and $k-1$ are, respectively,
\begin{equation}
\label{eqn:48}
\begin{aligned}
1+
\lambda_M m
\left(\frac{\rho_M}{\rho_H}\right)^k
e^{(\rho_H-\rho_M)\mu}
-
\lambda_L r
\left(\frac{\rho_L}{\rho_H}\right)^k
e^{(\rho_H-\rho_L)\mu}
&\geq0,\\
1+
\lambda_M m
\left(\frac{\rho_M}{\rho_H}\right)^{k-1}
e^{(\rho_H-\rho_M)\mu}
-
\lambda_L r
\left(\frac{\rho_L}{\rho_H}\right)^{k-1}
e^{(\rho_H-\rho_L)\mu}
&<0.
\end{aligned}
\end{equation}

Let
\[
z:=
\lambda_L a
\left(\frac{\rho_L}{\rho_H}\right)^k
e^{(\rho_H-\rho_L)\mu}.
\]
By \eqref{eqn:47},
\[
\lambda_M b
\left(\frac{\rho_M}{\rho_H}\right)^k
e^{(\rho_H-\rho_M)\mu}
=1-z.
\]
Substituting this equality into the first inequality in \eqref{eqn:48} gives
\[
1+\frac{m}{b}(1-z)-\frac{r}{a}z
\geq
1-\frac{r}{a}
\geq0.
\]
Thus the receiver's condition at $k$ is automatically satisfied whenever
\eqref{eqn:47} holds. The condition at $k-1$ is equivalent to
\[
1+
\frac{\rho_H}{\rho_M}\cdot\frac{m}{b}(1-z)
-
\frac{\rho_H}{\rho_L}\cdot\frac{r}{a}z
<0.
\]
Define
\[
z^\ast:=
\frac{
1+\frac{\rho_H}{\rho_M}\cdot\frac{m}{b}}
{
\frac{\rho_H}{\rho_M}\cdot\frac{m}{b}
+
\frac{\rho_H}{\rho_L}\cdot\frac{r}{a}
}.
\]
By \eqref{eqn:46}, $z^\ast\in(0,1)$, and the receiver's condition at $k-1$ is equivalent to
\[
z>z^\ast.
\]

Now define
\[
z^0_k:=
\lambda_L a
\left(\frac{\rho_L}{\rho_H}\right)^k,
\qquad
\bar z_k:=\max\{z^\ast,z^0_k\}.
\]
The restriction $\mu\geq0$ is equivalent to $z\geq z^0_k$. For $z\in(z^0_k,1)$, let
\[
\mu_k(z):=
\frac{1}{\rho_H-\rho_L}
\log\left(\frac{z}{z^0_k}\right).
\]
Substituting $\mu=\mu_k(z)$ into \eqref{eqn:47} gives
\begin{equation}
\label{eqn:49}
\lambda_M
=
\frac{1-z}
{b\left(\frac{\rho_M}{\rho_H}\right)^k}
\left(\frac{z^0_k}{z}\right)^\alpha,
\qquad
\alpha:=\frac{\rho_H-\rho_M}{\rho_H-\rho_L}\in(0,1).
\end{equation}
Hence define
\begin{equation}
\label{eqn:50}
\hat\lambda_{M,k}:=
\begin{cases}
\displaystyle
\frac{1-\bar z_k}
{b\left(\frac{\rho_M}{\rho_H}\right)^k}
\left(\frac{z^0_k}{\bar z_k}\right)^\alpha,
&\text{if }\bar z_k<1,\\[1.2em]
0,
&\text{if }\bar z_k\geq1.
\end{cases}
\end{equation}
For each fixed $k$, the right-hand side of \eqref{eqn:49} is continuous and strictly
decreasing in $z$ on $(z^0_k,1)$. Therefore, for each
$\lambda_M<\hat\lambda_{M,k}$, there exists $z\in(\bar z_k,1)$ such that
\eqref{eqn:47} and both inequalities in \eqref{eqn:48} hold, with the two receiver
inequalities strict.

Let
\[
\hat\lambda_M:=\sup_{k\geq1}\hat\lambda_{M,k}.
\]
This number is strictly positive because $z^0_k\to0$ and $z^\ast<1$, so
$\hat\lambda_{M,k}>0$ for all sufficiently large $k$. It is finite because, for all sufficiently
large $k$, $\bar z_k=z^\ast$ and
\[
\hat\lambda_{M,k}
=
C
\left[
\left(\frac{\rho_L}{\rho_H}\right)^\alpha
\frac{\rho_H}{\rho_M}
\right]^k
\]
for some constant $C>0$. Strict concavity of $\log$ implies
\[
\left(\frac{\rho_L}{\rho_H}\right)^\alpha
\frac{\rho_H}{\rho_M}
<1,
\]
so $\hat\lambda_{M,k}\to0$.

We now show persistence when $\lambda_M<\hat\lambda_M$. Choose $k$ such that
$\lambda_M<\hat\lambda_{M,k}$, and choose $z\in(\bar z_k,1)$ satisfying
\eqref{eqn:49}. Let $\mu=\mu_k(z)>0$. For large $N$, define
\[
\begin{aligned}
\Phi_N(\tilde\mu)
&:=
\lambda_M b
\left(\frac{\rho_M}{\rho_H}\right)^k
\left(
\frac{1-\rho_M\tilde\mu/N}{1-\rho_H\tilde\mu/N}
\right)^{N-k} \\
&\quad+
\lambda_L a
\left(\frac{\rho_L}{\rho_H}\right)^k
\left(
\frac{1-\rho_L\tilde\mu/N}{1-\rho_H\tilde\mu/N}
\right)^{N-k}
-1.
\end{aligned}
\]
Define the limiting function
\begin{equation}
\label{eqn:51}
\Phi(\tilde\mu)
:=
\lambda_M b
\left(\frac{\rho_M}{\rho_H}\right)^k
e^{(\rho_H-\rho_M)\tilde\mu}
+
\lambda_L a
\left(\frac{\rho_L}{\rho_H}\right)^k
e^{(\rho_H-\rho_L)\tilde\mu}
-1.
\end{equation}
By the choice of $z$ and $\mu=\mu_k(z)$, we have $\Phi(\mu)=0$. Moreover,
\begin{equation}
\label{eqn:52}
\Phi'(\tilde\mu)
=
(\rho_H-\rho_M)\lambda_M b
\left(\frac{\rho_M}{\rho_H}\right)^k
e^{(\rho_H-\rho_M)\tilde\mu}
+
(\rho_H-\rho_L)\lambda_L a
\left(\frac{\rho_L}{\rho_H}\right)^k
e^{(\rho_H-\rho_L)\tilde\mu}
>0.
\end{equation}
Thus $\Phi$ is strictly increasing. Choose $\varepsilon\in(0,\mu)$ small enough that
\[
\Phi(\mu-\varepsilon)<0<\Phi(\mu+\varepsilon).
\]
The functions $\Phi_N$ converge uniformly to $\Phi$ on
$[\mu-\varepsilon,\mu+\varepsilon]$. Hence, for all sufficiently large $N$,
\[
\Phi_N(\mu-\varepsilon)<0<\Phi_N(\mu+\varepsilon).
\]
Since $\Phi_N$ is continuous, the intermediate value theorem gives
$\mu_N\in(\mu-\varepsilon,\mu+\varepsilon)$ such that
\[
\Phi_N(\mu_N)=0.
\]
Choosing one such root for each sufficiently large $N$, we have $\mu_N\to\mu$: indeed,
every convergent subsequence of $\{\mu_N\}$ has a limit point $\mu^\ast$ satisfying
$\Phi(\mu^\ast)=0$, and the strict monotonicity of $\Phi$ implies $\mu^\ast=\mu$.

Set $x_N:=\mu_N/N$. Then $x_N\in(0,1)$ for all sufficiently large $N$, and
$\Phi_N(\mu_N)=0$ is exactly
\[
L_S(g,\mathrm{piv};x_N,k)=1.
\]
By strict MLRP,
\[
L_S(b,\mathrm{piv};x_N,k)<L_S(g,\mathrm{piv};x_N,k)=1,
\]
so the senders' strategy with $x_b=0$ and $x_g=x_N$ satisfies \eqref{eqn:4}. Since the
limiting receiver inequalities in \eqref{eqn:48} hold strictly at the chosen $z$, the exact
receiver expressions satisfy
\[
L_R(k;x_N)>1
\qquad\text{and}\qquad
L_R(k-1;x_N)<1
\]
for all sufficiently large $N$. By the monotonicity of $L_R(\cdot;x_N)$ and \eqref{eqn:6},
the receiver's best response is $\HT=k$. Thus, for all sufficiently large $N$, there exists
an informative equilibrium with cutoff $k$. Therefore information transmission persists
whenever $\lambda_M<\hat\lambda_M$.

We next show failure when $\lambda_M>\hat\lambda_M$. Suppose, toward a contradiction,
that informative equilibria exist for infinitely many $N$. By \cref{thm:3}, along this
sequence both $Nx_N$ and $\HT_N$ are bounded. Select an infinite subsequence on which
$\HT_N$ is constant and $Nx_N$ converges; to simplify notation, keep indexing this
subsequence by $N$. Thus,
\[
\HT_N=k
\qquad\text{for some fixed }k\geq1,
\qquad
Nx_N\to\mu\geq0.
\]
For large $N$, $x_N<1$, so \eqref{eqn:4} implies
\[
L_S(g,\mathrm{piv};x_N,k)=1.
\]
Define
\[
z_N:=
\lambda_L a
\left(\frac{\rho_L}{\rho_H}\right)^k
\left(
\frac{1-\rho_Lx_N}{1-\rho_Hx_N}
\right)^{N-k}.
\]
Then
\[
z_N\to z
=
z^0_k e^{(\rho_H-\rho_L)\mu}
\geq z^0_k.
\]
The sender indifference condition gives
\[
\lambda_M b
\left(\frac{\rho_M}{\rho_H}\right)^k
\left(
\frac{1-\rho_Mx_N}{1-\rho_Hx_N}
\right)^{N-k}
+
z_N
=1.
\]
Taking limits yields \eqref{eqn:47}. Since $\lambda_M>0$, the limiting $M$-term is strictly
positive, so $z<1$. Moreover, \eqref{eqn:6} implies $L_R(k-1;x_N)<1$. Taking limits in the
normalized receiver condition at $k-1$ gives the weak version of the second inequality in
\eqref{eqn:48}, so $z\geq z^\ast$. Therefore
\[
z\geq \bar z_k.
\]
If $\bar z_k\geq1$, this contradicts $z<1$. If $\bar z_k<1$, then the monotonicity of the
right-hand side of \eqref{eqn:49} gives
\[
\lambda_M\leq \hat\lambda_{M,k}\leq \hat\lambda_M,
\]
contradicting $\lambda_M>\hat\lambda_M$. Thus, for all sufficiently large $N$, no
informative equilibrium exists. Since babbling equilibria always exist, information
transmission fails.

Next, translate the threshold from $\lambda_M$ back to $q^0_M$. Since
\[
q^0_M=\frac{\lambda_M}{1+\lambda_L+\lambda_M},
\]
define
\begin{equation}
\label{eqn:53}
\hat q:=
\frac{\hat\lambda_M}{1+\lambda_L+\hat\lambda_M}.
\end{equation}
Because $q^0_M$ is strictly increasing in $\lambda_M$ when $\lambda_L$ is fixed,
information transmission persists if $q^0_M<\hat q$ and fails if $q^0_M>\hat q$.

It remains to verify the comparative statics of $\hat q$. Since $q^0_M$ is strictly
increasing in $\lambda_M$, it is enough to show that $\hat\lambda_M$ is weakly decreasing
in $a/r$. Fix $k$. If $\bar z_k\geq1$, then $\hat\lambda_{M,k}=0$. Otherwise,
$\hat\lambda_{M,k}$ is given by \eqref{eqn:50}. The map in \eqref{eqn:49} is strictly
decreasing in $z$, so increasing $\bar z_k$ weakly decreases $\hat\lambda_{M,k}$.

When $a/r$ increases through a decrease in $r$, $z^0_k$ is unchanged and $z^\ast$
increases, so $\hat\lambda_{M,k}$ weakly decreases. When $a/r$ increases through an
increase in $U_R(H)$, both $r$ and $m$ decrease proportionally, which also increases
$z^\ast$ and leaves $z^0_k$ unchanged; hence $\hat\lambda_{M,k}$ weakly decreases.

When $a/r$ increases through an increase in $a$, $z^0_k$ increases and $z^\ast$ also
increases. If $\bar z_k=z^0_k$, then \eqref{eqn:50} becomes
\[
\hat\lambda_{M,k}
=
\frac{1-z^0_k}
{b\left(\frac{\rho_M}{\rho_H}\right)^k},
\]
which decreases in $a$. If $\bar z_k=z^\ast$, write
\[
A:=\frac{\rho_H}{\rho_M}\cdot\frac{m}{b},
\qquad
D:=\frac{\rho_H}{\rho_L}\cdot r.
\]
Then $z^\ast=(1+A)/(A+D/a)$, and \eqref{eqn:50} is proportional to
\[
(D-a)(Aa+D)^{\alpha-1},
\]
which is decreasing in $a$ because $\alpha\in(0,1)$.

Finally, when $a/r$ increases through a decrease in $U_S(H)$, both $a$ and $b$ increase
proportionally. If $\bar z_k=z^0_k$, then \eqref{eqn:50} decreases directly. If
$\bar z_k=z^\ast$, writing $h:=U_S(H)$, \eqref{eqn:50} is proportional to
\[
\frac{d_0h-1}{(1+A_0h)^\alpha}
\]
for positive constants $d_0$ and $A_0$, and this expression is increasing in $h$ on the
relevant region. Hence it decreases when $h$ decreases. Therefore each
$\hat\lambda_{M,k}$, and hence $\hat\lambda_M=\sup_{k\geq1}\hat\lambda_{M,k}$, is weakly
decreasing in the conflict ratio $a/r$. Thus $\hat q$ is weakly decreasing in
\[
\frac{U_S(L)}{U_S(H)}\cdot
\frac{U_R(H)}{U_R(L)}.
\]

\subsection{Proof of Proposition~\ref{prop:3}}

If no informative equilibrium exists under the original, higher-conflict parameter vector, then $x_{\max}=0$ and the desired comparison is immediate. Suppose instead that an informative equilibrium exists under the original parameter vector. Use the notation $\lambda_L,\lambda_M,a,b,r,m$ from the proof of \cref{thm:4}. Thus
\[
\lambda_L:=\frac{q^0_L}{q^0_H},
\qquad
\lambda_M:=\frac{q^0_M}{q^0_H},
\]
and
\[
a:=-\frac{U_S(L)}{U_S(H)},
\qquad
b:=-\frac{U_S(M)}{U_S(H)},
\qquad
r:=-\frac{U_R(L)}{U_R(H)},
\qquad
m:=\frac{U_R(M)}{U_R(H)}.
\]
Then $a,b,r,m>0$, \eqref{eqn:2} gives $r\leq a$, and the conflict ratio is $a/r$. When $q^0_M$ varies, we hold $\lambda_L$ fixed. Since
\begin{equation}
\label{eqn:54}
q^0_M=\frac{\lambda_M}{1+\lambda_L+\lambda_M},
\end{equation}
$q^0_M$ is strictly increasing in $\lambda_M$.

It is enough to show that every informative equilibrium under a higher-conflict parameter vector generates an informative equilibrium under any lower-conflict parameter vector with weakly larger $x$. Fix $N$ and an informative equilibrium $(x,k)$ under the original parameter vector, where $k=\HT$. By \cref{lem:1}, $x_b=0$ and $x_g=x$.

For $y\in(0,1]$, define
\[
D_k(y)
:=
\lambda_M b
\left(\frac{\rho_M}{\rho_H}\right)^k
\left(\frac{1-\rho_M y}{1-\rho_H y}\right)^{N-k}
+
\lambda_L a
\left(\frac{\rho_L}{\rho_H}\right)^k
\left(\frac{1-\rho_L y}{1-\rho_H y}\right)^{N-k}.
\]
For each $i\in\{L,M\}$,
\begin{equation}
\label{eqn:55}
\frac{d}{dy}
\log\left(\frac{1-\rho_i y}{1-\rho_H y}\right)
=
\frac{\rho_H-\rho_i}{(1-\rho_i y)(1-\rho_H y)}
>0.
\end{equation}
Hence, for $k<N$, $D_k$ is strictly increasing in $y$; for $k=N$, $D_k$ is constant in $y$. Moreover,
\[
L_S(g,\mathrm{piv};y,k)=\frac{1}{D_k(y)}.
\]
Since $(x,k)$ is an informative equilibrium, \eqref{eqn:4} implies $L_S(g,\mathrm{piv};x,k)\geq1$, with equality whenever $x<1$. Hence
\[
D_k(x)\leq1.
\]

Now compare the original parameter vector with a lower-conflict parameter vector. Primes denote the objects under the lower-conflict parameter vector. The lower-conflict change is one of the following: $\lambda_M$ decreases, or $a/r$ decreases by varying one of $U_S(L)$, $U_S(H)$, $U_R(L)$, and $U_R(H)$ while holding the other three fixed, within the parameter region satisfying \eqref{eqn:2}. If $\lambda_M$ decreases, or if the decrease in $a/r$ comes through $U_S(L)$ or $U_S(H)$, then $D'_k(y)\leq D_k(y)$ for every $y\in(0,1]$. If the decrease in $a/r$ comes through $U_R(L)$ or $U_R(H)$, then $D'_k=D_k$. Hence
\[
D'_k(x)\leq D_k(x)\leq1.
\]

Choose $x'\geq x$ as follows. If there exists $y\in[x,1]$ such that $D'_k(y)=1$, let $x'$ be one such value. Otherwise, set $x'=1$. By the last display and continuity of $D'_k$, either
\[
D'_k(x')=1,
\]
or
\[
x'=1
\qquad\text{and}\qquad
D'_k(1)<1.
\]

We next verify the receiver's cutoff condition. For $t<N$,
\[
L_R(t;y)
=
\frac{1}{\lambda_L r}
\left(\frac{\rho_H}{\rho_L}\right)^t
\left(\frac{1-\rho_H y}{1-\rho_L y}\right)^{N-t}
+
\frac{\lambda_M m}{\lambda_L r}
\left(\frac{\rho_M}{\rho_L}\right)^t
\left(\frac{1-\rho_M y}{1-\rho_L y}\right)^{N-t}.
\]
For each $i\in\{M,H\}$,
\begin{equation}
\label{eqn:56}
\frac{d}{dy}
\log\left(\frac{1-\rho_i y}{1-\rho_L y}\right)
=
\frac{\rho_L-\rho_i}{(1-\rho_i y)(1-\rho_L y)}
<0.
\end{equation}
Therefore $L_R(t;y)$ is decreasing in $y$. It is also weakly lower under the lower-conflict parameter vector: lowering $\lambda_M$ lowers the second term; lowering $a/r$ through $U_S(L)$ or $U_S(H)$ leaves $L_R$ unchanged; lowering $a/r$ through $U_R(L)$ lowers both terms; and lowering $a/r$ through $U_R(H)$ lowers the first term and leaves the second term unchanged. Therefore, since $k-1<N$ and $x'\geq x$,
\begin{equation}
\label{eqn:57}
L'_R(k-1;x')\leq L_R(k-1;x)<1.
\end{equation}

First consider the case $D'_k(x')=1$. Then
\[
L'_S(g,\mathrm{piv};x',k)=1.
\]
For each $\theta\in\{L,M,H\}$,
\[
\rho_\theta\mathbb{P}[\mathrm{piv}\mid\theta;x',k]
=
\frac{k}{Nx'}\mathbb{P}[T=k\mid\theta;x'].
\]
The coefficient is positive and independent of $\theta$. Hence $L'_S(g,\mathrm{piv};x',k)=1$ implies
\[
q^{0\prime}_H\mathbb{P}[T=k\mid H;x']
=
a' q^{0\prime}_L\mathbb{P}[T=k\mid L;x']
+
b' q^{0\prime}_M\mathbb{P}[T=k\mid M;x'].
\]
Since $r'\leq a'$, we obtain
\[
q^{0\prime}_H\mathbb{P}[T=k\mid H;x']
-
r' q^{0\prime}_L\mathbb{P}[T=k\mid L;x']
+
m' q^{0\prime}_M\mathbb{P}[T=k\mid M;x']
\geq0.
\]
Thus $L'_R(k;x')\geq1$. Combining this with \eqref{eqn:57} and the monotonicity of $L'_R(\cdot;x')$, the receiver's best response is the cutoff $k$. Moreover, $L'_S(g,\mathrm{piv};x',k)=1$ and strict MLRP imply
\[
L'_S(b,\mathrm{piv};x',k)<1.
\]
Thus the senders' best-response condition in \eqref{eqn:4} is satisfied. Hence $(x',k)$ is an informative equilibrium under the lower-conflict parameter vector.

Now consider the case $x'=1$ and $D'_k(1)<1$. Then
\[
L'_S(g,\mathrm{piv};1,k)>1.
\]
Since
\[
\rho_\theta\mathbb{P}[\mathrm{piv}\mid\theta;1,k]
=
\frac{k}{N}\mathbb{P}[T=k\mid\theta;1],
\]
the senders strictly prefer the proposal conditional on $T=k$. Therefore,
\[
q^{0\prime}_H\mathbb{P}[T=k\mid H;1]
-
a' q^{0\prime}_L\mathbb{P}[T=k\mid L;1]
-
b' q^{0\prime}_M\mathbb{P}[T=k\mid M;1]
>0.
\]
Using $r'\leq a'$, $b'>0$, and $m'>0$, this implies
\[
q^{0\prime}_H\mathbb{P}[T=k\mid H;1]
-
r' q^{0\prime}_L\mathbb{P}[T=k\mid L;1]
+
m' q^{0\prime}_M\mathbb{P}[T=k\mid M;1]
>0.
\]
Hence $L'_R(k;1)>1$.

By \eqref{eqn:57}, the receiver strictly prefers the status quo at $T=k-1$. Equivalently,
\[
q^{0\prime}_H\mathbb{P}[T=k-1\mid H;1]
-
r' q^{0\prime}_L\mathbb{P}[T=k-1\mid L;1]
+
m' q^{0\prime}_M\mathbb{P}[T=k-1\mid M;1]
<0.
\]
Since $r'\leq a'$ and $m',b'>0$, this implies
\[
q^{0\prime}_H\mathbb{P}[T=k-1\mid H;1]
-
a' q^{0\prime}_L\mathbb{P}[T=k-1\mid L;1]
-
b' q^{0\prime}_M\mathbb{P}[T=k-1\mid M;1]
<0.
\]
Thus the senders strictly prefer the status quo conditional on $T=k-1$. Since
\[
(1-\rho_\theta)\mathbb{P}[\mathrm{piv}\mid\theta;1,k]
=
\frac{N-k+1}{N}\mathbb{P}[T=k-1\mid\theta;1],
\]
we have
\[
L'_S(b,\mathrm{piv};1,k)<1.
\]
Therefore the senders' best-response condition in \eqref{eqn:4} is satisfied. Together with $L'_R(k;1)>1$ and \eqref{eqn:57}, the receiver's best response is the cutoff $k$. Hence $(1,k)$ is an informative equilibrium under the lower-conflict parameter vector.

We have shown that every informative equilibrium under the original parameter vector generates an informative equilibrium under the lower-conflict parameter vector with weakly larger $x$. Applying this construction to the most informative equilibrium under the original parameter vector gives
\[
x'_{\max}\geq x'\geq x_{\max}.
\]
Therefore $x_{\max}$ is weakly decreasing in $q^0_M$ and in
\[
\frac{U_S(L)}{U_S(H)}\cdot\frac{U_R(H)}{U_R(L)}.
\]

\subsection{Proof of Proposition~\ref{prop:finite_sender_optimum}}

Write
\[
    w_\theta=q_\theta^0U_S(\theta),
    \qquad
    w_L,w_M<0<w_H.
\]
Recall that every informative symmetric equilibrium has the form
\((x,k)\), where \(x_b=0\), \(x_g=x\in(0,1]\), and the receiver
implements the proposal if and only if \(T\geq k\).

We first verify that a most informative equilibrium exists for every fixed
\(N\). For each \(k\leq N\), define the loss-to-gain ratio
\begin{equation}
\label{eqn:a1}
D_{N,k}(x)
:=
\frac{
    \displaystyle
    \sum_{\theta\in\{L,M\}}
    (-w_\theta)\rho_\theta^k
    (1-\rho_\theta x)^{N-k}
}{
    w_H\rho_H^k(1-\rho_Hx)^{N-k}
}.
\end{equation}
At every interior equilibrium with cutoff \(k\), the good-signal
indifference condition is equivalent to \(D_{N,k}(x)=1\). If \(k<N\),
each term in \eqref{eqn:a1} is strictly increasing in \(x\), so there is
at most one interior candidate for each cutoff; the only possible boundary
candidate is \(x=1\). If \(k=N\), the sender incentive conditions have
signs independent of \(x\). Moreover, whenever some \(x\) supports cutoff
\(N\), so does \(x=1\): the receiver's score at tally \(N\) has a sign
independent of \(x\), while her score at tally \(N-1\) weakly decreases
with \(x\). Since there are finitely many cutoffs, a largest feasible
\(x\) exists for every \(N\).

Let
\[
    \overline V
    :=
    \sup_{N\geq2}V_S^{\max}(N).
\]
Suppose, toward a contradiction, that \(\overline V\) is not attained at
any finite population. We can then choose distinct populations
\(N_j\to\infty\) such that
\[
    V_S^{\max}(N_j)\longrightarrow\overline V.
\]
The most informative equilibrium cannot be babbling along an infinite
subsequence, since the babbling payoff is independent of \(N\) and would
then attain \(\overline V\). Hence, after passing to a subsequence, write
the most informative equilibrium as
\[
    \Gamma_{\max}(N_j)=(x_j,k_j),
    \qquad x_j>0.
\]

By \cref{thm:3}, both \(N_jx_j\) and \(k_j\) are uniformly bounded.
Passing to a further subsequence, we may therefore assume that
\[
    k_j=k,
    \qquad
    \mu_j:=N_jx_j\longrightarrow\mu\geq0.
\]
Since \(N_j\to\infty\), eventually \(k<N_j\) and \(x_j\in(0,1)\), so the
good-signal type is indifferent and
\[
    D_{N_j,k}(x_j)=1.
\]

We claim that \(\mu>0\). If instead \(\mu=0\), then, for
\(\theta\in\{L,M\}\),
\[
    \left(
    \frac{1-\rho_\theta x_j}{1-\rho_Hx_j}
    \right)^{N_j-k}
    \longrightarrow1.
\]
Thus, \eqref{eqn:a1} and \(D_{N_j,k}(x_j)=1\) imply that
\(D_{N_j,k}(0)=1\). But \(D_{N_j,k}\) is strictly increasing and
\(x_j>0\), which would give \(D_{N_j,k}(x_j)>1\), a contradiction.
Therefore,
\[
    \mu>0.
\]

For fixed \(k\), define
\[
    Q_{N,k}(z)
    :=
    \Pr\!\left\{
        \operatorname{Bin}\left(N,\frac{z}{N}\right)\geq k
    \right\},
    \qquad
    Q_k(z)
    :=
    \Pr\{\operatorname{Poisson}(z)\geq k\},
\]
and
\[
    F_{N,k}(u)
    :=
    \sum_{\theta}w_\theta Q_{N,k}(\rho_\theta u),
    \qquad
    f_k(u)
    :=
    \sum_{\theta}w_\theta Q_k(\rho_\theta u).
\]
Then
\[
    V_S^{\max}(N_j)=F_{N_j,k}(\mu_j).
\]
Moreover, differentiation of a binomial tail shows that the exact
good-signal indifference condition is
\begin{equation}
\label{eqn:a2}
    F_{N_j,k}'(\mu_j)=0.
\end{equation}

Let
\[
    \pi_t(z):=e^{-z}\frac{z^t}{t!},
    \qquad
    \pi_t(z):=0\quad\text{for }t<0.
\]
For every fixed \(k\), uniformly for \(z\) in compact sets,
\begin{equation}
\label{eqn:a3}
    Q_{N,k}(z)
    =
    Q_k(z)
    +
    \frac{z^2}{2N}
    \bigl\{\pi_{k-1}(z)-\pi_{k-2}(z)\bigr\}
    +
    O(N^{-2}),
\end{equation}
and the expansion remains valid after differentiation. Indeed, for every
fixed \(t<k\),
\[
\Pr\!\left\{
    \operatorname{Bin}\left(N,\frac{z}{N}\right)=t
\right\}
=
\pi_t(z)
\left[
    1+
    \frac{tz-z^2/2-t(t-1)/2}{N}
    +O(N^{-2})
\right],
\]
and summing these expansions over \(t<k\) gives \eqref{eqn:a3}.

Since
\[
    \frac{d^2}{du^2}Q_k(\rho_\theta u)
    =
    \rho_\theta^2
    \bigl\{
        \pi_{k-2}(\rho_\theta u)
        -
        \pi_{k-1}(\rho_\theta u)
    \bigr\},
\]
equation \eqref{eqn:a3} implies
\begin{equation}
\label{eqn:a4}
    F_{N,k}(u)
    =
    f_k(u)
    -
    \frac{u^2}{2N}f_k''(u)
    +
    O(N^{-2}),
\end{equation}
uniformly on compact subsets of \((0,\infty)\), together with its first
derivative.

Ordinary binomial-to-Poisson convergence gives
\[
    \overline V=f_k(\mu),
\]
while \eqref{eqn:a2} and the differentiated version of
\eqref{eqn:a4} imply \(f_k'(\mu)=0\). Furthermore,
\[
    f_k'(u)
    =
    \frac{u^{k-1}}{(k-1)!}
    \sum_{\theta}w_\theta\rho_\theta^k e^{-\rho_\theta u}.
\]
At a root of \(f_k'\), the positive \(H\)-state term equals the sum of
the magnitudes of the \(L\)- and \(M\)-state terms. Since
\(\rho_H>\rho_M>\rho_L\),
\[
\begin{aligned}
\sum_\theta
w_\theta\rho_\theta^{k+1}e^{-\rho_\theta\mu}
={}&
(\rho_H-\rho_L)(-w_L)\rho_L^ke^{-\rho_L\mu}\\
&+
(\rho_H-\rho_M)(-w_M)\rho_M^ke^{-\rho_M\mu}
>0.
\end{aligned}
\]
Consequently,
\begin{equation}
\label{eqn:a5}
    f_k'(\mu)=0,
    \qquad
    f_k''(\mu)<0.
\end{equation}

The differentiated expansion in \eqref{eqn:a4}, together with
\eqref{eqn:a2} and \eqref{eqn:a5}, yields
\[
    \mu_j-\mu=O(N_j^{-1}).
\]
Using again \eqref{eqn:a4} and \(f_k'(\mu)=0\), we obtain
\[
\begin{aligned}
V_S^{\max}(N_j)
&=
F_{N_j,k}(\mu_j)\\
&=
f_k(\mu)
-
\frac{\mu^2}{2N_j}f_k''(\mu)
+
O(N_j^{-2})\\
&>
f_k(\mu)
=
\overline V
\end{aligned}
\]
for all sufficiently large \(j\), where the strict inequality follows
from \(\mu>0\) and \(f_k''(\mu)<0\). This contradicts the definition of
\(\overline V\).

Therefore, the supremum of \(V_S^{\max}(N)\) is attained at some finite
population size \(N^*\geq2\).

\subsection{Proof of Proposition~\ref{prop:4}}

\subsubsection{Step 1: Lowest-Signal Pooling}

\begin{lemma}
\label{lem:3}
In every non-babbling monotone equilibrium represented in the reduced message space,
senders who observe the lowest signal \(s_1\) send the lowest message \(z_1\) with
probability one:
\[
    p_{1,1}=1.
\]
\end{lemma}

\begin{proof}
Fix a non-babbling monotone equilibrium \((\boldsymbol P,\psi)\) represented in the reduced message
space introduced before \cref{prop:4}. Write
\[
    \pi_{j,\theta}:=\Pr[s_j\mid\theta],
    \qquad
    f_{\theta,k}:=\Pr[z_k\mid\theta;P]
    =
    \sum_{\ell=1}^J \pi_{\ell,\theta}p_{\ell,k}.
\]
Since inactive messages have been discarded and signals have full support, every remaining
message has positive probability in every state.

Suppose, toward a contradiction, that for some \(k>1\),
\[
    p_{1,k}>0.
\]
Then \(z_k\) is active. Since \(z_1\) is also active in the reduced message space and no two
active messages are equivalent, \(z_k\) is not equivalent to \(z_1\). For each
\(\tilde{\boldsymbol T}\in\Delta^K(N-1)\), define
\[
    \Delta_k(\tilde{\boldsymbol T})
    :=
    \psi(\tilde{\boldsymbol T}+e_k)
    -
    \psi(\tilde{\boldsymbol T}+e_1).
\]
By monotonicity of \(\psi\), \(\Delta_k(\tilde{\boldsymbol T})\ge0\) for every
\(\tilde{\boldsymbol T}\). Since \(z_k\) is not equivalent to \(z_1\),
\(\Delta_k(\tilde{\boldsymbol T})>0\) for some \(\tilde{\boldsymbol T}\).

Because \(p_{1,k}>0\), message \(z_k\) is a best response after signal \(s_1\). Hence
sending \(z_k\) gives a weakly higher payoff than sending \(z_1\):
\begin{equation}
    \sum_{\tilde{\boldsymbol T}\in\Delta^K(N-1)}
    \Delta_k(\tilde{\boldsymbol T})
    \sum_{\theta\in\{L,M,H\}}
    q^0_\theta\pi_{1,\theta}
    \Pr[\tilde{\boldsymbol T}\mid\theta;P]U_S(\theta)
    \ge 0.
    \label{eqn:58}
\end{equation}
For each state \(\theta\), set
\[
    A_\theta
    :=
    q^0_\theta\pi_{1,\theta}
    \sum_{\tilde{\boldsymbol T}\in\Delta^K(N-1)}
    \Delta_k(\tilde{\boldsymbol T})
    \Pr[\tilde{\boldsymbol T}\mid\theta;P].
\]
Then \(A_\theta\ge0\) for every \(\theta\), and \eqref{eqn:58} implies
\[
    \sum_{\theta\in\{L,M,H\}} A_\theta U_S(\theta)\ge0.
\]
Since \(U_S(M)<0\),
\[
    A_HU_S(H)+A_LU_S(L)
    \ge
    -A_MU_S(M)
    \ge0.
\]
Thus
\[
    A_H\ge -\frac{U_S(L)}{U_S(H)}A_L.
\]
By \eqref{eqn:2},
\[
    -\frac{U_R(L)}{U_R(H)}
    \le
    -\frac{U_S(L)}{U_S(H)}.
\]
Therefore
\[
    A_HU_R(H)+A_LU_R(L)\ge0.
\]
Using also \(U_R(M)>0\), we obtain
\begin{equation}
    \sum_{\theta\in\{L,M,H\}} A_\theta U_R(\theta)\ge0.
    \label{eqn:59}
\end{equation}

Now define
\[
    \omega_\theta:=\frac{f_{\theta,1}}{\pi_{1,\theta}}.
\]
Strict MLRP of the signal distribution implies
\[
    \omega_L\le\omega_M\le\omega_H.
\]
Indeed,
\[
    \omega_\theta
    =
    \sum_{\ell=1}^J p_{\ell,1}
    \frac{\pi_{\ell,\theta}}{\pi_{1,\theta}},
\]
and each ratio \(\pi_{\ell,\theta}/\pi_{1,\theta}\) is weakly increasing in \(\theta\).

Combining this monotonicity with \eqref{eqn:59},
\[
    \sum_{\theta}\omega_\theta A_\theta U_R(\theta)
    =
    \omega_L\sum_{\theta}A_\theta U_R(\theta)
    +(\omega_M-\omega_L)A_MU_R(M)
    +(\omega_H-\omega_L)A_HU_R(H)
    \ge0.
\]
Substituting the definitions of \(A_\theta\) and \(\omega_\theta\), this gives
\begin{equation}
    \sum_{\tilde{\boldsymbol T}\in\Delta^K(N-1)}
    \Delta_k(\tilde{\boldsymbol T})
    \sum_{\theta\in\{L,M,H\}}
    q^0_\theta f_{\theta,1}
    \Pr[\tilde{\boldsymbol T}\mid\theta;P]U_R(\theta)
    \ge0.
    \label{eqn:60}
\end{equation}

For every \(\tilde{\boldsymbol T}\in\Delta^K(N-1)\) and every state \(\theta\),
\begin{equation}
    f_{\theta,1}\Pr[\tilde{\boldsymbol T}\mid\theta;P]
    =
    \frac{\tilde T_1+1}{N}
    \Pr[\tilde{\boldsymbol T}+e_1\mid\theta;P].
    \label{eqn:61}
\end{equation}
Let
\[
    R(\boldsymbol T)
    :=
    \sum_{\theta\in\{L,M,H\}}
    q^0_\theta\Pr[\boldsymbol T\mid\theta;P]U_R(\theta).
\]
Using \eqref{eqn:61}, inequality \eqref{eqn:60} becomes
\begin{equation}
    \sum_{\tilde{\boldsymbol T}\in\Delta^K(N-1)}
    \Delta_k(\tilde{\boldsymbol T})
    \frac{\tilde T_1+1}{N}
    R(\tilde{\boldsymbol T}+e_1)
    \ge0.
    \label{eqn:62}
\end{equation}

Since \(\Delta_k(\tilde{\boldsymbol T})\ge0\) for every \(\tilde{\boldsymbol T}\), and
\(\Delta_k(\tilde{\boldsymbol T})>0\) for some \(\tilde{\boldsymbol T}\), \eqref{eqn:62}
implies that there exists \(\tilde{\boldsymbol T}\) such that
\[
    \Delta_k(\tilde{\boldsymbol T})>0
    \quad\text{and}\quad
    R(\tilde{\boldsymbol T}+e_1)\ge0.
\]
At the count \(\tilde{\boldsymbol T}+e_1\), the receiver weakly prefers the proposal.
By the receiver's tie-breaking rule,
\[
    \psi(\tilde{\boldsymbol T}+e_1)=1.
\]
But then
\[
    \Delta_k(\tilde{\boldsymbol T})
    =
    \psi(\tilde{\boldsymbol T}+e_k)
    -
    \psi(\tilde{\boldsymbol T}+e_1)
    \le0,
\]
contradicting \(\Delta_k(\tilde{\boldsymbol T})>0\).

Therefore \(p_{1,k}=0\) for every \(k>1\). Since the probabilities in the first row of
\(P\) sum to one, \(p_{1,1}=1\).
\end{proof}

\subsubsection{Step 2: Non-Crossing of Equilibrium Supports}

\begin{lemma}
\label{lem:4}
In every non-babbling monotone equilibrium represented in the reduced message space,
the supports of the senders' message distributions are non-crossing. That is, for every
\(j<j'\) and \(k<k'\),
\[
    p_{j,k'}p_{j',k}=0.
\]
Equivalently, if a lower signal sends a higher message with positive probability, then no
higher signal sends a strictly lower message with positive probability.
\end{lemma}

\begin{proof}
Fix a non-babbling monotone equilibrium \((\boldsymbol P,\psi)\) represented in the reduced message
space. Suppose, toward a contradiction, that there exist \(j<j'\) and \(k<k'\) such that
\[
    p_{j,k'}>0
    \quad\text{and}\quad
    p_{j',k}>0.
\]
Then both \(z_k\) and \(z_{k'}\) are active. Since the equilibrium is represented in the
reduced message space, no two active messages are equivalent. Hence \(z_k\) and
\(z_{k'}\) are not equivalent.

For each \(\tilde{\boldsymbol T}\in\Delta^K(N-1)\), define
\[
    \Delta_{k,k'}(\tilde{\boldsymbol T})
    :=
    \psi(\tilde{\boldsymbol T}+e_{k'})
    -
    \psi(\tilde{\boldsymbol T}+e_k).
\]
By monotonicity of \(\psi\), \(\Delta_{k,k'}(\tilde{\boldsymbol T})\ge0\) for every
\(\tilde{\boldsymbol T}\). Since \(z_k\) and \(z_{k'}\) are not equivalent,
\(\Delta_{k,k'}(\tilde{\boldsymbol T})>0\) for some \(\tilde{\boldsymbol T}\).

Write
\[
    \pi_{\ell,\theta}:=\Pr[s_\ell\mid\theta].
\]
For each state \(\theta\in\{L,M,H\}\), set
\[
    B_\theta
    :=
    q^0_\theta
    \sum_{\tilde{\boldsymbol T}\in\Delta^K(N-1)}
    \Delta_{k,k'}(\tilde{\boldsymbol T})
    \Pr[\tilde{\boldsymbol T}\mid\theta;P].
\]
Then \(B_\theta\ge0\) for every \(\theta\), and in particular \(B_L>0\) and \(B_H>0\).
Indeed, since all remaining messages are active and signals have full support, every
remaining message has positive probability in states \(L\) and \(H\).

For any signal \(s_\ell\), the payoff gain from sending \(z_{k'}\) rather than \(z_k\),
up to the positive normalizing constant \(\Pr[s_\ell]\), is
\begin{equation}
    \Xi_\ell
    :=
    \sum_{\theta\in\{L,M,H\}}
    \pi_{\ell,\theta}B_\theta U_S(\theta).
    \label{eqn:63}
\end{equation}
Since \(p_{j,k'}>0\), message \(z_{k'}\) is a best response after signal \(s_j\). Hence
\[
    \Xi_j\ge0.
\]
Since \(p_{j',k}>0\), message \(z_k\) is a best response after signal \(s_{j'}\). Hence
\[
    \Xi_{j'}\le0.
\]

Now define
\[
    R_\theta:=\frac{\pi_{j',\theta}}{\pi_{j,\theta}}.
\]
By strict MLRP of the signal distribution and \(j<j'\),
\[
    R_L<R_M<R_H.
\]
Using \eqref{eqn:63},
\begin{equation}
    \Xi_{j'}
    =
    R_M\Xi_j
    +(R_H-R_M)\pi_{j,H}B_HU_S(H)
    +(R_L-R_M)\pi_{j,L}B_LU_S(L).
    \label{eqn:64}
\end{equation}
The first term in \eqref{eqn:64} is nonnegative because \(\Xi_j\ge0\). The second term is
strictly positive because \(R_H>R_M\), \(\pi_{j,H}>0\), \(B_H>0\), and \(U_S(H)>0\).
The third term is also strictly positive because \(R_L<R_M\), \(\pi_{j,L}>0\), \(B_L>0\),
and \(U_S(L)<0\). Therefore
\[
    \Xi_{j'}>0,
\]
contradicting \(\Xi_{j'}\le0\).

Thus no such \(j<j'\) and \(k<k'\) can exist. Hence
\[
    p_{j,k'}p_{j',k}=0
    \quad
    \text{for every } j<j' \text{ and } k<k',
\]
which proves the non-crossing property.
\end{proof}

\subsubsection{Step 3: Large-Deviation Lemmas}

\begin{lemma}[Boundary large-deviation bounds]
\label{lem:5}
Let \(q^n\to q\in\Delta^K\), and let \(Y^n\) be multinomial with
\(n\) trials and cell probabilities \(q^n\). Use the convention
\[
    KL(\gamma,q)
    :=
    \begin{cases}
    \displaystyle \sum_{k:q_k>0}\gamma_k\log\frac{\gamma_k}{q_k},
        & \text{if }\gamma_k=0\text{ whenever }q_k=0,\\[1.2ex]
    +\infty,
        & \text{otherwise.}
    \end{cases}
\]
For every closed set \(C\subseteq\Delta^K\),
\begin{equation}
    \limsup_{n\to\infty}\frac1n
    \log\Pr\left[\frac{Y^n}{n}\in C\right]
    \le
    -\inf_{\gamma\in C}KL(\gamma,q).
    \label{eqn:65}
\end{equation}
Moreover, for every sequence \(y^n\in\Delta^K(n)\) such that
\(y^n_k>0\) only if \(q^n_k>0\),
\begin{equation}
    \Pr[Y^n=y^n]
    \ge
    (n+1)^{-K}
    \exp\left\{
        -n\,KL\left(\frac{y^n}{n},q^n\right)
    \right\}.
    \label{eqn:66}
\end{equation}
Consequently, if \(\gamma_k=0\) whenever \(q_k=0\), then there exist lattice
points \(y^n\in\Delta^K(n)\) with \(y^n/n\to\gamma\) such that
\begin{equation}
    \Pr[Y^n=y^n]
    \ge
    \exp\{-n(KL(\gamma,q)+o(1))\}.
    \label{eqn:67}
\end{equation}
\end{lemma}

\begin{proof}
The multinomial formula and Stirling's bounds imply, uniformly over lattice
points \(y\in\Delta^K(n)\),
\[
    \Pr[Y^n=y]
    \le
    \operatorname{poly}(n)
    \exp\left\{-n\,KL\left(\frac{y}{n},q^n\right)\right\},
\]
where coordinates with \(q^n_k=0<y_k/n\) have probability zero. Since the
number of empirical types is polynomial in \(n\), this gives \eqref{eqn:65}
after taking limits, using lower semicontinuity of relative entropy.

For the pointwise lower bound, write \(\widehat\gamma^n=y^n/n\). The standard
type-count lower bound gives
\[
    {n\choose y^n_1,\ldots,y^n_K}
    \ge
    (n+1)^{-K}\exp\{nH(\widehat\gamma^n)\},
\]
where \(H\) denotes entropy. Hence, whenever \(y^n_k>0\) only if \(q^n_k>0\),
\[
    \Pr[Y^n=y^n]
    =
    {n\choose y^n_1,\ldots,y^n_K}
    \prod_{k=1}^K(q^n_k)^{y^n_k}
    \ge
    (n+1)^{-K}
    \exp\left\{
        -n\,KL\left(\widehat\gamma^n,q^n\right)
    \right\},
\]
which proves \eqref{eqn:66}.

Finally, if \(\gamma\) is supported on \(\{k:q_k>0\}\), choose
\(y^n/n\to\gamma\) with \(y^n_k=0\) whenever \(q_k=0\). Then
\(KL(y^n/n,q^n)\to KL(\gamma,q)\) along those coordinates with positive
limiting mass, and \eqref{eqn:66} gives \eqref{eqn:67}.
\end{proof}

\begin{lemma}[Chernoff separation under MLRP]
\label{lem:6}
Let \(f_L,f_M,f_H\) be strictly positive distributions on a finite ordered set
\(S\). Suppose \(f_M/f_L\) and \(f_H/f_M\) are weakly increasing on \(S\), and
both comparisons are nontrivial. Define
\[
    F:=\left\{\gamma\in\Delta(S):
    KL(\gamma,f_L)=\min\{KL(\gamma,f_M),KL(\gamma,f_H)\}\right\}.
\]
Let \(C=C_{L,M}\) be the Chernoff information between \(f_L\) and \(f_M\),
written equivalently as
\[
    C_{L,M}
    =
    \min_{\gamma\in\Delta(S):\,KL(\gamma,f_L)=KL(\gamma,f_M)}
    KL(\gamma,f_L).
\]
Then
\begin{equation}
    C
    <
    \inf_{\gamma\in F}KL(\gamma,f_H).
    \label{eqn:68}
\end{equation}
\end{lemma}

\begin{proof}
Write
\[
    X_k:=\frac{f_{M,k}}{f_{L,k}},
    \qquad
    Z_k:=\frac{f_{H,k}}{f_{M,k}},
    \qquad
    Y_k:=\frac{f_{H,k}}{f_{L,k}}=X_kZ_k.
\]
Both \(X_k\) and \(Z_k\) are weakly increasing and nonconstant. Fix
\(\beta\in(0,1)\), and define
\[
    A_\beta:=\sum_{k\in S}f_{L,k}X_k^\beta,
    \qquad
    \mu_\beta(k):=\frac{f_{L,k}X_k^\beta}{A_\beta}.
\]
Then
\[
    \frac{f_{M,k}}{\mu_\beta(k)}=A_\beta X_k^{1-\beta},
\]
so \(f_M\) strictly MLR-dominates \(\mu_\beta\). Since \(Z_k\) is increasing
and nonconstant,
\[
    E_{\mu_\beta}[Z]<E_{f_M}[Z]=1.
\]
By concavity of \(t\mapsto t^\beta\),
\[
    E_{\mu_\beta}[Z^\beta]
    \le
    \bigl(E_{\mu_\beta}[Z]\bigr)^\beta
    <1.
\]
Therefore
\[
    \sum_{k\in S}f_{L,k}Y_k^\beta
    =
    A_\beta E_{\mu_\beta}[Z^\beta]
    <
    A_\beta
    =
    \sum_{k\in S}f_{L,k}X_k^\beta.
\]
Because \(f_L\) and \(f_M\) are distinct and strictly positive, the function
\[
    \chi_{L,M}(\beta)
    :=
    -\log\sum_{k\in S} f_{L,k}^{1-\beta}f_{M,k}^{\beta}
\]
is continuous on \([0,1]\), equals zero at \(\beta=0\) and \(\beta=1\), and is
strictly positive on \((0,1)\) by strict Hölder inequality. Hence the Chernoff
expression for the pair \((L,M)\) has a maximizer \(\beta^*\in(0,1)\). Applying
the strict inequality above at this \(\beta^*\) gives
\[
    C_{L,H}
    \ge
    -\log\sum_{k\in S} f_{L,k}^{1-\beta^*}f_{H,k}^{\beta^*}
    >
    -\log\sum_{k\in S} f_{L,k}^{1-\beta^*}f_{M,k}^{\beta^*}
    =
    C_{L,M}.
\]

Let \(\gamma^*=\gamma^*_{L,M}\) be the tilted distribution
\[
    \gamma^*_k
    =
    \frac{f_{L,k}^{1-\beta^*}f_{M,k}^{\beta^*}}
    {\sum_{\ell\in S}f_{L,\ell}^{1-\beta^*}f_{M,\ell}^{\beta^*}}.
\]
Since \(\beta^*\) is an interior maximizer,
\[
    \sum_{k\in S}\gamma^*_k\log\frac{f_{M,k}}{f_{L,k}}=0.
\]
Therefore
\[
    KL(\gamma^*,f_L)-KL(\gamma^*,f_M)
    =
    \sum_{k\in S}\gamma^*_k\log\frac{f_{M,k}}{f_{L,k}}
    =
    0.
\]
Moreover, the common value is
\[
    KL(\gamma^*,f_L)
    =
    KL(\gamma^*,f_M)
    =
    -\log\sum_{k\in S} f_{L,k}^{1-\beta^*}f_{M,k}^{\beta^*}
    =
    C.
\]
The argument above gives \(E_{\gamma^*}[Z]<1\). Hence
\[
    E_{\gamma^*}[\log Z]
    <
    \log E_{\gamma^*}[Z]
    <0,
\]
so
\[
    KL(\gamma^*,f_H)-KL(\gamma^*,f_M)
    =
    -E_{\gamma^*}[\log Z]
    >0.
\]
Thus \(\gamma^*\in F\).

Now take any \(\gamma\in F\). If
\(KL(\gamma,f_L)=KL(\gamma,f_M)\), then
\(KL(\gamma,f_L)\ge C_{L,M}=C\). If
\(KL(\gamma,f_L)=KL(\gamma,f_H)\), then
\(KL(\gamma,f_L)\ge C_{L,H}>C\). Hence
\(KL(\gamma,f_L)\ge C\) for every \(\gamma\in F\). Since
\(\gamma\in F\) also implies \(KL(\gamma,f_H)\ge KL(\gamma,f_L)\), we have
\(KL(\gamma,f_H)\ge C\). Equality would place \(\gamma\) on the
\((L,H)\)-Chernoff boundary with cost \(C\), contradicting \(C_{L,H}>C\).
Compactness of \(F\) gives the uniform strict inequality \eqref{eqn:68}.
\end{proof}

\begin{lemma}[Localization of pivotal counts]
\label{lem:7}
Consider a sequence \(P^N\to P\) of monotone sender strategies. Let
\[
    S:=\left\{k:\sum_{j=1}^Jp_{j,k}>0\right\},
\]
and suppose \(1,h\in S\), with \(z_1\) and \(z_h\) active under \(P^N\) for
all sufficiently large \(N\). Let \(\psi^N\) be monotone receiver best responses,
and define
\[
    \Delta_N(\tilde{\boldsymbol T})
    :=
    \psi^N(\tilde{\boldsymbol T}+e_h)
    -
    \psi^N(\tilde{\boldsymbol T}+e_1).
\]
Let
\[
    F:=\left\{\gamma\in\Delta(S):
    KL(\gamma,f_L)=\min\{KL(\gamma,f_M),KL(\gamma,f_H)\}\right\}.
\]
For every open neighborhood \(U\) of \(F\) in \(\Delta(S)\), there exists
\(\eta>0\) such that, for all sufficiently large \(N\),
\begin{equation}
    \Delta_N(\tilde{\boldsymbol T})>0
    \quad\Longrightarrow\quad
    r_N(\tilde{\boldsymbol T})>\eta
    \quad\text{or}\quad
    \widehat\gamma_N(\tilde{\boldsymbol T})\in U,
    \label{eqn:69}
\end{equation}
where
\[
    r_N(\tilde{\boldsymbol T})
    :=
    \frac1{N-1}\sum_{k\notin S}\tilde T_k,
\]
and, when \(r_N(\tilde{\boldsymbol T})<1\),
\(\widehat\gamma_N(\tilde{\boldsymbol T})\in\Delta(S)\) is the empirical
distribution conditional on the message lying in \(S\).
\end{lemma}

\begin{proof}
Since \(U\) contains \(F\), compactness gives \(a>0\) such that, for every
\(\gamma\in\Delta(S)\setminus U\),
\[
    \left|
    KL(\gamma,f_L)-\min\{KL(\gamma,f_M),KL(\gamma,f_H)\}
    \right|
    \ge a.
\]
Full support of signals gives a uniform one-message likelihood-ratio bound: there is
\(B<\infty\) such that, for every active message \(k\), every large \(N\), and every
pair of states \(\theta,\theta'\),
\[
    B^{-1}\le \frac{f^N_{\theta,k}}{f^N_{\theta',k}}\le B.
\]
Indeed, this follows from
\[
    f^N_{\theta,k}
    =
    \sum_j\Pr[s_j\mid\theta]p^N_{j,k}
    \le
    \left(\max_j\frac{\Pr[s_j\mid\theta]}{\Pr[s_j\mid\theta']}\right)
    f^N_{\theta',k},
\]
and the reverse inequality is analogous.

Choose \(\eta>0\) so small that \(\eta\log B<a/8\) and
\((1-\eta)a>a/2\). Suppose \(r_N(\tilde{\boldsymbol T})\le\eta\) and
\(\widehat\gamma_N(\tilde{\boldsymbol T})\notin U\). The contribution of messages
in \(S\) to every log-likelihood ratio is
\[
    (N-1)(1-r_N(\tilde{\boldsymbol T}))
    \left[
    KL(\widehat\gamma_N(\tilde{\boldsymbol T}),f_{\theta'})
    -
    KL(\widehat\gamma_N(\tilde{\boldsymbol T}),f_{\theta})
    +o(1)
    \right],
\]
uniformly over such \(\tilde{\boldsymbol T}\), because \(f^N_{\theta,k}\to f_{\theta,k}\)
for every \(k\in S\). The outside-\(S\) messages can change any state likelihood ratio
by at most \(\exp\{\eta(N-1)\log B\}\), and the remaining sender's own message changes
likelihood ratios by at most a constant independent of \(N\).

If
\[
    KL(\widehat\gamma_N,f_L)
    \le
    \min\{KL(\widehat\gamma_N,f_M),KL(\widehat\gamma_N,f_H)\}-a,
\]
then, after both corrections above, the posterior probabilities of \(M\) and \(H\) after
both \(\tilde{\boldsymbol T}+e_1\) and \(\tilde{\boldsymbol T}+e_h\) are exponentially small.
The receiver therefore strictly chooses the status quo at both counts for all large \(N\).
If the reverse inequality with \(+a\) holds, then the posterior probability of \(L\) after
both counts is exponentially small, and the receiver strictly chooses the proposal at both
counts for all large \(N\), because \(U_R(M)>0\) and \(U_R(H)>0\). In either case,
\(\Delta_N(\tilde{\boldsymbol T})=0\). This proves \eqref{eqn:69}.
\end{proof}

\begin{lemma}[Rounded monotone paths]
\label{lem:8}
Let \(S\) be a finite ordered set, let \(\gamma^0\in\Delta(S)\), and let \(V\)
be an open neighborhood of \(\gamma^0\) in \(\Delta(S)\). There exists
\(\rho>0\) such that the following holds.

Suppose \(\gamma^-,\gamma^+\in\Delta(S)\) satisfy
\[
    \|\gamma^- - \gamma^0\|_1<\rho,
    \qquad
    \|\gamma^+ - \gamma^0\|_1<\rho,
\]
and \(\gamma^+\) first-order stochastically dominates \(\gamma^-\) on the
ordered set \(S\). Then, for all sufficiently large \(N\), there exist count
vectors \(T_N^-,T_N^+\in\Delta^K(N)\), supported on \(S\), and a path
\[
    T_N^0,T_N^1,\ldots,T_N^{m_N}
\]
such that
\[
    T_N^0=T_N^-,
    \qquad
    T_N^{m_N}=T_N^+,
    \qquad
    \frac{T_N^-}{N}\to\gamma^-,
    \qquad
    \frac{T_N^+}{N}\to\gamma^+,
\]
and, for each \(r<m_N\),
\[
    T_N^{r+1}=T_N^r-e_a+e_b
    \quad
    \text{for some } a,b\in S \text{ with } a<b.
\]
Moreover, \(m_N\le N\), and all empirical frequencies \(T_N^r/N\) lie in a
compact subset of \(V\) for all sufficiently large \(N\).
\end{lemma}

\begin{proof}
Choose \(\rho>0\) small enough that the closed \(\ell^1\)-ball of radius
\(4\rho\) around \(\gamma^0\) is contained in \(V\).

Since \(\gamma^+\) first-order stochastically dominates \(\gamma^-\), there
exists a monotone transport plan \((m_{ab})_{a,b\in S}\) such that
\(m_{ab}\ge0\), \(m_{ab}=0\) whenever \(a>b\), and
\[
    \gamma^-=\sum_{a,b\in S}m_{ab}e_a,
    \qquad
    \gamma^+=\sum_{a,b\in S}m_{ab}e_b.
\]
For example, this plan is obtained by the usual greedy quantile coupling.
Matching common mass on the diagonal first, we may choose the plan so that the
total mass moved upward satisfies
\[
    \sum_{a<b}m_{ab}
    \le
    \frac12\|\gamma^+-\gamma^-\|_1
    < \rho.
\]
Thus every partial transport obtained from this plan differs from \(\gamma^-\)
by less than \(2\rho\) in \(\ell^1\)-distance, and hence lies within \(3\rho\)
of \(\gamma^0\).

For each large \(N\), choose integers \(M^N_{ab}\ge0\), supported on pairs
\(a\le b\), such that
\[
    \sum_{a\le b}M^N_{ab}=N,
    \qquad
    \frac{M^N_{ab}}{N}\to m_{ab}
    \quad\text{for every }a\le b.
\]
This is possible by simultaneous rounding of the finitely many numbers
\(Nm_{ab}\). Define
\[
    T_N^-:=\sum_{a\le b}M^N_{ab}e_a,
    \qquad
    T_N^+:=\sum_{a\le b}M^N_{ab}e_b.
\]
Then \(T_N^-/N\to\gamma^-\) and \(T_N^+/N\to\gamma^+\).

Starting from \(T_N^-\), perform \(M^N_{ab}\) replacements \(e_a\mapsto e_b\)
for each pair \(a<b\), in any order. This produces a path from \(T_N^-\) to
\(T_N^+\) using only upward unit replacements. The number of replacements is
\[
    m_N=\sum_{a<b}M^N_{ab}\le N.
\]
Every empirical frequency along the path is \(o(1)\)-close to a partial
transport distribution generated by \((m_{ab})\). Hence, for all sufficiently
large \(N\), every empirical frequency \(T_N^r/N\) lies in the closed
\(\ell^1\)-ball of radius \(4\rho\) around \(\gamma^0\), which is a compact
subset of \(V\).
\end{proof}

\begin{lemma}[Pivotal likelihood bounds]
\label{lem:9}
In the setting of \cref{lem:7}, suppose the limiting message distributions on \(S\)
are weakly MLR ordered, nonidentical, and satisfy the hypotheses of \cref{lem:6}. Let
\(C=C_{L,M}\). Choose \(\delta>0\) and an open neighborhood \(U\) of \(F\) such that
\[
    \inf_{\gamma\in\overline U}KL(\gamma,f_H)>C+3\delta,
\]
where the closure is taken in \(\Delta(S)\). For
\[
    W_{\theta,N}
    :=
    \sum_{\tilde{\boldsymbol T}\in\Delta^K(N-1)}
    \Delta_N(\tilde{\boldsymbol T})
    \Pr[\tilde{\boldsymbol T}\mid\theta;P^N],
\]
we have
\begin{equation}
    W_{H,N}
    \le
    \exp\{-(N-1)(C+2\delta)+o(N)\},
    \label{eqn:70}
\end{equation}
and, for each \(\theta\in\{L,M\}\),
\begin{equation}
    W_{\theta,N}
    \ge
    \exp\{-(N-1)(C+\delta)+o(N)\}.
    \label{eqn:71}
\end{equation}
\end{lemma}

\begin{proof}
By \cref{lem:7}, there exists \(\eta_0>0\) such that pivotal counts must either
have outside-\(S\) frequency above \(\eta_0\), or have
\(\widehat\gamma_N\in U\). Fix \(\eta\in(0,\eta_0]\), to be chosen smaller
below. Then
\[
    \Delta_N(\tilde{\boldsymbol T})>0
    \quad\Longrightarrow\quad
    r_N(\tilde{\boldsymbol T})>\eta
    \quad\text{or}\quad
    \widehat\gamma_N(\tilde{\boldsymbol T})\in U.
\]

Since \(\sum_{k\notin S}f^N_{H,k}\to0\), a binomial Chernoff bound gives, for
every \(A>0\),
\[
    \Pr[r_N(\tilde{\boldsymbol T})>\eta\mid H;P^N]
    \le e^{-A(N-1)}
\]
for all sufficiently large \(N\). On the event
\(r_N(\tilde{\boldsymbol T})\le\eta\) and
\(\widehat\gamma_N(\tilde{\boldsymbol T})\in U\), condition on the number and
composition of outside-\(S\) messages. The conditional empirical distribution
over \(S\) is multinomial with cell probabilities
\[
    \bar f^N_{H,k}
    :=
    \frac{f^N_{H,k}}{\sum_{\ell\in S}f^N_{H,\ell}},
    \qquad k\in S,
\]
and \(\bar f^N_H\to f_H\). Since \(U\subseteq\overline U\), \(\overline U\) is
closed, and
\[
    \inf_{\gamma\in\overline U}KL(\gamma,f_H)>C+3\delta,
\]
\cref{lem:5} implies that, uniformly over the conditional sample sizes
\(n_S\ge(1-\eta)(N-1)\), the probability that the conditional empirical
distribution belongs to \(U\) is at most
\[
    \operatorname{poly}(N)
    \exp\{-(1-\eta)(N-1)(C+3\delta+o(1))\}.
\]
Summing over the polynomially many possible outside-\(S\) counts only changes
the polynomial factor. Choose \(\eta>0\) small enough that
\[
    (1-\eta)(C+3\delta)>C+2\delta,
\]
and then choose \(A>C+2\delta\). Since \(0\le\Delta_N\le1\), these bounds prove
\eqref{eqn:70}.

It remains to prove \eqref{eqn:71}. Let \(\gamma^*=\gamma^*_{L,M}\) be the
Chernoff point between \(f_L\) and \(f_M\). Choose an open neighborhood \(V\)
of \(\gamma^*\) such that
\[
    \sup_{\gamma\in\overline V}KL(\gamma,f_L)<C+\delta,
    \qquad
    \sup_{\gamma\in\overline V}KL(\gamma,f_M)<C+\delta.
\]
For \(\beta\) near the Chernoff parameter \(\beta^*\), write
\[
    \gamma(\beta)_k
    :=
    \frac{f_{L,k}^{1-\beta}f_{M,k}^{\beta}}
    {\sum_{\ell\in S}f_{L,\ell}^{1-\beta}f_{M,\ell}^{\beta}}.
\]
Choose \(\beta^-<\beta^*<\beta^+\), both sufficiently close to \(\beta^*\), and set
\[
    \gamma^-:=\gamma(\beta^-),
    \qquad
    \gamma^+:=\gamma(\beta^+).
\]
Because
\[
    \frac{\gamma^+_k}{\gamma^-_k}
    =
    \left(
        \frac{\sum_{\ell\in S}f_{L,\ell}^{1-\beta^-}f_{M,\ell}^{\beta^-}}
             {\sum_{\ell\in S}f_{L,\ell}^{1-\beta^+}f_{M,\ell}^{\beta^+}}
    \right)
    \left(\frac{f_{M,k}}{f_{L,k}}\right)^{\beta^+-\beta^-},
\]
and \(f_M/f_L\) is weakly increasing on \(S\), \(\gamma^+\) MLR-dominates
\(\gamma^-\). Hence \(\gamma^+\) first-order stochastically dominates
\(\gamma^-\).

By choosing \(\beta^-\) and \(\beta^+\) sufficiently close to \(\beta^*\), we also
ensure that \(\gamma^-\) and \(\gamma^+\) are close enough to \(\gamma^*\) for
\cref{lem:8} to apply with the neighborhood \(V\). Thus, for all
sufficiently large \(N\), there are count vectors
\(T_N^-,T_N^+\in\Delta^K(N)\), supported on \(S\), and a path
\[
    T_N^0,T_N^1,\ldots,T_N^{m_N}
\]
such that
\[
    T_N^0=T_N^-,
    \qquad
    T_N^{m_N}=T_N^+,
    \qquad
    \frac{T_N^-}{N}\to\gamma^-,
    \qquad
    \frac{T_N^+}{N}\to\gamma^+,
\]
each step is an upward unit replacement,
\[
    T_N^{r+1}=T_N^r-e_a+e_b
    \quad\text{for some }a,b\in S\text{ with }a<b,
\]
and all empirical frequencies \(T_N^r/N\) lie in a compact subset of \(V\). In
particular, \(m_N\le N\).

By continuity and the strict inequalities on both sides of the Chernoff
point, \(\beta^-\) and \(\beta^+\) can also be chosen so that
\[
    KL(\gamma^-,f_L)
    <
    \min\{KL(\gamma^-,f_M),KL(\gamma^-,f_H)\},
\]
and
\[
    KL(\gamma^+,f_M)
    <
    \min\{KL(\gamma^+,f_L),KL(\gamma^+,f_H)\}.
\]
Therefore, at the full count \(T_N^-\), the posterior probability of state
\(L\) converges to one, so the receiver strictly chooses the status quo for all
large \(N\). Similarly, at the full count \(T_N^+\), the posterior probability
of state \(M\) converges to one, so the receiver strictly chooses the proposal
for all large \(N\). Hence
\[
    \psi^N(T_N^0)=0,
    \qquad
    \psi^N(T_N^{m_N})=1.
\]

Each path step is upward, so monotonicity of \(\psi^N\) implies
\(\psi^N(T_N^{r+1})-\psi^N(T_N^r)\ge0\). Since the total increase along the
path is one and \(m_N\le N\), there exists \(r_N<m_N\) such that
\[
    \psi^N(T_N^{r_N+1})-\psi^N(T_N^{r_N})
    \ge
    \frac1N
    \ge
    \frac1{KN}.
\]
Write
\[
    T_N^{r_N+1}=T_N^{r_N}-e_a+e_b
    \quad\text{for some }a,b\in S\text{ with }a<b,
\]
and set
\[
    \tilde{\boldsymbol T}_N^*:=T_N^{r_N}-e_a.
\]
Then
\[
    \tilde{\boldsymbol T}_N^*+e_a=T_N^{r_N},
    \qquad
    \tilde{\boldsymbol T}_N^*+e_b=T_N^{r_N+1}.
\]
Since the path is supported on \(S\), and \(1=\min S\) and \(h=\max S\), we
have \(1\le a<b\le h\). By monotonicity of \(\psi^N\),
\[
    \psi^N(\tilde{\boldsymbol T}_N^*+e_1)
    \le
    \psi^N(\tilde{\boldsymbol T}_N^*+e_a)
    \le
    \psi^N(\tilde{\boldsymbol T}_N^*+e_b)
    \le
    \psi^N(\tilde{\boldsymbol T}_N^*+e_h).
\]
Therefore
\[
    \Delta_N(\tilde{\boldsymbol T}_N^*)
    =
    \psi^N(\tilde{\boldsymbol T}_N^*+e_h)
    -
    \psi^N(\tilde{\boldsymbol T}_N^*+e_1)
    \ge
    \psi^N(\tilde{\boldsymbol T}_N^*+e_b)
    -
    \psi^N(\tilde{\boldsymbol T}_N^*+e_a)
    \ge
    \frac1{KN}.
\]

The empirical frequency
\(\tilde{\boldsymbol T}_N^*/(N-1)\) is \(o(1)\)-close to \(T_N^{r_N}/N\).
Since the latter lies in a compact subset of \(V\), the former lies in \(V\)
for all large \(N\). Hence, for each \(\theta\in\{L,M\}\),
\[
    KL\left(\frac{\tilde{\boldsymbol T}_N^*}{N-1},f_\theta^N\right)
    \le
    C+\delta+o(1).
\]
Applying the pointwise lower bound \eqref{eqn:66} from \cref{lem:5} gives
\[
    \Pr[\tilde{\boldsymbol T}_N^*\mid\theta;P^N]
    \ge
    \exp\{-(N-1)(C+\delta)+o(N)\}.
\]
Therefore, for \(\theta\in\{L,M\}\),
\[
    W_{\theta,N}
    \ge
    \Delta_N(\tilde{\boldsymbol T}_N^*)
    \Pr[\tilde{\boldsymbol T}_N^*\mid\theta;P^N]
    \ge
    \frac1{KN}
    \exp\{-(N-1)(C+\delta)+o(N)\}.
\]
The polynomial factor \(1/(KN)\) is absorbed into \(o(N)\), giving
\eqref{eqn:71}.
\end{proof}

\subsubsection{Completion of the Proof of Proposition~\ref{prop:4}}

We work throughout with equilibria represented in the reduced message space introduced
before \cref{prop:4}. Suppose, toward a contradiction, that \cref{prop:4} fails. If a
reduced equilibrium has only one active message, then the conclusion is immediate. Hence
there exists a sequence of reduced non-babbling monotone equilibria \((P^N,\psi^N)\),
with \(N\to\infty\), such that either
\[
    p^N_{j,1}<1
    \quad\text{for some }j<J,
\]
or there exists \(\epsilon_0>0\) such that
\[
    p^N_{J,1}\le 1-\epsilon_0.
\]
After passing to a subsequence, the reduced message space has a fixed finite cardinality
\(K\), and \(P^N\to P\).

By \cref{lem:3}, \(p^N_{1,1}=1\) for every \(N\). If \(p^N_{j,1}<1\) for some \(j<J\),
then there exists \(k>1\) with \(p^N_{j,k}>0\). By \cref{lem:4}, this implies
\(p^N_{J,1}=0\). Hence, in either case, there is a constant \(c>0\) such that
\[
    \sum_{k>1}p^N_{J,k}\ge c
    \quad\text{for all }N.
\]
Passing to a further subsequence, choose \(r>1\) such that \(p^N_{J,r}\) has a positive
limit. Let
\[
    S:=\left\{k:\sum_{j=1}^Jp_{j,k}>0\right\}
\]
be the set of messages with positive limiting probability, and let \(h:=\max S\). Then
\(h>1\), and \(p^N_{J,h}\) has a positive limit. Indeed, if \(h=r\), this is immediate.
If \(h>r\), then \(p^N_{J,r}>0\) for all large \(N\), so \cref{lem:4} implies that no
signal \(j<J\) can send \(z_h\) with positive probability. Since \(h\in S\), the positive
limiting mass on \(z_h\) must therefore come from signal \(s_J\).

For \(\theta\in\{L,M,H\}\), write
\[
    f^N_{\theta,k}:=\Pr[z_k\mid\theta;P^N],
    \qquad
    f_{\theta,k}:=\Pr[z_k\mid\theta;P].
\]
Set \(d^N_j:=p^N_{j,1}\) and \(d_j:=p_{j,1}\). By \cref{lem:3,lem:4},
\(d^N_1=1\), and \((d^N_1,\ldots,d^N_J)\) is weakly decreasing. Hence
\((d_1,\ldots,d_J)\) is weakly decreasing. Since \(p^N_{J,h}\) has a positive limit,
\(d_J<1=d_1\), so \(d\) is not constant.

Strict MLRP of the signal distribution gives
\[
    f_{L,1}>f_{M,1}>f_{H,1}.
\]
To see this, let \(F_\theta(m):=\sum_{\ell=1}^m\Pr[s_\ell\mid\theta]\). Strict signal
MLRP implies \(F_L(m)>F_M(m)>F_H(m)\) for every \(m<J\). Summation by parts gives
\[
    f_{\theta,1}-f_{\theta',1}
    =
    \sum_{m=1}^{J-1}(d_m-d_{m+1})
    \bigl(F_\theta(m)-F_{\theta'}(m)\bigr).
\]
Because \(d\) is weakly decreasing and nonconstant, the inequalities are strict. Thus the
limiting message distributions are nonidentical. Monotonicity of \(P\) implies that
\((f_L,f_M,f_H)\) are weakly ordered by MLRP on \(S\).

For each \(\tilde{\boldsymbol T}\in\Delta^K(N-1)\), define
\[
    \Delta_N(\tilde{\boldsymbol T})
    :=
    \psi^N(\tilde{\boldsymbol T}+e_h)
    -
    \psi^N(\tilde{\boldsymbol T}+e_1).
\]
By monotonicity, \(\Delta_N(\tilde{\boldsymbol T})\ge0\). Define
\[
    W_{\theta,N}
    :=
    \sum_{\tilde{\boldsymbol T}\in\Delta^K(N-1)}
    \Delta_N(\tilde{\boldsymbol T})
    \Pr[\tilde{\boldsymbol T}\mid\theta;P^N].
\]
Let
\[
    F:=\left\{\gamma\in\Delta(S):
    KL(\gamma,f_L)=\min\{KL(\gamma,f_M),KL(\gamma,f_H)\}\right\},
\]
and let \(C=C_{L,M}\). By \cref{lem:6},
\[
    C<\inf_{\gamma\in F}KL(\gamma,f_H).
\]
Choose \(\delta>0\) and an open neighborhood \(U\) of \(F\) such that
\[
    \inf_{\gamma\in\overline U}KL(\gamma,f_H)>C+3\delta,
\]
where the closure is taken in \(\Delta(S)\).
Applying \cref{lem:9} gives \eqref{eqn:70} and \eqref{eqn:71}. Therefore
\begin{equation}
    \frac{W_{H,N}}{W_{L,N}}\to0,
    \qquad
    \frac{W_{H,N}}{W_{M,N}}\to0.
    \label{eqn:72}
\end{equation}

Now consider the payoff gain from sending \(z_h\) rather than \(z_1\) after signal \(s_j\):
\begin{equation}
    \Omega_{j,N}
    :=
    \sum_{\theta\in\{L,M,H\}}
    q^0_\theta\Pr[s_j\mid\theta]W_{\theta,N}U_S(\theta).
    \label{eqn:73}
\end{equation}
The only positive term in \eqref{eqn:73} is the \(H\)-term. Since no signal rules out any
state, the private-signal likelihood ratios
\(\Pr[s_j\mid H]/\Pr[s_j\mid L]\) and
\(\Pr[s_j\mid H]/\Pr[s_j\mid M]\) are finite. By \eqref{eqn:72}, the positive
\(H\)-term in \eqref{eqn:73} is negligible relative to the negative \(L\)- and \(M\)-terms.
Because \(q^0_L,q^0_M>0\), \(U_S(L)<0\), and \(U_S(M)<0\), we have
\[
    \Omega_{j,N}<0
    \quad\text{for every }j
\]
for all sufficiently large \(N\). Thus every sender strictly prefers \(z_1\) to \(z_h\),
regardless of his signal.

But \(p^N_{J,h}\) has a positive limit, so \(z_h\) is sent with positive probability after
\(s_J\) for all sufficiently large \(N\). This contradicts equilibrium optimality. Therefore
\cref{prop:4} cannot fail. Hence, for every \(\epsilon>0\), there exists \(N_\epsilon\)
such that, for every \(N>N_\epsilon\), every reduced monotone equilibrium satisfies
\[
    p_{j,1}=1\quad\forall j<J,
    \qquad
    p_{J,1}>1-\epsilon.
\]
This proves \cref{prop:4}.

\subsection{\texorpdfstring{Proof of \cref{thm:5}}{Proof of Theorem 5}}

\begin{lemma}[Local moderate deviations for binomial point probabilities]
\label{lem:10}
Fix $\rho\in(0,1)$ and let $\sigma:=\sqrt{\rho(1-\rho)}$. Let $Z_n\sim\operatorname{Bin}(n,\rho)$. If $z_n\to\infty$, $z_n=o(n^{1/6})$, and
\[
    m_n=\left\lfloor n\rho-\sigma\sqrt n\,z_n\right\rfloor,
\]
then
\begin{equation}
\label{eqn:74}
    \mathbb{P}[Z_n=m_n]
    =
    \frac{1+o(1)}{\sigma\sqrt{2\pi n}}
    \exp\left\{-\frac{z_n^2}{2}\right\}.
\end{equation}
Moreover, if $a_n$ is any integer sequence with $a_n/n\to a\neq\rho$, then there exist $c>0$ and $n_0$ such that
\begin{equation}
\label{eqn:75}
    \mathbb{P}[Z_n=a_n]\leq e^{-cn}
    \qquad\text{for all }n\geq n_0.
\end{equation}
\end{lemma}

\begin{proof}
Write $x_n:=m_n/n$. Then
\[
    x_n=\rho-\frac{\sigma z_n}{\sqrt n}+O(n^{-1}),
\]
so $x_n\to\rho$ and, for all large $n$, $x_n$ lies in a compact subset of $(0,1)$. Stirling's formula gives
\[
    \mathbb{P}[Z_n=m_n]
    =
    \frac{1+o(1)}{\sqrt{2\pi n x_n(1-x_n)}}
    \exp\{-nD(x_n\|\rho)\},
\]
where
\[
    D(x\|\rho)
    :=
    x\log\frac{x}{\rho}+(1-x)\log\frac{1-x}{1-\rho}.
\]
A Taylor expansion around $\rho$ gives
\[
    D(\rho+\delta\|\rho)
    =
    \frac{\delta^2}{2\sigma^2}+O(\delta^3).
\]
With $\delta_n:=x_n-\rho$, we have
\[
    nD(x_n\|\rho)
    =
    \frac{z_n^2}{2}
    +O\!\left(\frac{z_n}{\sqrt n}\right)
    +O\!\left(\frac{z_n^3}{\sqrt n}\right)
    +o(1)
    =
    \frac{z_n^2}{2}+o(1),
\]
because $z_n=o(n^{1/6})$. Also $x_n(1-x_n)\to\rho(1-\rho)=\sigma^2$, proving \eqref{eqn:74}.

For the second claim, if $a_n/n\to a\neq\rho$, the same Stirling representation gives
\[
    \mathbb{P}[Z_n=a_n]
    \leq
    C\sqrt n\,\exp\{-nD(a_n/n\|\rho)\}
\]
for some constant $C<\infty$ and all large $n$. Since $D(a\|\rho)>0$, choose $c\in(0,D(a\|\rho))$. Then \eqref{eqn:75} holds for all sufficiently large $n$.
\end{proof}

\begin{proof}[Proof of \cref{thm:5}]
The case $q^0_M=0$ is immediate from the discussion following \cref{thm:5}. We therefore prove the case $q^0_M>0$.

For $\theta\in\{L,M,H\}$, let
\[
    \sigma_\theta:=\sqrt{\rho_\theta(1-\rho_\theta)}.
\]
Fix any $\bar p\in[0,1)$ such that
\begin{equation}
\label{eqn:76}
    q^0_LU_R(L)+\bar p\,q^0_MU_R(M)<0.
\end{equation}
We first construct a sequence of feasible mechanisms that asymptotically recommends the proposal with probabilities
\[
    0,\qquad 1-\bar p,\qquad 1
\]
in states $L,M,H$, respectively.

Let $h_N:=N^{1/8}$. Define
\[
    \TT_N^\alpha
    :=
    1+\left\lfloor
        (N-1)\rho_M-\sigma_M\sqrt{N-1}\,h_N
    \right\rfloor .
\]
If $\bar p=0$, set $k_N:=h_N/2$. If $\bar p>0$, set
\[
    \tau:=\frac{\bar p}{1-\bar p},
\]
and define
\[
    k_N
    :=
    \left[
        h_N^2
        +2\log\left(
            \tau\,
            \frac{q^0_H\rho_HU_S(H)\sigma_M}
                 {q^0_M\rho_M[-U_S(M)]\sigma_H}
        \right)
    \right]^{1/2}.
\]
For all sufficiently large $N$, $k_N$ is well-defined, $k_N\to\infty$, and $k_N=o(N^{1/6})$. Define
\[
    \TT_N^\beta
    :=
    1+\left\lfloor
        (N-1)\rho_H-\sigma_H\sqrt{N-1}\,k_N
    \right\rfloor .
\]
Because $\rho_M<\rho_H$ and both deviations are $o(N)$, for all sufficiently large $N$,
\[
    1\leq \TT_N^\alpha<\TT_N^\beta\leq N,
    \qquad
    \frac{\TT_N^\alpha}{N}\to\rho_M,
    \qquad
    \frac{\TT_N^\beta}{N}\to\rho_H.
\]

For $\theta\in\{L,M,H\}$, write
\[
    p^\alpha_{\theta,N}
    :=
    \mathbb{P}\!\left[
        \operatorname{Bin}(N-1,\rho_\theta)=\TT_N^\alpha-1
    \right],
    \qquad
    p^\beta_{\theta,N}
    :=
    \mathbb{P}\!\left[
        \operatorname{Bin}(N-1,\rho_\theta)=\TT_N^\beta-1
    \right].
\]
For $s\in\{g,b\}$, define
\[
    A_{s,N}
    :=
    \sum_{\theta\in\{L,M,H\}}
        q^0_\theta\mathbb{P}[s\mid\theta]p^\alpha_{\theta,N}U_S(\theta),
\]
and
\[
    B_{s,N}
    :=
    \sum_{\theta\in\{L,M,H\}}
        q^0_\theta\mathbb{P}[s\mid\theta]p^\beta_{\theta,N}U_S(\theta).
\]
By \cref{lem:10}, applied with $n=N-1$, $\rho=\rho_M$, and $z_n=h_N$, and then with $\rho=\rho_H$ and $z_n=k_N$,
\begin{equation}
\label{eqn:77}
    p^\alpha_{M,N}
    =
    \frac{1+o(1)}{\sigma_M\sqrt{2\pi N}}
    e^{-h_N^2/2},
    \qquad
    p^\beta_{H,N}
    =
    \frac{1+o(1)}{\sigma_H\sqrt{2\pi N}}
    e^{-k_N^2/2}.
\end{equation}
The replacement of $N-1$ by $N$ in the square-root factor is absorbed by $1+o(1)$. By \eqref{eqn:75}, all the other state terms in $A_{s,N}$ and $B_{s,N}$ are exponentially smaller, because the corresponding empirical frequencies converge to values different from the relevant state means. Hence
\begin{equation}
\label{eqn:78}
\begin{aligned}
    A_{g,N}
    &=
    -\frac{q^0_M\rho_M[-U_S(M)]}{\sigma_M\sqrt{2\pi N}}
    e^{-h_N^2/2}(1+o(1))<0,\\
    B_{g,N}
    &=
    \frac{q^0_H\rho_HU_S(H)}{\sigma_H\sqrt{2\pi N}}
    e^{-k_N^2/2}(1+o(1))>0.
\end{aligned}
\end{equation}
Moreover,
\begin{equation}
\label{eqn:79}
    \frac{A_{b,N}}{A_{g,N}}\to\frac{1-\rho_M}{\rho_M},
    \qquad
    \frac{B_{b,N}}{B_{g,N}}\to\frac{1-\rho_H}{\rho_H}.
\end{equation}
Indeed, replacing signal $g$ by signal $b$ changes the dominant $M$-term in $A_{s,N}$ by the factor $(1-\rho_M)/\rho_M$, and changes the dominant $H$-term in $B_{s,N}$ by the factor $(1-\rho_H)/\rho_H$.

By \eqref{eqn:78} and the definition of $k_N$,
\begin{equation}
\label{eqn:80}
    -\frac{A_{g,N}}{B_{g,N}}\to
    \begin{cases}
        0, &\text{if }\bar p=0,\\[0.3em]
        \displaystyle \frac{\bar p}{1-\bar p}, &\text{if }\bar p>0.
    \end{cases}
\end{equation}
For all sufficiently large $N$, define
\[
    \mu_N
    :=
    \frac{B_{g,N}}{B_{g,N}-A_{g,N}}
    \in(0,1).
\]
Then
\[
    \mu_NA_{g,N}+(1-\mu_N)B_{g,N}=0,
\]
and \eqref{eqn:80} implies
\[
    \mu_N\to 1-\bar p.
\]

Define the step mechanism
\begin{equation}
\label{eqn:81}
    \phi_N(t)
    =
    \begin{cases}
        0,
        & t<\TT_N^\alpha,\\
        \mu_N,
        & \TT_N^\alpha\leq t<\TT_N^\beta,\\
        1,
        & t\geq \TT_N^\beta.
    \end{cases}
\end{equation}
Given truthful reporting by all other senders, a sender changes the recommendation probability only in two pivotal events: when the other $N-1$ reports contain $\TT_N^\alpha-1$ reports of $g$, or when they contain $\TT_N^\beta-1$ reports of $g$. Hence the payoff gain from reporting $g$ rather than $b$ after signal $s$ is
\begin{equation}
\label{eqn:82}
    \Delta_{s,N}:=\mu_NA_{s,N}+(1-\mu_N)B_{s,N}.
\end{equation}
By construction, $\Delta_{g,N}=0$. Thus a sender with signal $g$ is willing to report $g$. For signal $b$, \eqref{eqn:79} and \eqref{eqn:82} give
\[
\begin{aligned}
    \Delta_{b,N}
    &=
    \mu_NA_{g,N}
    \left(\frac{1-\rho_M}{\rho_M}+o(1)\right)
    +(1-\mu_N)B_{g,N}
    \left(\frac{1-\rho_H}{\rho_H}+o(1)\right)\\
    &=
    \mu_NA_{g,N}
    \left(
        \frac{1-\rho_M}{\rho_M}
        -
        \frac{1-\rho_H}{\rho_H}
        +o(1)
    \right)
    <0.
\end{aligned}
\]
The last inequality follows because $A_{g,N}<0$ and $\rho\mapsto(1-\rho)/\rho$ is strictly decreasing, so
\[
    \frac{1-\rho_M}{\rho_M}>\frac{1-\rho_H}{\rho_H}.
\]
Therefore truthful reporting is incentive compatible.

We next verify the receiver's obedience constraints. Let
\[
    Q_{\theta,N}:=\mathbb{E}[\phi_N(T)\mid\theta],
\]
where $T\sim\operatorname{Bin}(N,\rho_\theta)$ is the total number of reports of $g$ under truthful reporting. Under state $L$, since $\TT_N^\alpha/N\to\rho_M>\rho_L$, the law of large numbers gives $Q_{L,N}\to0$. Under state $M$, the lower cutoff is $h_N\to\infty$ standard deviations below the $M$-mean up to an $O(1)$ term, while $\TT_N^\beta/N\to\rho_H>\rho_M$; hence
\[
    \mathbb{P}_M[T<\TT_N^\alpha]\to0,
    \qquad
    \mathbb{P}_M[T\geq\TT_N^\beta]\to0.
\]
Under state $H$, the upper cutoff is $k_N\to\infty$ standard deviations below the $H$-mean up to an $O(1)$ term, so $\mathbb{P}_H[T<\TT_N^\beta]\to0$. Therefore
\begin{equation}
\label{eqn:83}
    Q_{L,N}\to0,
    \qquad
    Q_{M,N}\to1-\bar p,
    \qquad
    Q_{H,N}\to1.
\end{equation}

For a proposal recommendation, the receiver's posterior expected payoff from choosing the proposal has numerator
\[
    \sum_{\theta\in\{L,M,H\}} q^0_\theta Q_{\theta,N}U_R(\theta),
\]
and the probability of a proposal recommendation is positive for all sufficiently large $N$. By \eqref{eqn:83}, this numerator converges to
\begin{equation}
\label{eqn:84}
    q^0_HU_R(H)+(1-\bar p)q^0_MU_R(M)>0.
\end{equation}
For a status-quo recommendation, the receiver is willing to follow the recommendation if her posterior expected payoff from deviating to the proposal is nonpositive. The relevant numerator is
\[
    \sum_{\theta\in\{L,M,H\}} q^0_\theta(1-Q_{\theta,N})U_R(\theta),
\]
and the probability of a status-quo recommendation is positive for all sufficiently large $N$. By \eqref{eqn:83}, this numerator converges to
\begin{equation}
\label{eqn:85}
    q^0_LU_R(L)+\bar p\,q^0_MU_R(M)<0,
\end{equation}
where the strict inequality is \eqref{eqn:76}. Thus, for all sufficiently large $N$, the receiver strictly follows both recommendations. Hence the mechanisms in \eqref{eqn:81} are feasible for all sufficiently large $N$ and satisfy
\[
    \mathbb{P}[\mathrm{proposal}\mid L;\phi_N]\to0,
    \qquad
    \mathbb{P}[\mathrm{proposal}\mid M;\phi_N]\to1-\bar p,
    \qquad
    \mathbb{P}[\mathrm{proposal}\mid H;\phi_N]\to1.
\]

For the finitely many small values of $N$ not covered by the construction, use a constant mechanism. If
\[
    \sum_{\theta\in\{L,M,H\}}q^0_\theta U_R(\theta)\geq0,
\]
set $\phi_N(t)\equiv1$; otherwise set $\phi_N(t)\equiv0$. Reports do not affect a constant recommendation, so sender incentive compatibility is immediate. The receiver follows the unique on-path recommendation by the choice of the constant. This finite-prefix patch does not affect the preceding limits.

It remains to obtain the desired value $p\in[0,p^*]$. If $p=0$, take $\bar p=0$ in the construction above. Since $q^0_LU_R(L)<0$, condition \eqref{eqn:76} holds.

If $p>0$, choose a sequence $\bar p_m\uparrow p$ with $\bar p_m<p$ for every $m$. Since $p\leq p^*$,
\[
    q^0_LU_R(L)+p\,q^0_MU_R(M)\leq0.
\]
Because $q^0_MU_R(M)>0$,
\[
    q^0_LU_R(L)+\bar p_m q^0_MU_R(M)<0
    \qquad\text{for every }m.
\]
For each $m$, the construction above gives a feasible sequence $\{\phi_N^m\}_{N>1}$ satisfying
\[
    \mathbb{P}[\mathrm{proposal}\mid L;\phi_N^m]\to0,
    \qquad
    \mathbb{P}[\mathrm{proposal}\mid M;\phi_N^m]\to1-\bar p_m,
    \qquad
    \mathbb{P}[\mathrm{proposal}\mid H;\phi_N^m]\to1.
\]
Choose a strictly increasing sequence of integers $N_m$ such that, for every $N\geq N_m$, all three probabilities for $\phi_N^m$ are within $1/m$ of their respective limits. Define
\[
    \phi_N:=\phi_N^m
    \qquad
    \text{whenever }N_m\leq N<N_{m+1}.
\]
For $N<N_1$, use the constant finite-prefix mechanism described above. Then every $\phi_N$ is feasible. As $N\to\infty$, the block index $m\to\infty$, and therefore
\[
    \mathbb{P}[\mathrm{proposal}\mid L;\phi_N]\to0,
    \qquad
    \mathbb{P}[\mathrm{proposal}\mid M;\phi_N]\to1-p,
    \qquad
    \mathbb{P}[\mathrm{proposal}\mid H;\phi_N]\to1.
\]
This proves \cref{thm:5}.
\end{proof}

\subsection{Proof of Proposition~\ref{prop:5}}

For each \(\lambda_L>0\), let \(\hat\lambda_M(\lambda_L)\) denote the threshold
\(\hat\lambda_M\) constructed in \cref{ap:d2} when the ratio \(q^0_L/q^0_H\) is fixed
at \(\lambda_L\). When a value of \(\lambda_L\) is fixed, we keep the notation
\(\hat q\) for the corresponding threshold in prior probabilities:
\begin{equation}
\label{eqn:b1}
    \hat q
    :=
    \frac{\hat\lambda_M(\lambda_L)}
    {1+\lambda_L+\hat\lambda_M(\lambda_L)} .
\end{equation}
This is the convention stated before \cref{prop:5}; no additional threshold is being
introduced.

We use the notation from \cref{ap:d2}:
\[
    a:=-\frac{U_S(L)}{U_S(H)},
    \qquad
    b:=-\frac{U_S(M)}{U_S(H)},
    \qquad
    r:=-\frac{U_R(L)}{U_R(H)},
    \qquad
    m:=\frac{U_R(M)}{U_R(H)}.
\]
Then \(a,b,r,m>0\), and
\[
    -\frac{U_R(L)}{U_R(M)}
    =
    \frac{r}{m}.
\]
Moreover, \eqref{eqn:30} is equivalent to
\[
    \frac{a}{r}<\frac{\rho_H}{\rho_L}.
\]

\begin{lemma}
\label{lem:b1}
Suppose that \eqref{eqn:30} holds. There exist constants \(\bar\lambda>0\) and
\(C>0\) such that, for every \(\lambda_L\in(0,\bar\lambda)\),
\begin{equation}
\label{eqn:b2}
    \hat\lambda_M(\lambda_L)
    =
    C\lambda_L^\alpha,
    \qquad
    \alpha:=\frac{\rho_H-\rho_M}{\rho_H-\rho_L}\in(0,1).
\end{equation}
In particular, \(\hat\lambda_M\) is strictly concave near zero, and
\begin{equation}
\label{eqn:b3}
    \lim_{\lambda_L\downarrow0}
    \frac{\hat\lambda_M(\lambda_L)}{\lambda_L}
    =
    +\infty .
\end{equation}
\end{lemma}

\begin{proof}
Since \(0<\rho_L<\rho_M<\rho_H<1\),
\[
    \alpha:=\frac{\rho_H-\rho_M}{\rho_H-\rho_L}
\]
satisfies \(\alpha\in(0,1)\).

We first verify explicitly why \eqref{eqn:30} implies the condition \(z^\ast\in(0,1)\)
used in \cref{ap:d2}. Recall that
\[
    z^\ast
    =
    \frac{
        1+\frac{\rho_H}{\rho_M}\cdot\frac{m}{b}
    }{
        \frac{\rho_H}{\rho_M}\cdot\frac{m}{b}
        +
        \frac{\rho_H}{\rho_L}\cdot\frac{r}{a}
    }.
\]
All terms are positive, so \(z^\ast>0\). Moreover, \(z^\ast<1\) holds if and only if
\[
    1+\frac{\rho_H}{\rho_M}\cdot\frac{m}{b}
    <
    \frac{\rho_H}{\rho_M}\cdot\frac{m}{b}
    +
    \frac{\rho_H}{\rho_L}\cdot\frac{r}{a},
\]
which is equivalent to
\[
    \frac{a}{r}<\frac{\rho_H}{\rho_L}.
\]
Thus \(z^\ast\in(0,1)\) under \eqref{eqn:30}.

For each \(k\geq1\), recall from \cref{ap:d2} that
\[
    z_k^0(\lambda_L)
    =
    \lambda_L a
    \left(\frac{\rho_L}{\rho_H}\right)^k,
    \qquad
    \bar z_k(\lambda_L)
    =
    \max\{z^\ast,z_k^0(\lambda_L)\}.
\]
Choose
\[
    \bar\lambda
    :=
    \frac{z^\ast}
    {a\left(\frac{\rho_L}{\rho_H}\right)}
    >0.
\]
If \(\lambda_L\in(0,\bar\lambda)\), then, because
\(0<\rho_L/\rho_H<1\),
\[
    z_k^0(\lambda_L)
    =
    \lambda_L a
    \left(\frac{\rho_L}{\rho_H}\right)^k
    \leq
    \lambda_L a
    \left(\frac{\rho_L}{\rho_H}\right)
    <z^\ast
    \qquad
    \text{for every } k\geq1.
\]
Hence \(\bar z_k(\lambda_L)=z^\ast\) for every \(k\geq1\). By \eqref{eqn:50},
\[
\begin{aligned}
    \hat\lambda_{M,k}(\lambda_L)
    &=
    \frac{1-z^\ast}
    {b\left(\frac{\rho_M}{\rho_H}\right)^k}
    \left(
        \frac{
        \lambda_L a\left(\frac{\rho_L}{\rho_H}\right)^k
        }
        {z^\ast}
    \right)^\alpha  \\
    &=
    C_0
    \left[
        \left(\frac{\rho_L}{\rho_H}\right)^\alpha
        \frac{\rho_H}{\rho_M}
    \right]^k
    \lambda_L^\alpha,
\end{aligned}
\]
where
\[
    C_0
    :=
    \frac{1-z^\ast}{b}
    \left(\frac{a}{z^\ast}\right)^\alpha
    >0.
\]

It remains to show that the geometric factor is below one. Let
\[
    B:=
    \left(\frac{\rho_L}{\rho_H}\right)^\alpha
    \frac{\rho_H}{\rho_M}.
\]
The inequality \(B<1\) is equivalent to
\[
    \alpha\log\rho_L+(1-\alpha)\log\rho_H<\log\rho_M.
\]
Since
\[
    \rho_M=\alpha\rho_L+(1-\alpha)\rho_H,
\]
and since \(\log\) is strictly concave on \((0,\infty)\), we have
\[
    \alpha\log\rho_L+(1-\alpha)\log\rho_H
    <
    \log\bigl(\alpha\rho_L+(1-\alpha)\rho_H\bigr)
    =
    \log\rho_M.
\]
Thus \(B<1\).

Therefore, for every \(\lambda_L\in(0,\bar\lambda)\),
\[
    \hat\lambda_M(\lambda_L)
    =
    \sup_{k\geq1}\hat\lambda_{M,k}(\lambda_L)
    =
    C_0B\lambda_L^\alpha.
\]
Setting
\[
    C:=C_0B>0
\]
gives \eqref{eqn:b2}. Since \(\alpha\in(0,1)\), the map
\(\lambda_L\mapsto C\lambda_L^\alpha\) is strictly concave on
\((0,\bar\lambda)\). Moreover,
\[
    \frac{\hat\lambda_M(\lambda_L)}{\lambda_L}
    =
    C\lambda_L^{\alpha-1}
    \to+\infty
    \qquad
    \text{as }\lambda_L\downarrow0,
\]
which proves \eqref{eqn:b3}.
\end{proof}

We now prove \cref{prop:5}. By \cref{lem:b1}, choose
\(\lambda_L\in(0,\bar\lambda)\) small enough that
\[
    \frac{r}{m}\lambda_L
    <
    \hat\lambda_M(\lambda_L).
\]
Then choose \(\lambda_M\) satisfying
\begin{equation}
\label{eqn:b4}
    \frac{r}{m}\lambda_L
    <
    \lambda_M
    <
    \hat\lambda_M(\lambda_L).
\end{equation}
Define
\begin{equation}
\label{eqn:b5}
    q^0_L
    :=
    \frac{\lambda_L}{1+\lambda_L+\lambda_M},
    \qquad
    q^0_M
    :=
    \frac{\lambda_M}{1+\lambda_L+\lambda_M},
    \qquad
    q^0_H
    :=
    \frac{1}{1+\lambda_L+\lambda_M}.
\end{equation}
Then \(q^0=(q^0_L,q^0_M,q^0_H)\in\Delta^3\), and all three prior probabilities are
strictly positive.

The left inequality in \eqref{eqn:b4} implies
\[
    q^0_M
    =
    \lambda_M q^0_H
    >
    \frac{r}{m}\lambda_L q^0_H
    =
    \frac{r}{m}q^0_L
    =
    -\frac{U_R(L)}{U_R(M)}q^0_L.
\]
Thus the lower bound in \cref{prop:5} holds.

It remains to verify the upper bound. For the prior in \eqref{eqn:b5},
\[
    \frac{q^0_L}{q^0_H}=\lambda_L.
\]
By the convention stated before \cref{prop:5}, the symbol \(\hat q\) in the upper bound
denotes the threshold associated with this value of \(\lambda_L\), namely the number in
\eqref{eqn:b1}. For fixed \(\lambda_L\), the map
\[
    \lambda
    \mapsto
    \frac{\lambda}{1+\lambda_L+\lambda}
\]
is strictly increasing because
\[
    \frac{d}{d\lambda}
    \left(
        \frac{\lambda}{1+\lambda_L+\lambda}
    \right)
    =
    \frac{1+\lambda_L}{(1+\lambda_L+\lambda)^2}
    >0.
\]
Therefore, the right inequality in \eqref{eqn:b4} gives
\[
    q^0_M
    =
    \frac{\lambda_M}{1+\lambda_L+\lambda_M}
    <
    \frac{\hat\lambda_M(\lambda_L)}
    {1+\lambda_L+\hat\lambda_M(\lambda_L)}
    =
    \hat q.
\]
Hence \(q^0\) satisfies both inequalities in \cref{prop:5}. The set in
\cref{prop:5} is therefore nonempty.

\end{appendix}

\end{document}